\documentclass[fleqn,usenatbib]{mnras}
\usepackage{newtxtext,newtxmath}
\usepackage[T1]{fontenc}
\usepackage{mathtools}
\usepackage{cuted}

\title[The Aswan cliff collapse on Comet 67P]{Cliff collapse on Comet 67P/Churyumov--Gerasimenko -- I. Aswan}

\author[Bj\"{o}rn J. R. Davidsson]{
Bj\"{o}rn J. R. Davidsson,$^{1}$\thanks{E-mail: bjorn.davidsson@jpl.nasa.gov}
\\
$^{1}$Jet Propulsion Laboratory, California Institute of Technology,  M/S 183--601, 4800 Oak Grove Drive, Pasadena, CA 91109, USA
}

\date{Accepted 2023 October 2. Received 2023 October 2; in original form 2023 August 29}

\pubyear{2023}

\begin{document}
\label{firstpage}
\pagerange{\pageref{firstpage}--\pageref{lastpage}}
\maketitle

%
\begin{abstract}
The Aswan cliff on Comet 67P/Churyumov--Gerasimenko collapsed on 2015 July 10. Thereby, relatively pristine comet material from a depth of $\sim 12\,\mathrm{m}$ was 
exposed at the surface. Observations of the collapse site by the microwave instrument \emph{Rosetta}/MIRO have been retrieved from 8 months prior to collapse, as well as from 5, 7, and 11 months post--collapse. 
The MIRO data are analysed with thermophysical and radiative transfer models. The pre--collapse observations are consistent with a 30 MKS thermal inertia dust mantle with a 
thickness of at least $3\,\mathrm{cm}$. The post--collapse data are consistent with:  1) a dust/water--ice mass ratio of $0.9\pm 0.5$ and a molar $\mathrm{CO_2}$ abundance of $\sim 30$ 
per cent relative to water; 2) formation of a dust mantle after $\sim 7$ months, having a thickness of a few millimetres or a fraction thereof; 3) a $\mathrm{CO_2}$ ice sublimation 
front at 0.4 cm that withdrew to 2.0 cm and later to $20\pm 6\,\mathrm{cm}$; 4) a thermal inertia ranging 10--45 MKS; 5) a gas diffusivity that decreased from $0.1\,\mathrm{m^2\,s^{-1}}$ to 
$0.001\,\mathrm{m^2\,s^{-1}}$; 6) presence of a solid--state greenhouse effect parts of the time. The data and the analysis provide a first empirical glimpse of how ice--rich cometary 
material ages and evolves when exposed to solar heating.
\end{abstract}


\section{Introduction} \label{sec_intro}

Comet nuclei are rich in ices, as evidenced by their copious production of volatiles such as $\mathrm{H_2O}$, $\mathrm{CO_2}$, and CO near 
perihelion \citep[e.~g.,][]{bockeleemorvanetal04,ahearnetal12,luspaykutietal15,hansenetal16}. Yet, spacecraft missions have demonstrated that 
the nucleus surfaces of Comets 1P/Halley \citep{kelleretal86}, 19P/Borrelly \citep{soderblometal02}, 81P/Wild~2  \citep{brownleeetal04}, 9P/Tempel~1 \citep{ahearnetal05b}, 
103P/Hartley~2 \citep{ahearnetal11}, and 67P/Churyumov--Gerasimenko \citep[hereafter 67P;][]{sierksetal15} consist of a dark (geometric albedo $0.04\pm 0.02$) mixture of 
organics and silicates \citep{capaccionietal15,quiricoetal16,mennellaetal20}, with rare exposures of $\mathrm{H_2O}$ ice \citep{sunshineetal06a,baruccietal16,filacchioneetal16c,oklay_2016b} 
and even less common $\mathrm{CO_2}$ ice \citep{filacchioneetal16}.  Clearly, the volatile component of comet nucleus material is hidden from view under a desiccated layer called 
the dust mantle. 

There is evidence to suggest that the dust mantle is thin, and that the material underneath is ice--rich. While bouncing on the surface during landing on 67P, the \emph{Philae} 
probe cracked open the top $\sim 0.25\,\mathrm{m}$ of a boulder and revealed a bright (normal albedo $\sim 0.5$) interior with an estimated dust/water--ice mass ratio $\mu=2.3_{-0.16}^{+0.2}$ 
\citep{orourkeetal20}. Collimated dust features are ubiquitous in the 67P coma, and their switch--off within $\sim 1\,\mathrm{h}$ after sunset is consistent with a source of 
water--driven activity at a depth of $0.6\,\mathrm{cm}$ \citep{shietal16}. Furthermore, thermophysical model reproduction of the $\mathrm{H_2O}$ and $\mathrm{CO_2}$ production rate 
curves of 67P show that the dust mantle thickness is typically $\sim 2\,\mathrm{cm}$ and that $1\stackrel{<}{_{\sim}}\mu\stackrel{<}{_{\sim}} 2$ \citep{davidssonetal22}.

Gaining access to the ice--rich material underneath the dust mantle is clearly a scientific priority. Being able to characterise the less evolved material under the substantially 
more processed dust mantle is necessary to understand comet formation and the chemical composition of the Solar nebula. An opportunity for such a study presented itself 
unexpectedly, when a $81\times 65\,\mathrm{m^2}$ segment of the Aswan cliff on 67P \citep[for region definitions and names see][]{thomasetal18} crumbled on 2015 July 10, and abruptly exposed the bright (normal albedo $>0.4$) interior that 
previously had been located $\sim 12\,\mathrm{m}$ below the surface \citep{pajolaetal17}. 

\begin{figure*}
\centering
\begin{tabular}{cc}
\scalebox{0.262}{\includegraphics{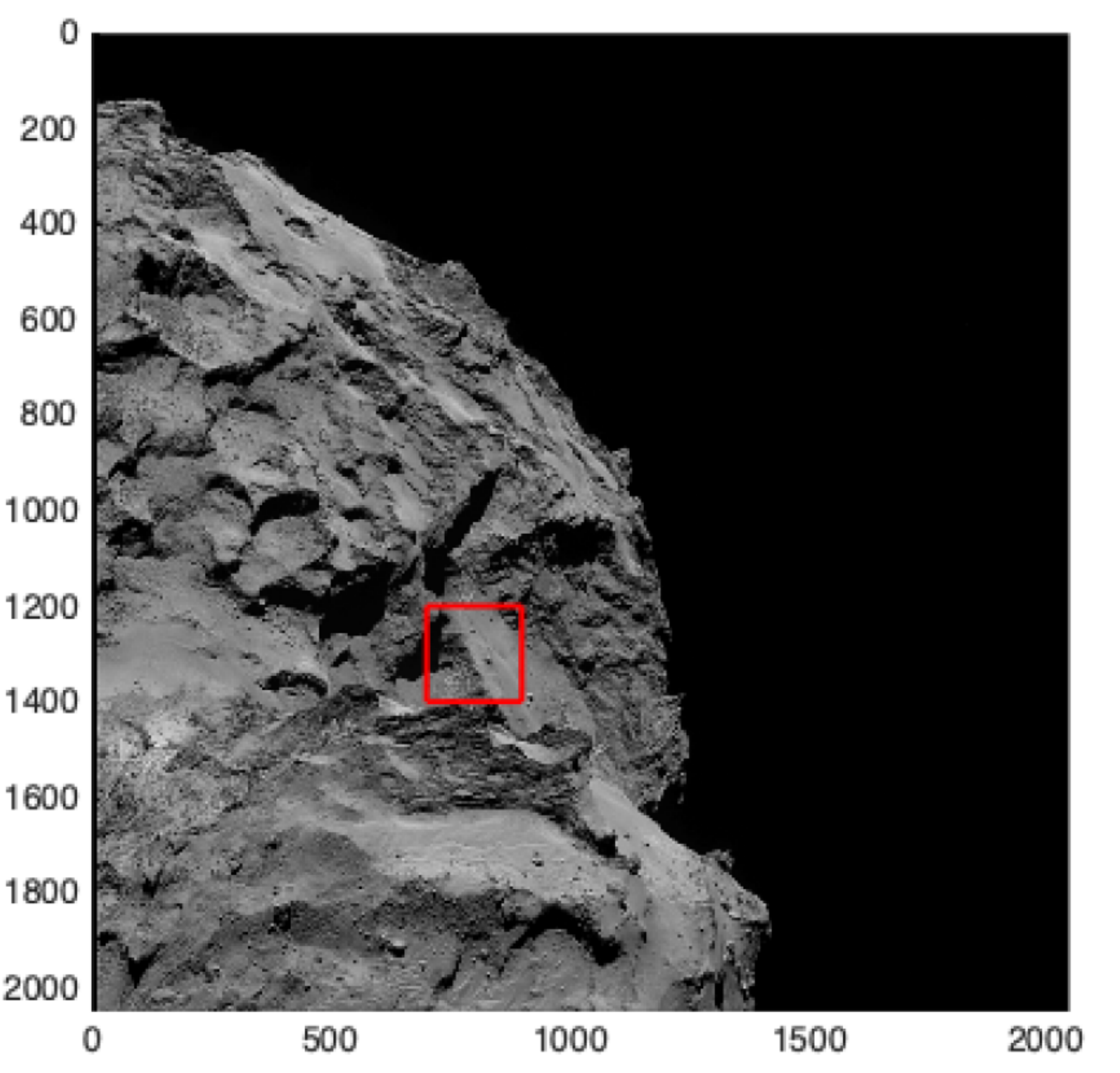}} & \scalebox{0.3}{\includegraphics{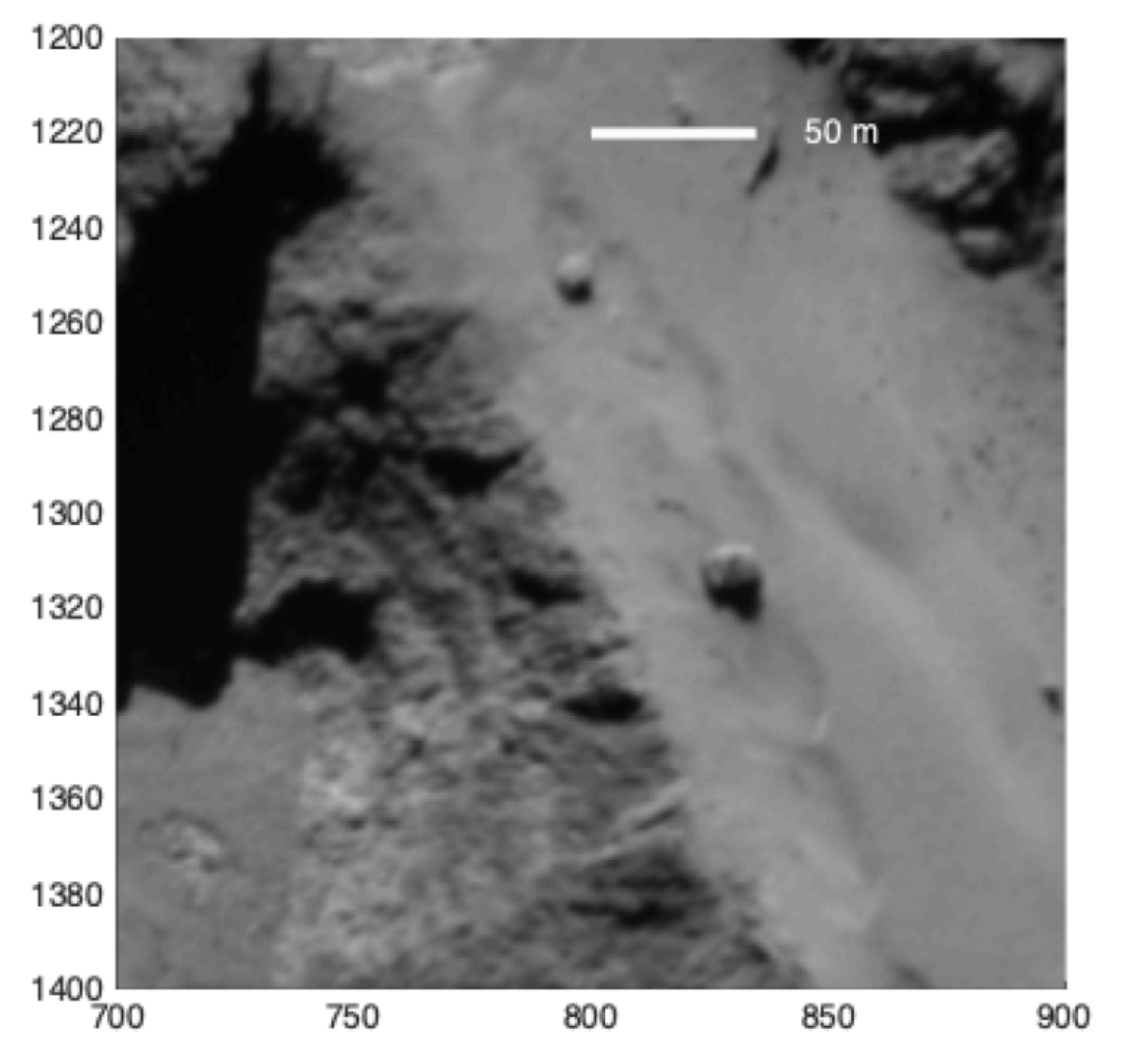}}\\
\end{tabular}
     \caption{\emph{Left:} Context image of the nucleus of Comet 67P showing parts of the Aswan cliff and plateau at the red square. The big lobe is in the background, the smooth Hapi valley is to the centre--left, 
and the small lobe is in the foreground. \emph{Right:} a close--up of the region within the red square in the left panel, showing the Aswan cliff wall prior to its collapse (note the smooth airfall material covering 
the horizontal plateau above the Aswan cliff, and the large boulders). Both panels show image MTP014/20150320t011747593id4df22.img \protect\citep{sierksetal18a} that was acquired on 2015 March 20 with $\sim 1.4\,\mathrm{m\,px^{-1}}$ resolution when \emph{Rosetta} was $80.7\,\mathrm{km}$ from the comet. Axis labels show pixel ID numbers.}
     \label{fig_precollapse}
\end{figure*}

\begin{figure*}
\centering
\begin{tabular}{cc}
\scalebox{0.352}{\includegraphics{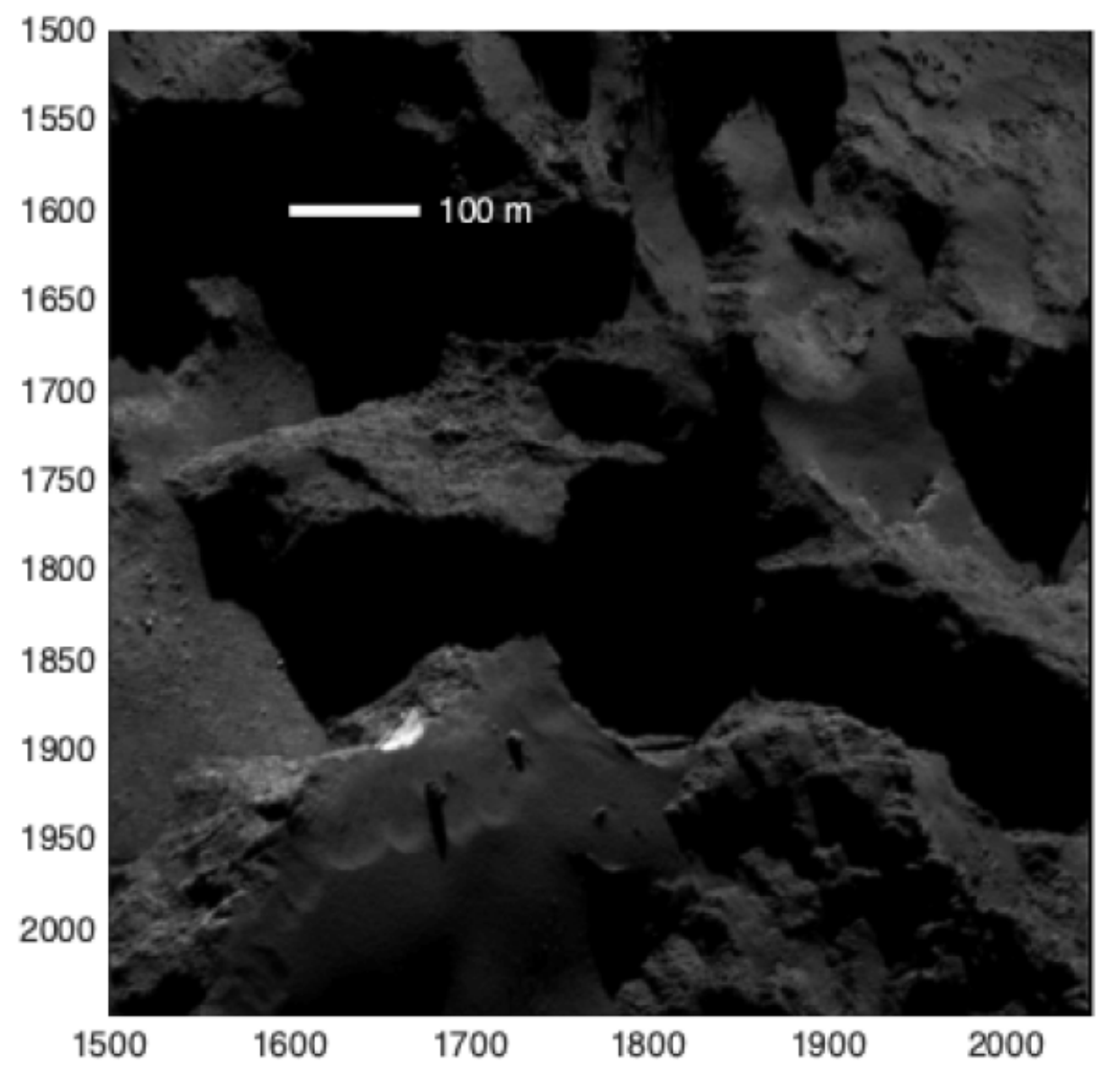}} & \scalebox{0.35}{\includegraphics{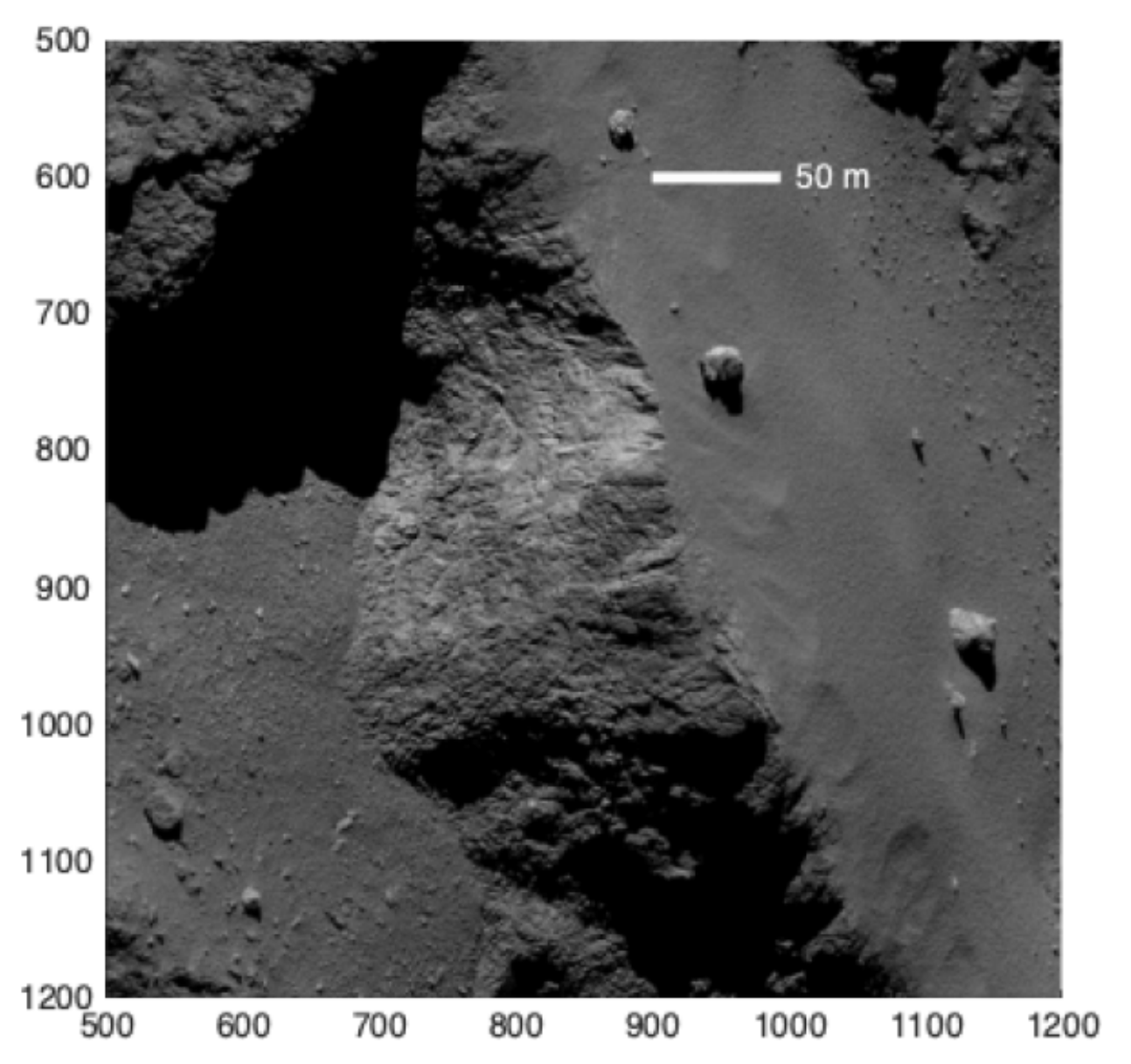}}\\
\end{tabular}
     \caption{\emph{Left:} This image shows an oblique view of the Aswan collapse site. Five months after the collapse it clearly stands out as a substantially brighter region. 
The normal albedo is $0.17\pm 0.01$ \protect\citep{pajolaetal17} and the current modelling suggests that dust and water ice near the intrinsic nucleus mixing ratio was exposed at this 
time (but no $\mathrm{CO_2}$ ice). Image MTP024~/~n20151226t170504370id4df22.img \protect\citep{sierksetal18b} was acquired on 2015 December 26 with $\sim 1.4\,\mathrm{m\,px^{-1}}$ resolution when \emph{Rosetta} 
was $77.0\,\mathrm{km}$ from the comet. \emph{Right:} This is a face--on view of the cliff showing that the brightness of the scar is significantly reduced, yet somewhat higher than for the surroundings 
eleven months after collapse. At this point the current modelling suggests a $\sim 0.5\,\mathrm{mm}$ dust mantle, i.~e., so thin that some reflection of the water ice still might reach the surface. Image MTP030~/~n20160608t143426745id4df22.img \protect\citep{sierksetal18c} was acquired on 2016 June 8 with $\sim 0.5\,\mathrm{m\,px^{-1}}$ resolution when \emph{Rosetta} was $29.7\,\mathrm{km}$ from the comet.}
     \label{fig_postcollapse}
\end{figure*}

\begin{figure}
\scalebox{0.4}{\includegraphics{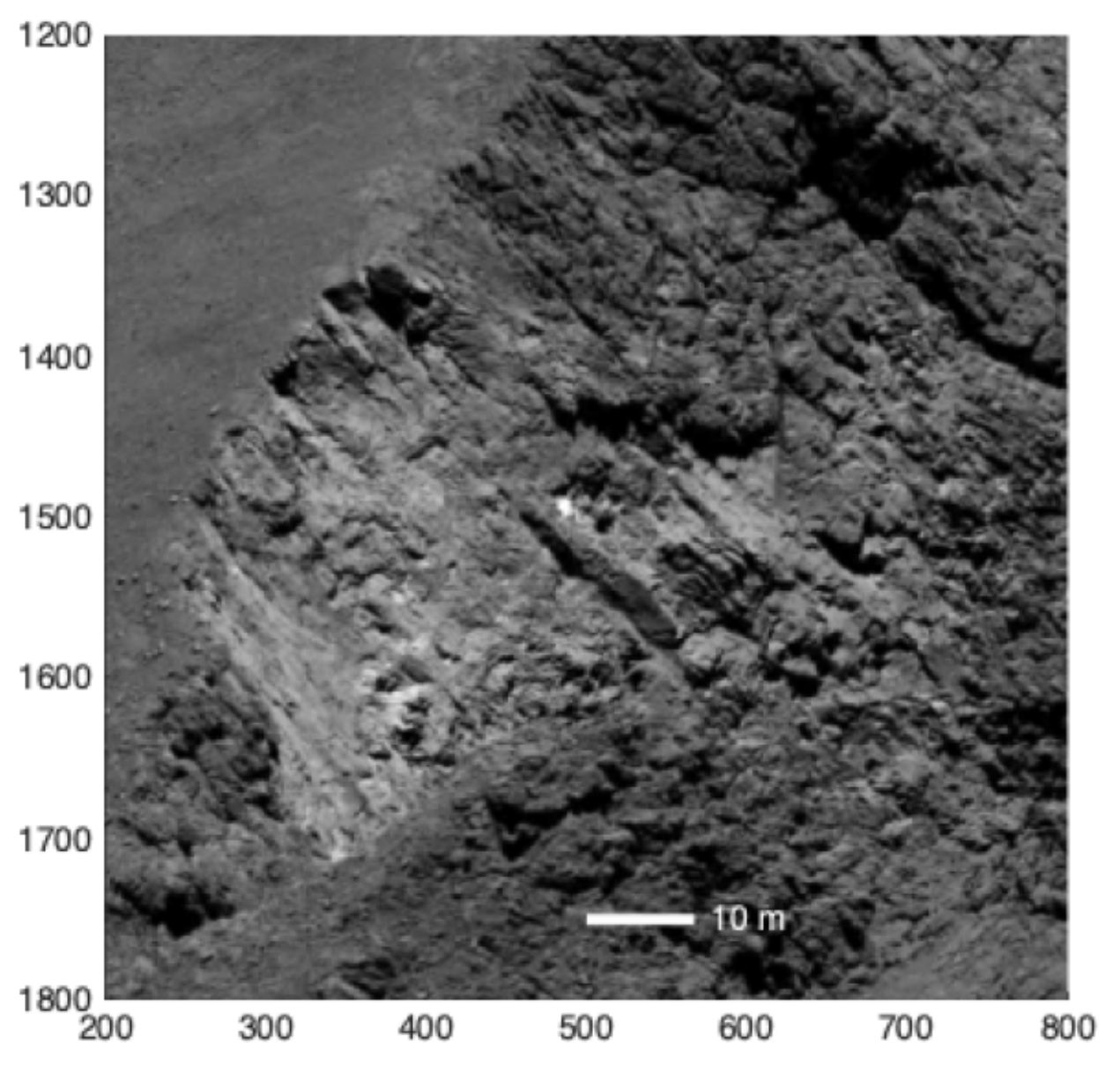}}
     \caption{This is arguably the best existing view of the Aswan collapse site, 13 months after the event, though some foreground topography obscures parts of the region. 
\protect\citet{pajolaetal17} report a normal albedo $\sim 0.11$ at this point, though some particularly bright spots still reached $\sim 0.18$. Image MTP032~/~n20160806t070848224id4ff22.img 
\protect\citep{sierksetal18d} was acquired on 2016 August 6 with $\sim 0.1\,\mathrm{m\,px^{-1}}$ resolution when \emph{Rosetta} was $8.4\,\mathrm{km}$ from the comet.}
     \label{fig_poststudy}
\end{figure}

Figure~\ref{fig_precollapse} shows \emph{Rosetta}/OSIRIS \citep{kelleretal07} images (downloaded from the NASA Planetary Data System, 
PDS\footnote{https://pds--smallbodies.astro.umd.edu/data\_sb/missions/rosetta/\\index\_OSIRIS.shtml}) of the location and pre--collapse appearance of the Aswan cliff. 
Figure~\ref{fig_postcollapse} (left panel) illustrates that the site remained a bright beacon in an otherwise dark landscape in the last days of 2015, almost half a year after the collapse. 
However, 11 months after the event the contrast in brightness between the collapse site and its surroundings is much reduced (Fig.~\ref{fig_postcollapse}, right panel), and another 
two months later the cliff is essentially back to normal (Fig.~\ref{fig_poststudy}).

Thus far, I am not aware of any publication except \citet{pajolaetal17} that has attempted to characterise the Aswan cliff collapse site. That paper focuses on OSIRIS imaging and the 
derived morphometric and photometric properties of the region. I here assemble observations of Aswan acquired by the Microwave Instrument for Rosetta Orbiter \citep[MIRO;][]{gulkisetal07} 
before (2014 November) and after (2015 November/December, 2016 February, 2016 June) the cliff collapse. Thermophysical and radiative transfer models are used to analyse the 
MIRO data and to place constraints on the composition, stratification, thermal inertia, gas diffusivity, opacity, extinction coefficients, and single--scattering albedo of the exposed 
material, and how they evolved with time. 

Importantly, this study provides a first account of the dust mantle formation time--scale and growth rate, as well as of the gradual withdrawal of the $\mathrm{CO_2}$ sublimation 
front in a comet nucleus, following the exposure of relatively pristine cometary material to solar heating. Such empirical input is needed in order to better understand comet activity, 
particularly in objects that have widespread exposure of fresh surfaces, such as recently fragmented or split nuclei, and potentially centaurs and/or dynamically new comets.

For previous work on analysing MIRO data, see e.~g., \citet{gulkisetal15}, \citet{schloerbetal15}, \citet{choukrounetal15},
\citet{leeetal15}, \citet{biveretal19}, \citet{marshalletal18}, \citet{rezacetal19,rezacetal21}, and \citet{davidssonetal22b}.

\section{Methodology} \label{sec_method}

The methods applied in the current paper are very similar to those of \citet{davidssonetal22b}. The applied models have been 
described in detail by \citet{davidsson21} and \citet{davidssonetal21,davidssonetal22b}. Therefore, the information provided in this  
section is kept to a minimum, in order to avoid unnecessary repetition.

\subsection{Overall approach} \label{sec_method_approach}

Millimetre (MM; wavelength $\lambda=1.594\,\mathrm{mm}$) and sub--millimetre (SMM; $\lambda=0.533\,\mathrm{mm}$)  antenna temperature diurnal curves are built by 
assembling MIRO observations of the Aswan collapse site acquired during a few weeks but time--shifted to a common 'master period'. The goal is then to reproduce these 
empirical curves with synthetic ones obtained by modelling the near--surface physical temperature as function of time and depth for different types of media and calculating the 
corresponding microwave thermal radiation emitted towards MIRO, expressed as antenna temperatures. Specifically, this includes: 1) calculating the illumination conditions at Aswan during nucleus 
rotation and orbital motion, accounting for shadowing and self--heating effects caused by the irregular nucleus (see section~\ref{sec_method_illum}); 
2) feeding the illumination conditions to a thermophysical model, here the Basic Thermophysical Model \citep[\textsc{btm};][see section~\ref{sec_method_BTM}]{davidssonetal21} or 
the more advanced Numerical Icy Minor Body evolUtion Simulator \citep[\textsc{nimbus};][see section~\ref{sec_method_NIMBUS}]{davidsson21} that provide the corresponding 
physical temperature profiles; 3) feeding the physical temperature profiles to the radiative transfer code \textsc{themis} (see section~\ref{sec_method_THEMIS}) 
that calculates the emitted microwave irradiance and the corresponding MM and SMM antenna temperatures. The model medium resulting in simultaneous 
MM and SMM antenna temperature curve reproduction is considered similar to that of the Aswan region itself. 

The \textsc{btm} and \textsc{nimbus} consider media with different levels of structural, physical, and compositional complexity, each regulated by 
a given set of physical parameters. When attempting to reproduce a certain empirical diurnal curve, the most simple medium is considered first, 
only moving on to more elaborate media if no satisfactory solutions can be found. The starting point is always an ice--free medium consisting of 
porous dust.  \textsc{btm} necessarily considers fixed densities $\rho$, specific heat capacities $c$ and solid--state conductivities $\kappa$. 
\textsc{nimbus} nominally considers $c$ and $\kappa$ functions of temperature, additionally includes radiative heat conductivity, and may treat 
$\rho$ as a function of depth. These parameters are typically reported in terms of values or ranges of the thermal inertia $\Gamma=\sqrt{\rho c\kappa}$ 
(media with the same $\Gamma$ have the same surface temperature curve, regardless of individual heat capacity and heat conductivity combinations). Optically, all media are 
characterised by their microwave extinction coefficients ($E_{\rm MM}$ and $E_{\rm SMM}$) and single--scattering albedos ($w_{\rm MM}$ and $w_{\rm SMM}$). 
However, $w_{\rm MM}=0$ even for rocky material \citep{garykeihm78} and $w_{\rm SMM}=0$ is nominally assumed as well, unless SMM scattering (suggestive of millimetre--scale 
lumpiness) is required to fit data \citep{davidssonetal22b}. If no thermal inertia can be found, that yield simultaneous MM and SMM fits for some combination 
of $E_{\rm MM}$, $E_{\rm SMM}$, and $w_{\rm SMM}$ it means that a refractories--only material is not a good analogue of that region on the comet.

The next step is therefore to introduce water ice. The \textsc{btm} necessarily assumes surface water ice, while \textsc{nimbus} considers water ice a 
finite resource underneath a potentially eroding dust mantle. \textsc{btm} considers sublimation cooling only, while \textsc{nimbus} includes sublimation 
cooling, energy and mass transport by vapour diffusion, and recondensation heating. The free \textsc{btm} parameters are now thermal inertia $\Gamma$ and 
volumetric ice fraction $f_{\rm i}$. For \textsc{nimbus} the parameters are thermal inertia, dust/water--ice mass fraction $\mu$, the dust mantle thickness $h_{\rm m}$ 
(all typically time--dependent) and diffusivity modelled by the length $L_{\rm p}$, radius $r_{\rm p}$, and tortuosity $\xi$ of pore spaces \citep[see equation~46 in][]{davidsson21}.   

If the nucleus is not heated sufficiently at any point during the day, the addition of water ice may not have a significant effect because its sublimation 
cooling is too weak. Yet, the data may suggest that efficient cooling is necessary. If so, a more volatile substance is required. Carbon dioxide is second to water in terms of abundance and volatility 
\citep{yamamoto85,bockeleemorvanetal04,rubinetal20} and is therefore the obvious candidate. \textsc{btm} does not include $\mathrm{CO_2}$ but \textsc{nimbus} does. 
Additional free parameters are the molar $\mathrm{CO_2/H_2O}$ abundance and the depth of the sublimation front of this supervolatile.

\subsection{Illumination conditions} \label{sec_method_illum}

The current work applies the nucleus shape model SHAP5 version 1.5 \citep{jordaetal16}, degraded from $3.1\cdot 10^6$ to $5\cdot 10^4$ facets. 
On this model, the Aswan collapse site is represented by 17 facets. The illumination conditions, including shadowing and self--heating due to the 
complex nucleus topography, are calculated for each of the 17 facets. This is done every $20\,\mathrm{min}$ (roughly $10^{\circ}$ of nucleus 
rotation) throughout the orbital arc stretching from aphelion (2012 May 23) up to the end of 2016 June. All 17 flux curves have similar shapes and were mutually 
in--phase. The vertical flux dispersion is at most $\pm 15\%$. The curves were deemed so similar that an area--weighted average flux was applied for a single 
representative facet, having its normal vector close to that of the average for the 17 facets. At any given moment, the thermophysical codes 
are fed with that area--weighted average flux. This simplification is further warranted because the detailed geometry of the region changed as 
a result of the collapse, i.~e., the shape model representation of the region is approximate to begin with. 

Shadowing and self--heating are calculated with the \citet{davidssonandrickman14} model code, which was validated against an independent 
implementation by \citet{gutierrezetal01}. Because of the huge workload, two time--saving simplifications had to be made with respect to the 
original code: 1) sunlit facets that are visible from the collapse site are set to the radiation equilibrium temperature, i.~e., a zero thermal inertia 
is assumed; 2) shadowed facets that are in the field of view from the collapse site are set to $0\,\mathrm{K}$, i.~e., they do not contribute to the self--heating of Aswan.

The level of error introduced by assuming $\Gamma=0$ for the surroundings when calculating self--heating at Aswan can be exemplified as follows. 
The self--heating flux primarily originates from the large Hathor cliff on the small lobe that faces Aswan. That is because Hathor typically is illuminated 
when Aswan lies in darkness. On 2014 November 9, the diurnal peak infrared flux emitted from Hathor was found to be 10 per cent lower when applying $\Gamma=30\,\mathrm{MKS}$ 
compared to the nominal $\Gamma=0$ assumption. That translates to a 10 per cent reduction of the self--heating calculated at Aswan (i.~e., a lowering from 
the typical self--heating flux of $\sim 20\,\mathrm{J\,m^{-2}\,s^{-1}}$ to about $\sim 18\,\mathrm{J\,m^{-2}\,s^{-1}}$).

To better understand the self--heating contributions from shadowed facets, consider the effect of re--setting their temperatures from the nominal $0\,\mathrm{K}$ to 
$130\,\mathrm{K}$ (a typical nighttime temperature of dust mantle material). A surface at $130\,\mathrm{K}$ emits $16\,\mathrm{J\,m^{-2}\,s^{-1}}$. Seen from the 
Aswan location, about half of the upper hemisphere is sky while the other half is nucleus. If all nucleus area is in darkness, there would be an additional $\sim 8\,\mathrm{J\,m^{-2}\,s^{-1}}$ 
self--heating flux that currently is not accounted for. However, most of the time a fraction of the nucleus visible from Aswan is illuminated (providing the $\sim 20\,\mathrm{J\,m^{-2}\,s^{-1}}$ flux 
that is accounted for). Typically, no more than 2/3 of the nucleus in the field of view would be in darkness, reducing the unaccounted flux to $\sim 5\,\mathrm{J\,m^{-2}\,s^{-1}}$. 
This flux fully compensates for the loss due to the thermal inertia assumption, and suggest that the applied self--heating flux is too small by typically $\sim 3\,\mathrm{J\,m^{-2}\,s^{-1}}$. 
Both approximations therefore have insignificant influence on the results.

\subsection{The basic thermophysical model \textsc{btm}} \label{sec_method_BTM}

The governing equation of the \textsc{btm} \citep[see][]{davidssonetal21} is the spatially one--dimensional heat conduction equation. Its upper boundary 
condition balances absorbed solar radiation, thermal emission, heat conduction, and energy consumption by sublimating water ice located on the surface. 
The lower boundary condition is a vanishing temperature gradient. The differential equation is solved numerically, using a finite element method. 

The \textsc{btm} is suitable to study non--sublimating media (i.~e., thick dust mantles), or sublimating media where the water ice is 
located sufficiently close to the surface to control its temperature. It assumes that the heat conductivity and specific heat capacity are 
constant. The thermal inertia $\Gamma$ and the volumetric fraction of water ice $f_{\rm i}$ are the only free parameters in this model.

\subsection{The advanced thermophysical model \textsc{nimbus}} \label{sec_method_NIMBUS}

\textsc{nimbus} \citep{davidsson21} considers a porous medium consisting of dust, $\mathrm{H_2O}$ ice (that can have amorphous, cubic, or hexagonal structure), 
$\mathrm{CO_2}$ ice, and CO ice. All water phases are capable of storing $\mathrm{CO_2}$ and CO, and CO can also be stored in $\mathrm{CO_2}$ ice. A number of 
phase transitions are considered: segregation of $\mathrm{CO_2:CO}$ mixtures that release CO vapour from $\mathrm{CO_2}$ ice; crystallisation of 
amorphous water ice that forms cubic water ice; transformation of cubic water ice into hexagonal water ice; sublimation of CO, $\mathrm{CO_2}$, and hexagonal $\mathrm{H_2O}$ 
ices; condensation of CO, $\mathrm{CO_2}$, and $\mathrm{H_2O}$ vapours (ice formation). Any of the three water ice phase transitions results in partial or 
full release of the trapped $\mathrm{CO_2}$ and CO. The phase transitions proceed at rates that are strongly temperature--dependent (additionally, sublimation and condensation 
rates depend on the partial vapour pressure of the species in question). The current study includes a subset of species available in \textsc{nimbus}: dust, hexagonal (crystalline) $\mathrm{H_2O}$ ice, 
and $\mathrm{CO_2}$ ice. 

Energy sources include absorbed solar radiation, radioactive decay (not considered in the current work), the exothermic crystallisation process, and release of latent heat 
during vapour condensation into ice. Energy sinks include $\mathrm{CO_2:CO}$ segregation, sublimation of ices into vapour, and thermal radiative loss of energy to space. 
Energy is transported within the body by solid--state and radiative conduction, and by diffusing vapour (advection). The mass flux rate due to gas diffusion depends on local 
temperature and vapour pressure gradients, along with geometric parameters that regulate diffusivity. Inflow or outflow of vapour causes transient deviations from 
the local temperature--dependent vapour saturation pressure. Excess pressure causes condensation, while pressure deficiency triggers sublimation. These processes 
are governed by the energy conservation equation and the mass conservation equations for solids and vapours. The core task of \textsc{nimbus} is to solve this 
system of coupled differential equations numerically, thereby providing temperatures, pressures, and masses of ices and gases of all types, as functions of spatial position and time. 

Most physical properties in \textsc{nimbus} are temperature--dependent and species--specific, and are based on laboratory measurements as far as possible. 
This includes the solid--state heat conductivity, the specific heat capacity, latent heats of sublimation, and the saturation pressure of vapours.

Nominally, \textsc{nimbus} considers a spherical, rotating, and orbiting body, that is divided into a large number of latitudinal and radial cells. 
Energy and mass transport processes take place both radially and latitudinally. However, in the current application the latitudinal energy and 
mass transport are switched off, which makes it possible to also consider erosion of solids at the surface in \textsc{nimbus}. Furthermore, latitude--dependent 
illumination is not applied here, but all latitudes are fed the same illumination flux described in section~\ref{sec_method_illum}.

If the medium initially is a chemically homogeneous mixture of dust, $\mathrm{H_2O}$ ice, and $\mathrm{CO_2}$ ice, the onset of solar illumination will 
cause a relatively slow sublimation of $\mathrm{H_2O}$ ice, and a relatively fast sublimation of $\mathrm{CO_2}$ ice. Because mass conservation is 
considered, the finite reservoirs of the ices in the top cells will eventually run out. As a result, the medium becomes chemically stratified, with an ice--free dust 
mantle located on top of a layer that only contains dust and $\mathrm{H_2O}$ ice, in turn located on top of the dust$+\mathrm{H_2O}+\mathrm{CO_2}$ mixture. 
The $\mathrm{H_2O}$ sublimation front is therefore located at the top of the middle layer, while the $\mathrm{CO_2}$ sublimation front is located at its bottom 
(deeper because the sublimation rate of $\mathrm{CO_2}$ exceeds that of $\mathrm{H_2O}$). Roughly half of the vapour released at any of the fronts diffuses 
upward and may eventually escape to space. The rest of the gas diffuses downwards where it eventually will recondense. These processes therefore contribute to 
shaping the stratification. Gases that are released at depth during nighttime might recondense temporarily within the dust mantle or elsewhere if the near--surface 
region has cooled off sufficiently. Erosion will act to make the dust mantle thinner or remove it entirely. Vivid $\mathrm{CO_2}$ sublimation may also erode 
the mixture of dust and $\mathrm{H_2O}$ ice, thereby bringing the $\mathrm{CO_2}$ sublimation front towards the surface. All these events take place naturally 
during a \textsc{nimbus} simulation. Because the composition, stratification, physical properties evolve continuously it is sometimes difficult to `force' a certain 
set of properties during the master period. Therefore, it is often necessary to gradually adjust the initial conditions until a desirable configuration emerges during 
the master period. 

Nominally, \textsc{nimbus} assumes that all incoming radiation is absorbed in the top cell. The default is also to keep the same diffusivity at all times, and 
let the medium evolve from some initially uniform state. However, if such models are incapable of providing a temperature (versus depth and time during nucleus rotation) 
that leads to reproduction of the data, those default assumptions are relaxed. If so, incoming radiation may be allowed to be absorbed over many cells according to Beer's 
law (resulting in a solid--state greenhouse effect), the diffusivity may be forced to change at specific rotational phases, or the initial conditions may consider e.~g., the 
bulk density and/or porosity to vary strongly with depth.

\subsection{The radiative transfer model \textsc{THEMIS}} \label{sec_method_THEMIS}

The physical temperatures provided by \textsc{btm} and \textsc{nimbus} are not immediately useful in the current context, because 
MIRO does not observe the radiation coming merely from the comet surface, but receives contributions from within a near--surface 
slab that may measure several millimetres or centimetres across. The physical temperature may vary drastically over such distances. 
The radiative transfer problem needs to be solved, to calculate the antenna temperature observed by MIRO, based on the 
temperature profiles generated by the thermophysical models. Here, this is done with the radiative transfer code \textsc{themis} 
\citep[for a full description, see][]{davidssonetal22b}.

\textsc{themis} is a Monte Carlo--based solver, that tracks large numbers of test--particle `photons' from creation, via transport and 
scattering (regulated by wavelength--dependent extinction coefficients and single--scattering albedos), to the eventual absorption or escape to space. 
The photon production rate at various depths depends on the local temperature. By collecting statistics on the fate of all photons, it is possible to calculate 
the radiance emerging from the medium, as function of the emergence angle (measured from the surface normal). 

Comparison between \textsc{themis} radiances \citep[re--calculated to antenna temperature as described by][]{davidssonetal22b} and MIRO 
observations are made at the emergence angle prevailing at the time of MIRO observation. 

The quality of the synthetic antenna temperature curve with respect to the data is here measured by the $Q$ parameter \citep[][see section~3.4.3]{davidssonetal22b}. 
Error bars of $\pm 2.5\,\mathrm{K}$ are used for the data, and the same number of model parameters when evaluating $Q$ are used here as by \citet{davidssonetal22b}. 
$Q\geq 0.01$ is considered a statistically good fit to the data, while orders of magnitude smaller values generally means that the model curve is in poor agreement with 
the measurements. Alternatives to $Q$ as measure of goodness--of--fit, and exceptions to the $Q\geq 0.01$ criterion, are discussed on a case--by--case basis.

\section{Results} \label{sec_results}

\subsection{November 2014: pre--collapse} \label{sec_results_nov14}

\subsubsection{November 2014: \textsc{btm} results} \label{sec_results_nov14_BTM}

A total of 17 facets were marked on the SHAP5 version 1.5 shape model, degraded to $5\cdot 10^4$ facet, as a representation of the 
Aswan collapse site. When the MIRO archive \citep[PDS website\footnote{https://pds-smallbodies.astro.umd.edu/holdings/ro-c-miro-3-prl\\-67p-v3.0/dataset.shtml 
and \\https://pds-smallbodies.astro.umd.edu/holdings/ro-c-miro-3-esc1\\-67p-v3.0/dataset.shtml};][]{hofstadteretal18a,hofstadteretal18b}  was searched for SMM and MM beam--centre interceptions at the longitudes, latitudes, and 
radial distances corresponding to these 17 facets, a total of 309 observations (one--second integrations) with emergence angle $e\leq 70^{\circ}$ were found, concentrated 
on nine of the facets. Time is measured in units of `day numbers', starting with $d_{\rm n}=1$ for 2014 January 1, $00:00:00\,\mathrm{UTC}$ and then incrementing by unity every $24\,\mathrm{h}$. 
Some observations were acquired when \emph{Rosetta} was $\sim 18\,\mathrm{km}$ from the comet, while most occurred at 
$\sim 30\,\mathrm{km}$, resulting in SMM and MM Full Width Half Maximum (FWHM) beam widths of $\stackrel{<}{_{\sim}} 65\,\mathrm{m}$ and   
$\stackrel{<}{_{\sim}} 207\,\mathrm{m}$ respectively. This should be compared to the final dimensions of the 57--$81\,\mathrm{m}$ wide (narrowing towards the top) 
and $65\,\mathrm{m}$ high collapse site. It suggests that the SMM beam was dominated by emission from the collapse area, while the MM beam may contain contributions from nearby areas. 
However, at this time there was no visual difference between the future collapse area and surrounding parts of the cliff. Because the entire cliff 
presumably had similar properties, it is reasonable that the MM measurements are representative of the region of interest as well. 

The observations were shifted in time to a common master period, starting at $d_{\rm n}=313.75$, corresponding to 2014 November 9, 18:00~UTC. 
When binned in $2.4\,\mathrm{min}$ intervals (roughly $1^{\circ}$ of nucleus rotation), this resulted in 10 temperature bins. Visual inspection of the Rosetta viewing conditions at the time of 
observation for the bins revealed that five were centred on the cliff side that was in darkness, but that the beams also contained parts of the plateau 
that was illuminated. These bins were deemed to be skewed towards too high temperatures, and were rejected. The remaining five bins sampled 
the late morning and late night of the diurnal temperature curve.

\begin{figure*}
\centering
\begin{tabular}{cc}
\scalebox{0.43}{\includegraphics{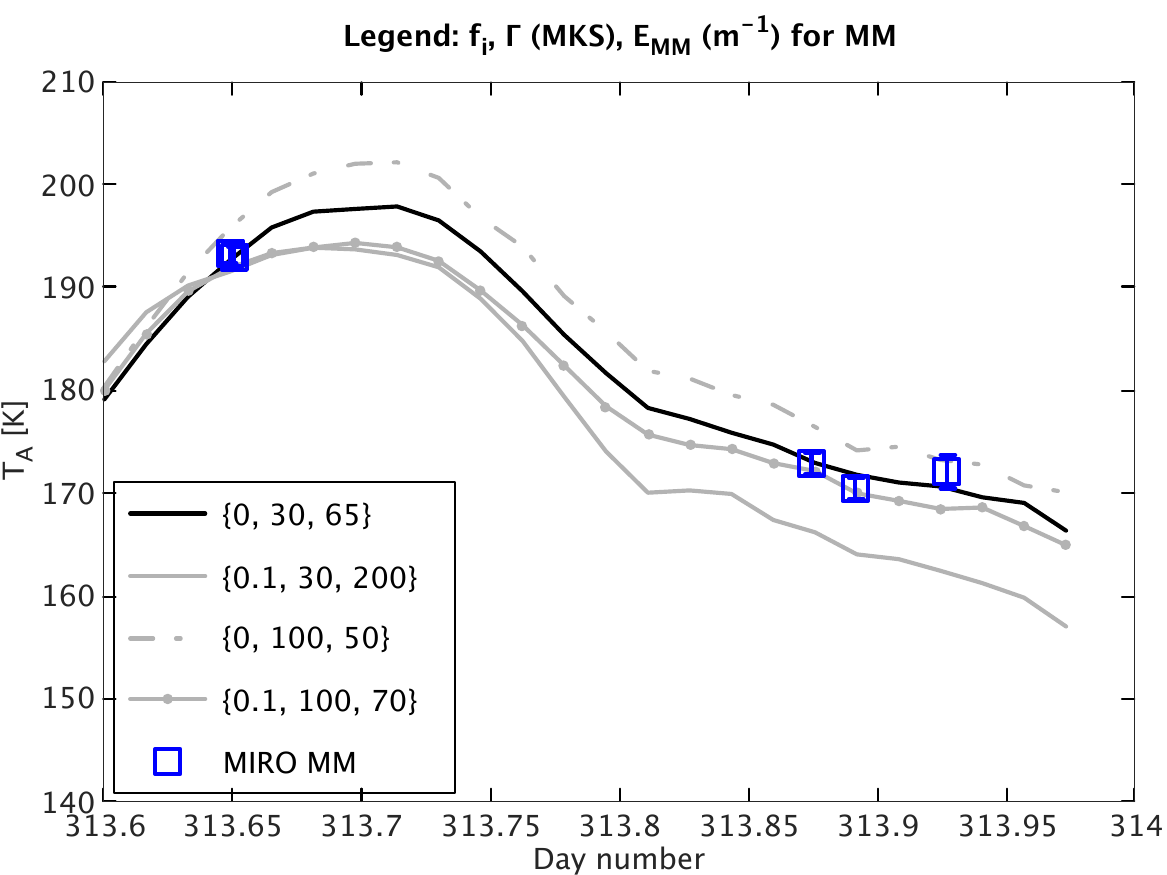}} & \scalebox{0.43}{\includegraphics{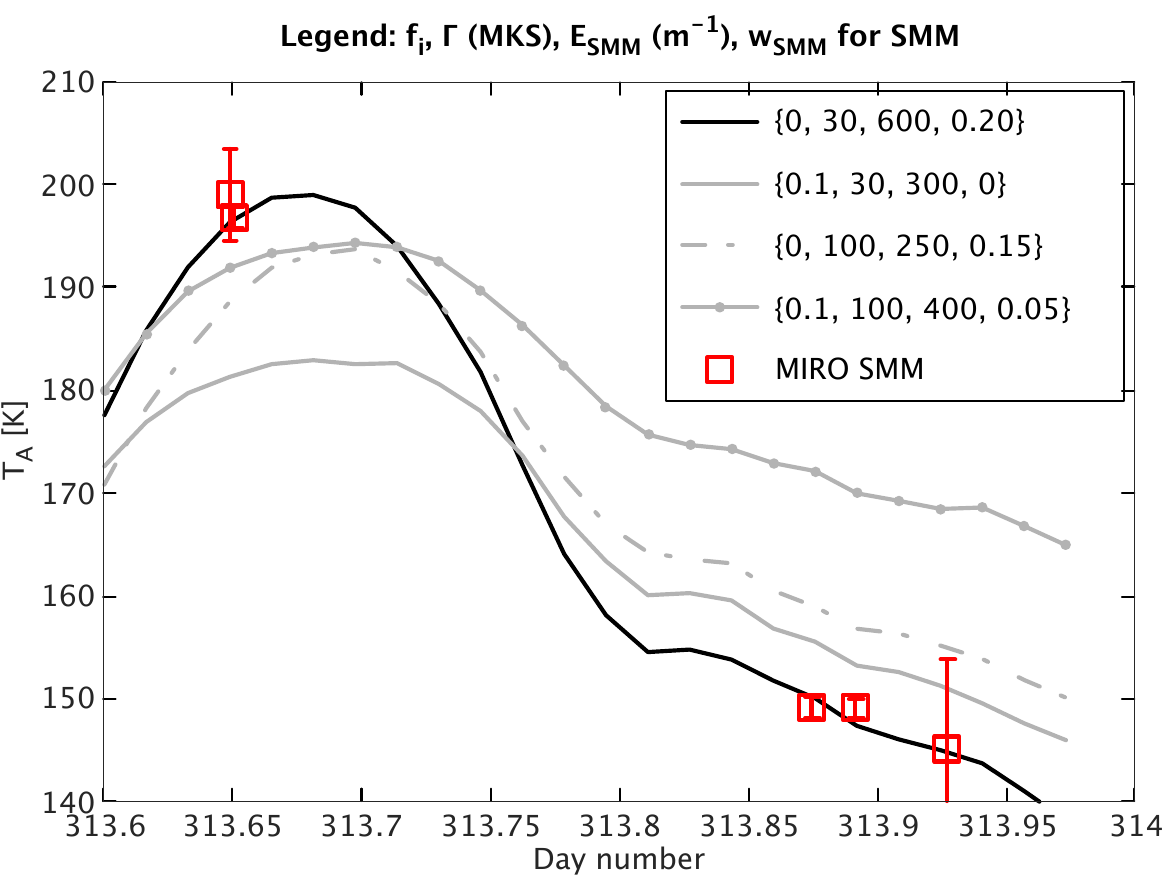}}\\
\end{tabular}
     \caption{\emph{Left:} Measured and \textsc{btm}--based synthetic MM antenna temperatures for the 2014 November pre--collapse case. The model with 
$\{f_{\rm i},\,\Gamma,\,E_{\rm MM}\}=\{0,\,30\,\mathrm{MKS},\,65\,\mathrm{m^{-1}}\}$ is consistent with the data ($Q_{\rm MM}=0.196$). \emph{Right:} Measured and \textsc{btm}--based 
synthetic SMM antenna temperatures for the November 2014 pre--collapse case. The model with $\{f_{\rm i},\,\Gamma,\,E_{\rm SMM},\,w_{\rm SMM}\}=\{0,\,30\,\mathrm{MKS},\,600\,\mathrm{m^{-1}},\,0.20\}$ 
is consistent with the data ($Q_{\rm SMM}=0.386$).}
     \label{fig_BTM_Nov14}
\end{figure*}

The \textsc{btm} model assumes a certain set of physical parameters (densities, specific heat capacities, heat conductivities) for refractories and water ice, 
given by Table~2 in \citet{davidssonetal22b}. Only two parameters are varied: the volumetric fraction of water ice $f_{\rm i}$ and the Hertz factor (a porosity--correction 
of the effective heat conductivity, set in order to achieve a given thermal inertia $\Gamma$ of the medium). As a starting point, the \textsc{btm} model was run for a grid with 
every combination of $f_{\rm i}=\{0,\,0.1\}$ and $\Gamma=\{30,\, 50,\, 80,\, 100,\, 130\}\,\mathrm{MKS}$, 
i.~e., 10 models. Next, \textsc{themis} was run in order to calculate the antenna temperatures corresponding to these thermophysical models. The total number of \textsc{themis} 
simulations were 58, first focusing on the MM while assuming $w_{\rm MM}=0$. The solution with $\{f_{\rm i},\,\Gamma\}=\{0,\,30\,\mathrm{MKS}\}$ was highly 
compatible with the observations for $E_{\rm MM}=65\,\mathrm{m^{-1}}$, having $Q_{\rm MM}=0.196$. An alternative but less convincing solution was found at 
$\{f_{\rm i},\,\Gamma\}=\{0.1,\,100\,\mathrm{MKS}\}$ with $E_{\rm MM}=70\,\mathrm{m^{-1}}$ and $Q_{\rm MM}=0.018$.  Figure~\ref{fig_BTM_Nov14} (left) shows 
the synthetic MM antenna temperature curves for several parameter combinations, compared with the MIRO measurements.

Next, the SMM data (Figure~\ref{fig_BTM_Nov14}, right) were considered. They are characterised by higher antenna temperature values and a bigger amplitude, compared to the MM. 
For the 10 combinations of $\{f_{\rm i},\,\Gamma\}$, a total of 81 \textsc{themis} models were run for different values of $\{E_{\rm SMM},\,w_{\rm SMM}\}$. It is evident that 
the amplitudes of the synthetic curves are too small, unless the model medium is ice--free and has low thermal inertia. The best fit was obtained for 
$\{f_{\rm i},\,\Gamma,\,E_{\rm SMM},\,w_{\rm SMM}\}=\{0,\,30\,\mathrm{MKS},\,600\,\mathrm{m^{-1}},\,0.20\}$, having  $Q_{\rm SMM}=0.386$. 
Note that the $\{f_{\rm i},\,\Gamma\}=\{0.1,\,100\,\mathrm{MKS}\}$ case, which had a borderline similarity with the MM data, is not anywhere near the SMM data.

Therefore, the combined MM and SMM measurements, when interpreted with the \textsc{btm} model, systematically indicates that the pre--collapse Aswan cliff side 
consisted of a thick dust mantle with low thermal inertia. To investigate exactly how thick the mantle needs to be in order not to display measurable signatures of 
sub--surface sublimation cooling, \textsc{nimbus} models are considered in the following.

\subsubsection{November 2014: \textsc{nimbus} results} \label{sec_results_nov14_NIMBUS}

A model medium was built by applying a refractories/water--ice mass ratio $\mu=2$, a 5.5\% molar $\mathrm{CO_2}$ abundance relative to water, a 70\% porosity, 
and compacted densities $\varrho_1=3000\,\mathrm{kg\,m^{-3}}$, $\varrho_4=917\,\mathrm{kg\,m^{-3}}$, and $\varrho_6=1500\,\mathrm{kg\,m^{-3}}$ for refractories, 
$\mathrm{H_2O}$, and $\mathrm{CO_2}$, respectively.  The resulting bulk density was $\rho_{\rm bulk}=510\,\mathrm{kg\,m^{-3}}$, marginally lower than that of 67P. A dust mantle was 
formed by removing all volatiles in an upper layer, which increased the local porosity to 89\% and reduced the local bulk density to $\rho_{\rm bulk}=330\,\mathrm{kg\,m^{-3}}$. 
The dust specific heat capacity $c_1=1260\,\mathrm{J\,kg^{-1}\,K^{-1}}$ was used to obtain the same volumetric heat capacity as in the \textsc{btm}, despite using different porosities 
and compacted dust densities. The porosity--corrected heat conductivity was set 
such that the dust mantle had a $30\,\mathrm{MKS}$ thermal inertia. It was verified that a thick ($11\,\mathrm{cm}$) dust mantle obtained the same near--surface temperature 
solution as the dust--only \textsc{btm}, when sublimation was artificially switched off at depth. 

At that point, sublimation was re--activated and a series of eight \textsc{nimbus} simulations with different dust mantle thicknesses in the range $0.21$--$11\,\mathrm{cm}$ were considered. The models 
assumed the same fixed $\kappa$ and $c$ as before, a diffusivity consistent with $\{L,\,r_{\rm p}\}=\{100,\,10\}\,\mathrm{\mu m}$ and $\xi=1$ \citep[see][for a motivation]{davidssonetal22}, and a 
non--eroding mantle. Each model was run for nine days (roughly 17 nucleus rotations) to achieve thermal relaxation, prior to running the master period.

\begin{table}
\begin{center}
\begin{tabular}{||l|r|r|r|r|r||}
\hline
\hline
Thickness $h_{\rm m}$ & $E_{\rm MM}$  & $Q_{\rm MM}$ & $E_{\rm SMM}$  & $w_{\rm SMM}$ & $Q_{\rm SMM}$\\
$\mathrm{(cm)}$ & $\mathrm{(m^{-1})}$  & & $\mathrm{(m^{-1})}$  &  & \\
\hline
11.0 & 65 & 0.239 & 400 & 0.20 & 0.0139\\
4.67 & 65 & 0.264 & 600 & 0.17 & 0.097\\
3.12 & 70 & 0.091 & 600 & 0.17 & 0.140\\
2.01 & 70 & $7.9\cdot 10^{-5}$ & 600 & 0.17 & 0.145\\
0.99 & 70 & $4.5\cdot 10^{-5}$ & 600 & 0.18 & 0.180\\
0.63 & 90 & 0.078 & 1000 & 0.20 & 0.008\\
0.47 & 100 & 0.462 & 1000 & 0.20 & 0.008\\
0.21 & 120 & 0.010 & 1000 & 0.20 & $2.6\cdot 10^{-11}$\\ 
\hline 
\hline
\end{tabular}
\caption{The best achievable reproductions of the MIRO November 2014 observations of Aswan prior to collapse, for \textsc{nimbus} models with different 
dust mantle thicknesses $h_{\rm m}$, in terms of the extinction coefficients $E_{\rm MM}$ and $E_{\rm SMM}$, and the SMM single--scattering albedo $w_{\rm SMM}$ (note 
that $w_{\rm MM}$ was applied at all times). A high--quality fit requires that $Q_{\rm MM}\geq 0.01$ and $Q_{\rm SMM}\geq 0.01$ simultaneously.}
\label{tab1}
\end{center}
\end{table}

\begin{figure*}
\centering
\begin{tabular}{cc}
\scalebox{0.43}{\includegraphics{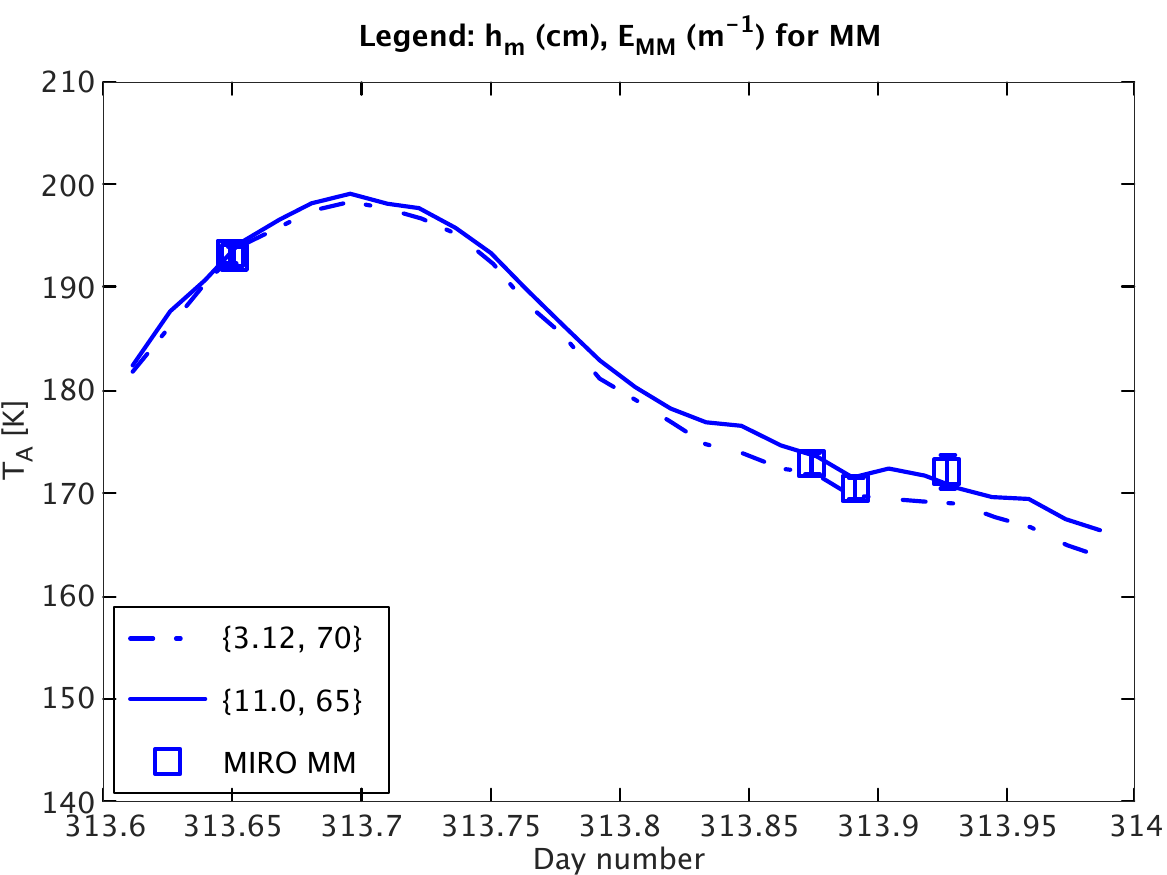}} & \scalebox{0.43}{\includegraphics{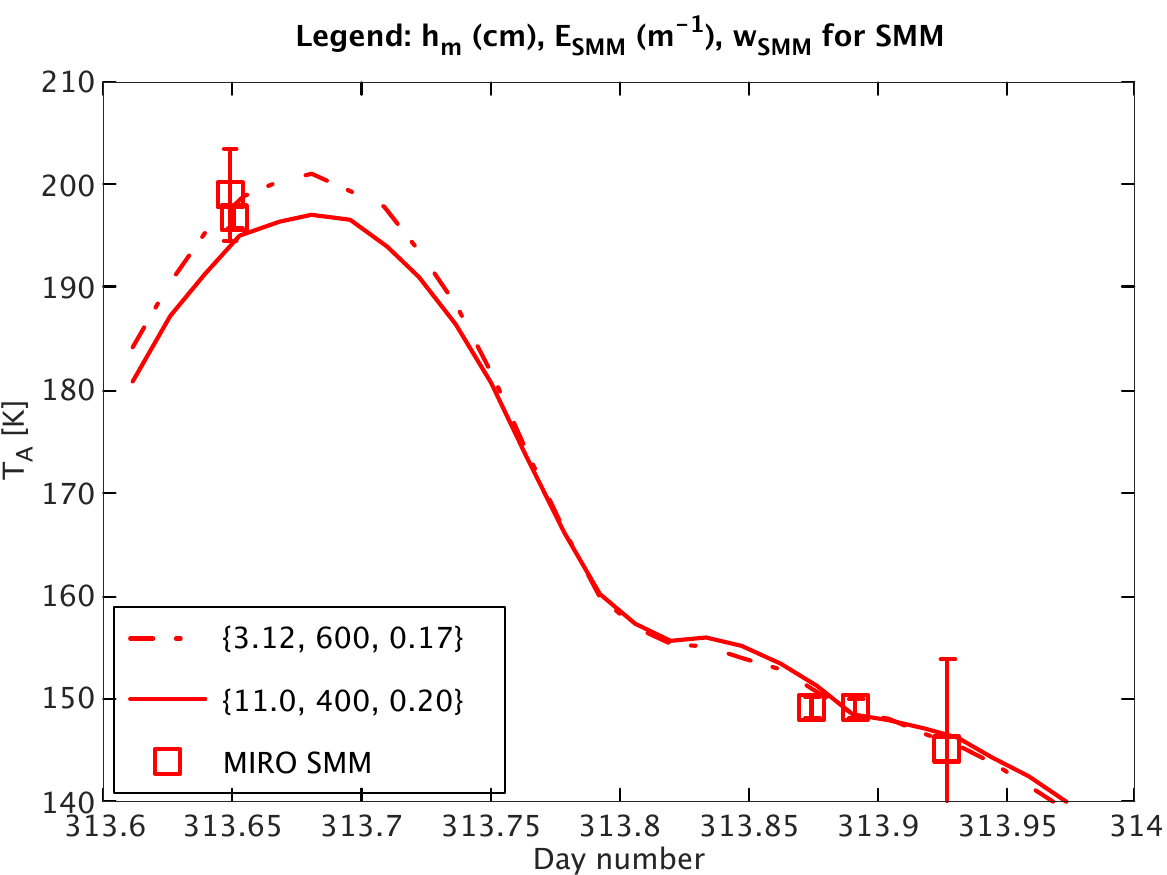}}\\
\end{tabular}
     \caption{\emph{Left:} Measured and \textsc{nimbus}--based synthetic MM antenna temperatures for the November 2014 pre--collapse case. 
\emph{Right:} Measured and \textsc{btm}--based synthetic SMM antenna temperatures for the November 2014 pre--collapse case. Both models 
are formally consistent with the data simultaneously at MM and SMM.}
     \label{fig_NIMBUS_Nov14}
\end{figure*}

Once the nucleus temperature (as function of depth and time) had been provided by \textsc{nimbus}, it was passed to \textsc{themis}, which first calculated the MM antenna 
temperature $T_{\rm A}$. $\mathrm{E_{\rm MM}}$ was adjusted until the synthetic $T_{\rm A}$ matched the measured one near $d_{\rm n}=313.65$. The $Q_{\rm MM}$ 
value therefore measures how well that solution also matched the data at $313.85\stackrel{<}{_{\sim}}d_{\rm n}\stackrel{<}{_{\sim}}  313.95$. The results are summarised in Table~\ref{tab1}.

When the dust mantle gets thinner, $E_{\rm MM}$ grows systematically. Good fits are obtained for $h_{\rm m}\geq 3\,\mathrm{cm}$. For thinner mantles, $Q_{\rm MM}$ first 
degrades, then recovers, only to degrade anew. This is because the modelled late--night $T_{\rm A}$ first is too cold, then about right, and finally too warm, compared with the data. 
This can be understood as follows. When $h_{\rm m}\geq 3\,\mathrm{cm}$, the cooling from sublimation is weak, and the quality of the fit is good (consistent with the \textsc{btm} ice--free solution). 
As $h_{\rm m}$ decreases, the lower part of the dust mantle experiences an increasing degree of cooling. Matching the day--time data requires a concentration of the escaping 
radiation to a thinning top layer of the dust mantle (situated above the cooled portion). This is achieved through stronger extinction, i.~e., a higher $E_{\rm MM}$. But the same particles 
responsible for absorption are also emitting radiation. For this reason, an increasing $E_{\rm MM}$ tends to increase $T_{\rm A}$. At $1\stackrel{<}{_{\sim}} h_{\rm m}\stackrel{<}{_{\sim}}2\,\mathrm{cm}$, 
the top layer still does not contain a sufficient number of emitters, thus $T_{\rm A}$ is too low at night. However, as $h_{\rm m}\rightarrow 0.2\,\mathrm{cm}$, very strong extinction 
($E_{\rm MM}\rightarrow 120\,\mathrm{m^{-1}}$) is needed to concentrate emission to the hottest top of the mantle at day. The number density of absorbers and emitters is now so high 
that the night--time $T_{\rm A}$ becomes too high. The transition between the two regimes is responsible for the brief recovery at $0.47\stackrel{<}{_{\sim}} h_{\rm m}\stackrel{<}{_{\sim}}0.63\,\mathrm{cm}$.

The non--monotonic behaviour of $Q_{\rm MM}$ when decreasing $h_{\rm m}$, and the unusually high $E_{\rm MM}$--values for thin mantles, lead to the suspicion that these 
solutions are nonphysical. That hypothesis can be tested by considering the SMM solutions. Because the extinction efficiency increases when the wavelength decreases, it is 
expected that $E_{\rm SMM}>E_{\rm MM}$. Because the observed sub--millimetre radiation emanates from an even thinner near--surface slab than the millimetre radiation, 
the dust mantle must become very thin before the SMM channel detects the cooling at the sublimation front. The SMM solutions start to degrade irreversibly at $h_{\rm m}<1\,\mathrm{cm}$. 

For a solution to be physically plausible, it must reproduce the MM and SMM antenna temperature curves simultaneously. Looking at the $Q_{\rm MM}$ and $Q_{\rm SMM}$ values, 
we only have clearly convincing solutions at $h_{\rm m}\geq 3\,\mathrm{cm}$. I therefore draw the conclusion that the thickness of the dust mantle covering the Aswan cliff was at 
least $3\,\mathrm{cm}$ in November 2014. Two examples are shown in Fig.~\ref{fig_NIMBUS_Nov14}.

\subsection{November--December 2015: five months after the collapse} \label{sec_results_novdec15}

Because of the relatively large comet/\emph{Rosetta} distance around the time of the 2015 August perihelion passage, and the 
difficult viewing geometry, the first useful MIRO observations of Aswan took place several months after the collapse. Between 
November and December 2015 a total of 18 MIRO 1--$\mathrm{s}$ continuum SMM observations of Aswan were made \citep[PDS website\footnote{https://pds-smallbodies.astro.umd.edu/holdings/ro-c-miro-3-esc4-\\67p-v3.0/dataset.shtml};][]{hofstadteretal18c}, distributed 
amongst eight bins when shifted to a common master period (starting on 2015 December 4, 00:36:03~UTC or at $d_{\rm n}=703.025$). 
However, two bins had to be excluded because the beam contained illuminated background terrain while Aswan was in darkness. 
The remaining six bins comprised three daytime or sunset observations acquired when \emph{Rosetta} was 78--$141\,\mathrm{km}$ from the comet, 
and three nighttime observations acquired when \emph{Rosetta} was 76--$200\,\mathrm{km}$ from the comet. The first nighttime bin has a very low 
antenna temperature, after Aswan spent roughly six hours in darkness. Shortly after, Aswan experienced a brief episode of illumination, as sunlight entered 
through a gap between the lobes. The following two nighttime bins have elevated temperatures, presumably because of the temporary heating. 

Because of the rather large distances only the higher--resolution SMM observations were analysed. Projected onto the nucleus, the SMM FWHM 
footprints ranged 170--$440\,\mathrm{m}$ across, to be compared with the maximum dimension of the Aswan collapse site of $\sim 100\,\mathrm{m}$. Therefore, the collapse site only 
constitutes a fraction of the registered signal, particularly in cases where the emergence angle is substantial. If the signal is dominated by the 
surrounding dust mantle, one would expect the best--fit solutions to be skewed towards that of section~\ref{sec_results_nov14_NIMBUS}, i.~e., 
a medium with $\Gamma\approx 30\,\mathrm{MKS}$ and no strong signs of sublimation cooling. However, if the collapse site has fundamentally 
different thermal properties than the surrounding dust mantle, e.~g., caused by strong sublimation of exposed ices, these may introduce characteristic 
distortions of the signal. The purpose of this work is to investigate whether such distortions exist.

\subsubsection{November--December 2015: \textsc{btm} results} \label{sec_results_novdec15_BTM}

A total of 22 combinations of volumetric ice abundances $0\leq f_{\rm i}\leq 0.9$ and thermal inertias $5\leq \Gamma\leq 100\,\mathrm{MKS}$ 
were considered with the \textsc{btm}, and a total of 74 \textsc{themis} simulations for different extinction coefficients and single--scattering albedos were made. 

\begin{figure}
\scalebox{0.4}{\includegraphics{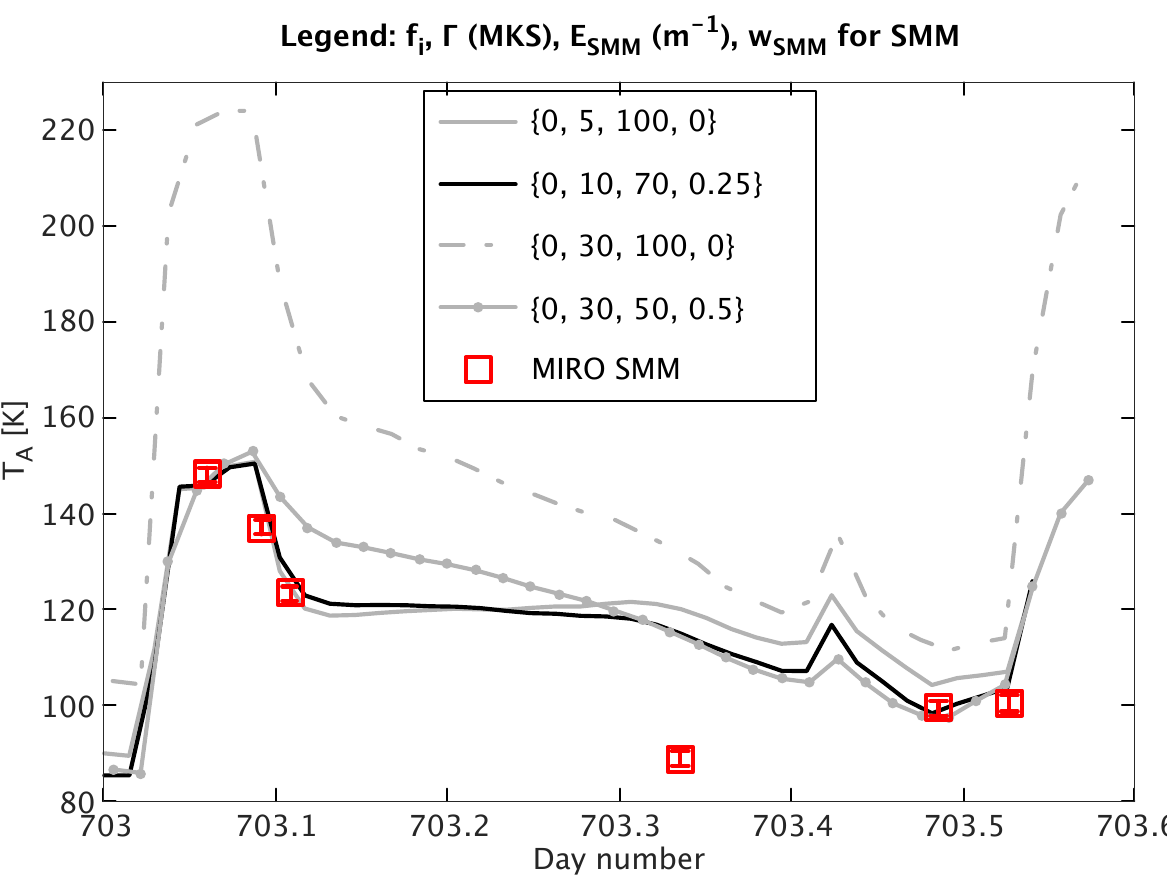}}
    \caption{November and December 2015 SMM data and ice--free \textsc{btm} models. Such models do not manage to reach 
the $T_{\rm A}=89\,\mathrm{K}$ minimum during nighttime.}
     \label{fig_BTM_NovDec15_dust}
\end{figure}

First, an ice--free medium with porosity $\psi=0.7$, bulk density $\rho_{\rm bulk}=975\,\mathrm{kg\,m^{-3}}$, and specific heat capacity 
$c_{\rm d}=420\,\mathrm{J\,kg^{-1}\,K^{-1}}$ was considered. Figure~\ref{fig_BTM_NovDec15_dust} shows examples of synthetic 
SMM antenna temperatures obtained with \textsc{btm} and \textsc{themis}, compared with the MIRO observations. Matching the 
daytime ($d_{\rm n}\approx 703.05$--703.10) and the late--night ($d_{\rm n}\approx 703.45$--703.55) bins require a low thermal inertia, 
$\Gamma=5$--$10\,\mathrm{MKS}$. Increasing $\Gamma$ has the effect of increasing both the amplitude and the overall antenna temperature. 
The amplitude increase can be quenched by reducing $E_{\rm SMM}$, but once the temperature difference between the daytime and late--night 
bins is obtained, the model curve is typically too warm (despite some temperature reduction when $E_{\rm SMM}$ is lowered). The curve can be lowered to the level of the data by introducing a non--zero single--scattering 
albedo. This is illustrated by Fig.~\ref{fig_BTM_NovDec15_dust}, where the parameter combinations $\{\Gamma,\,E_{\rm SMM},\,w_{\rm SMM}\}=\{5\,\mathrm{MKS},\,100\,\mathrm{m^{-1}},\,0\}$ 
and $\{10\,\mathrm{MKS},\,70\,\mathrm{m^{-1}},\,0.25\}$ give rise to similar curves. The $\Gamma$--dependence of the curve is also illustrated by the increase from $\Gamma=5\,\mathrm{MKS}$ 
to $30\,\mathrm{MKS}$ while keeping $E_{\rm SMM}=100\,\mathrm{m^{-1}}$ and $w_{\rm MM}=0$ constant. When $\Gamma=30\,\mathrm{MKS}$ the curve can be pushed back by lowering 
the extinction coefficient to $E_{\rm SMM}=50\,\mathrm{m^{-1}}$ and increasing the single--scattering albedo to $w_{\rm MM}=0.5$. However, for $\Gamma \stackrel{>}{_{\sim}}10\,\mathrm{MKS}$ 
the rapid cooling near sunset can no longer be matched.

The biggest difficulty is the mid--night bin near $d_{\rm n}=703.35$. At that point the antenna temperatures of all ice--free models are 
$\sim 30\,\mathrm{K}$ too high. The inability to reach that data bin, combined with the unusually low thermal inertia needed to fit 
other parts of the curve, suggests that the post--collapse Aswan surface material is substantially cooler than achievable with an 
ice--free model. Note that these problems cannot be fixed by changing the assumed heat capacity of the material -- the exact same curves are 
obtained as long as the quantity $\rho_{\rm bulk}c_{\rm d}/E_{\rm SMM}$ is held constant \citep{schloerbetal15}.

\begin{figure}
\scalebox{0.4}{\includegraphics{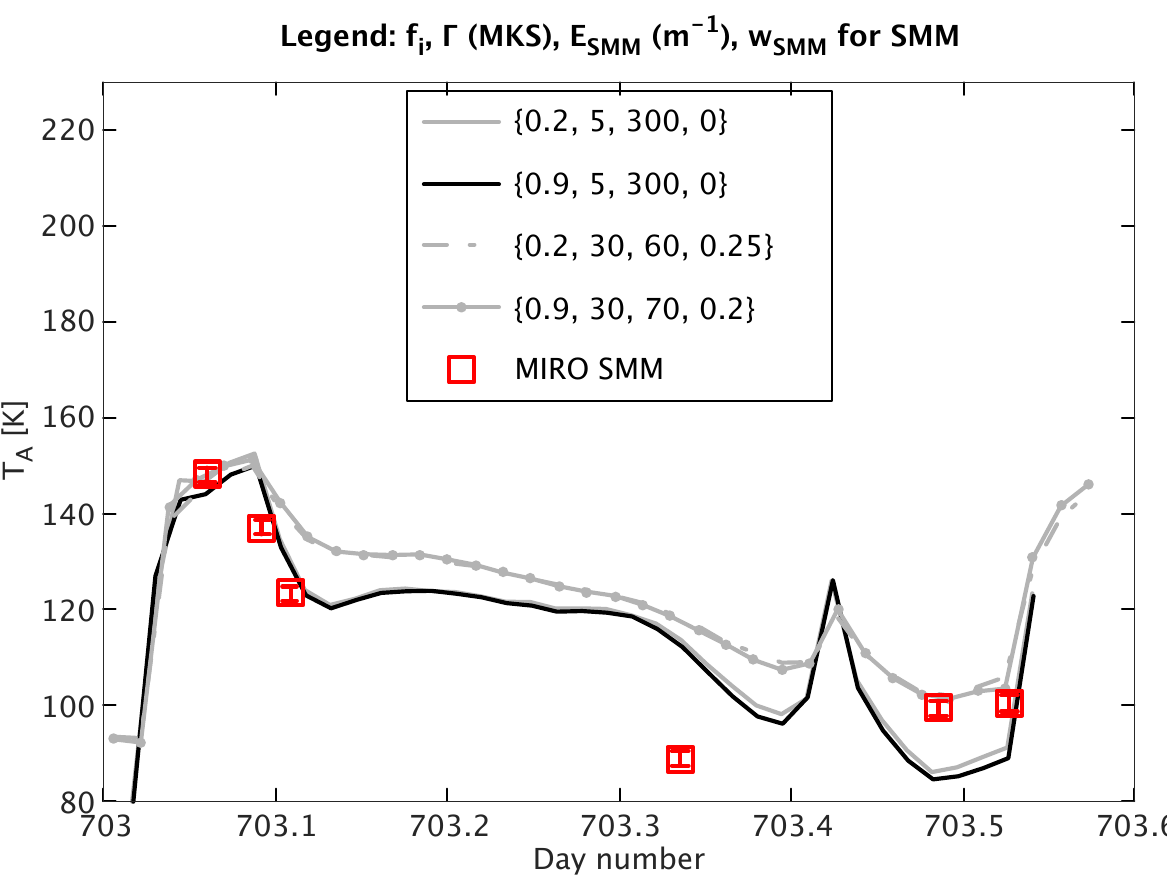}}
    \caption{November and December 2015 SMM data and \textsc{btm} models that include sublimation cooling by surface water ice. Also these models fail 
to reach $T_{\rm A}=89\,\mathrm{K}$.}
     \label{fig_BTM_NovDec15_dustH2O}
\end{figure}

Introducing cooling from sublimating surface water ice in the \textsc{btm} ($f_{\rm i}>0$) lowers the physical temperature of the 
medium. Reaching the same daytime antenna temperature as before is possible by increasing $E_{\rm SMM}$. Examples are 
shown in Fig.~\ref{fig_BTM_NovDec15_dustH2O}. At $\Gamma=5\,\mathrm{MKS}$ the daytime temperatures are reproduced if 
the extinction coefficient is increased from $E_{\rm SMM}=100\,\mathrm{m^{-1}}$ to $E_{\rm SMM}=300\,\mathrm{m^{-1}}$ 
when $f_{\rm i}=0.2$. The nighttime temperatures are reduced, but not sufficiently to match the coldest bin, even when an 
almost dust--free medium ($f_{\rm i}=0.9$) is considered. Increasing the thermal inertia only worsen this problem, and 
additionally destroys the decent fit of the daytime peak. It does not seem that the \textsc{btm} is capable of reproducing the 
SMM data for 2015 November and December. Particularly, water ice sublimation does not seem capable of reducing the 
antenna temperature to the coldest nighttime bin at $T_{\rm A}=89\,\mathrm{K}$.

It is noteworthy that the data associated with one of the removed bins was acquired two days after the data of the coldest nighttime bin, at a very 
similar nucleus rotational phase, but with a somewhat different viewing geometry. The antenna temperature of the removed bin 
($T_{\rm A}=91\,\mathrm{K}$) is somewhat elevated with respect to the coldest nighttime bin (this seems to be caused by the small portions of 
faintly illuminated terrain within the MM beam, that triggered its removal). However, this is still far below the modelled antenna temperatures of 
$T_{\rm A}\approx 110$--$120\,\mathrm{K}$. The confirmation of a very low nighttime antenna temperature is important, as it demonstrates that 
the $T_{\rm A}=89\,\mathrm{K}$ bin is not a fluke. The inability to reach temperatures this low even with large amounts of exposed water ice on 
the surface shows that very special conditions are required, as discussed in the following.

\subsubsection{November--December 2015: \textsc{nimbus} results} \label{sec_results_novdec15_NIMBUS}

Due to the complexity of understanding this data set, modelling was performed in stages with different focus: 1) expected evolution between 2015 July and November/December; 
2) investigating the conditions that yield very low night--time temperatures for fixed erosion rates; 3) understanding the reason for the pre--dawn temperature elevation; 4) introducing a time--variable erosion rate and attempting to 
constrain the composition; 5) introducing a time--dependent diffusivity. These stages, involving about 60 \textsc{nimbus} and 440 \textsc{themis} simulations are now described in turn. 

First, \textsc{nimbus} was used to investigate the level of processing expected at Aswan from the time of collapse 
(2015 July 10) to the master period (2015 December 4). The model assumed $\mu=1$ (crystalline water ice), molar $\nu_6=\mathrm{CO_2/H_2O}=0.32$, 
$\psi=0.7$ (resulting in $\rho_{\rm bulk}=435\,\mathrm{kg\,m^{-3}}$), tube dimensions $\{L_{\rm p},\,r_{\rm p}\}=\{100,\,10\}\,\mathrm{\mu m}$, and tortuosity $\xi=1$. 
The applied values of $\mu$, $\nu_6$,  $\{L_{\rm p},\,r_{\rm p}\}$, and $\xi$ are consistent with the observed pre--perihelion $\mathrm{H_2O}$ and $\mathrm{CO_2}$ 
production rates according to \citet{davidssonetal22}. The model material was considered homogeneous up to the surface (to imitate a suddenly exposed mixture of dust, $\mathrm{H_2O}$, and $\mathrm{CO_2}$ 
at the time of collapse), and erosion of solids was switched off (in order to obtain the largest expected distances from the surface to the $\mathrm{H_2O}$ and $\mathrm{CO_2}$ sublimation 
fronts at the end of the simulation).

\begin{figure}
\scalebox{0.4}{\includegraphics{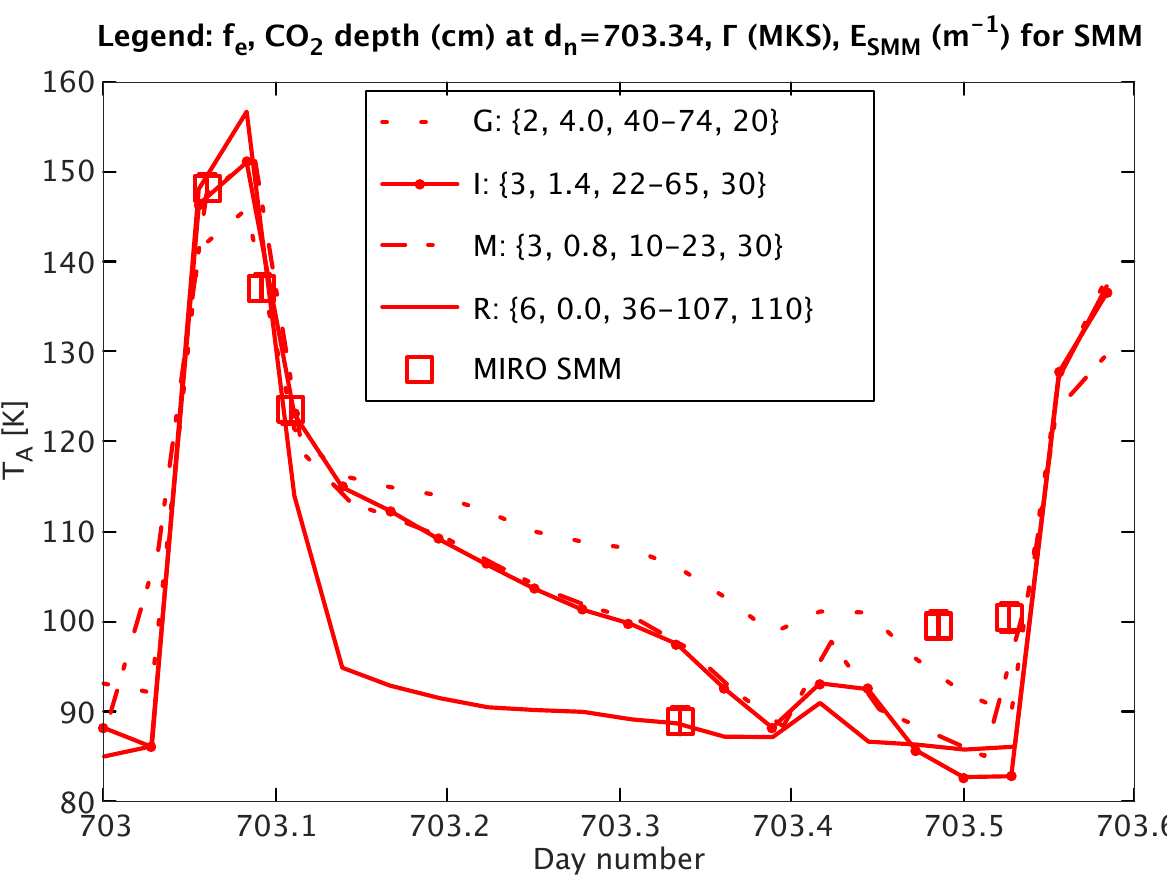}}
    \caption{2015 November and December SMM data and \textsc{nimbus} models with constant erosion rate 
($f_{\rm e}\times 3.06\cdot 10^{-5}\,\mathrm{kg\,m^{-2}\,s^{-1}}$). Models G and I had 
$\{L_{\rm p},\,r_{\rm p}\}=\{10,\,1\}\,\mathrm{cm}$. However, to reach very low thermal inertia ($10\leq\Gamma\leq 23\,\mathrm{MKS}$) for model M 
it was necessary to reduce radiative heat transport by lowering $r_{\rm p}$ and $\{L_{\rm p},\,r_{\rm p}\}=\{1,\,0.1\}\,\mathrm{cm}$ was used for models M and R. The coldest 
data point is only reproduced when $\mathrm{CO_2}$ ice is at the surface during nighttime.}
     \label{fig_NIMBUS_SMM_NovDec15_01}
\end{figure}

During the first month the $\mathrm{H_2O}$ and $\mathrm{CO_2}$ withdrew to $8.5\,\mathrm{mm}$ and $5.8\,\mathrm{cm}$ depths, respectively. 
Then followed two months of polar night at Aswan, during which no changes of either depth were observed because of very small activity levels. 
At that point, faint illumination resumed but was not sufficient to change the dust mantle thickness during two months. However, the $\mathrm{CO_2}$ 
front withdrew to $9.4\,\mathrm{cm}$ in that time. This simulation was used as initial condition for the others, typically considering a hand--over $\sim 3$ 
rotational revolutions prior to the master period to allow for thermal relaxation in the top few centimetres (sensed by MIRO) after changes to the model parameters.

The first priority was to understand under which conditions the lowest measured antenna temperature of $T_{\rm A, SMM}=89\,\mathrm{K}$ near $d_{\rm n}=703.35$ 
could be reproduced, because no \textsc{btm} model reached such low values (section~\ref{sec_results_novdec15_BTM}).  Erosion of solids at the surface was 
introduced to investigate whether a thinning of the dust mantle, or reduction of the $\mathrm{CO_2}$ depth, are necessary. A first series of 11 \textsc{nimbus} simulations considered a 
constant erosion rate of $3.06\cdot 10^{-5}\,\mathrm{kg\,m^{-2}\,s^{-1}}$ (removal of $8\,\mathrm{mm}$ in $12\,\mathrm{h}$), and multiples $f_{\rm e}$ thereof. 
Initially, a large diffusivity corresponding to $\{L_{\rm p},\,r_{\rm p}\}=\{10,\,1\}\,\mathrm{cm}$ and $\xi=1$ was considered to maximise the cooling due to water ice sublimation. 
Various combinations of $f_{\rm e}$ and thermal inertia were applied \citep[by lowering the Hertz factor with respect to the nominal value obtained by applying the method of][]{shoshanyetal02}.

Figure~\ref{fig_NIMBUS_SMM_NovDec15_01} shows a sub--set of the models, all of which perform reasonably well on the dayside. However, models G and I with 
$\mathrm{CO_2}$ at $\geq 1.4$--$4.0\,\mathrm{cm}$ depth with thermal inertia diurnal minima in the range $22\leq \Gamma\leq 40\,\mathrm{MKS}$ were still 
too warm at $d_{\rm n}=703.35$. Because low thermal inertia favours low temperatures in the absence of solar heating, attempts were made to lower $\Gamma$ further. 
That was only possible if the radiative contribution to heat transport was reduced by lowering $r_{\rm p}$. Therefore, $\{L_{\rm p},\,r_{\rm p}\}=\{1,\,0.1\}\,\mathrm{cm}$ was 
introduced and applied throughout the remainder of the November and December 2015 investigation, unless stated otherwise. That allowed for a lowering of the diurnal minimum thermal inertia to $\Gamma=10\,\mathrm{MKS}$, 
and additionally, the $\mathrm{CO_2}$ front was just $0.8\,\mathrm{cm}$ below the surface at night. Still, that did not lower the temperature sufficiently (model M).

Progress was only achieved when the erosion rate was increased sufficiently ($f_{\rm e}=6$) to force $\mathrm{CO_2}$ ice up to the 
surface itself at night (model R). It should be noted that $\mathrm{CO_2}$ temporarily withdrew $\sim 8\,\mathrm{mm}$ during daytime, so the 
Aswan collapse site would only expose a dust and $\mathrm{H_2O}$ ice mixture when illuminated (and imaged by \emph{Rosetta}/OSIRIS).

At this point attention turned to the relatively warm late--night bins near $d_{\rm n}\approx 703.45$--703.55 that no longer could be fitted 
when the coldest bin was reproduced. The effects of brief illumination are clearly visible at $d_{\rm n}=703.4$--703.5 in 
Fig.~\ref{fig_NIMBUS_SMM_NovDec15_01}. However, the temperature elevation is not sufficiently large or long--lived. A total of 15 \textsc{nimbus} 
simulations were performed with the purpose of trying to elevate the late--night temperatures. 
Variations in thermal inertia, ice abundance, erosion characteristics, and diffusivity were considered, with rationale as follows.

\begin{trivlist}
\item $\bullet$ \emph{Thermal inertia:} A higher thermal inertia would allow temperature elevations (caused by the brief late--night illumination pulse) to last longer. 
Therefore, the nominal Hertz factor was considered but also $\times 4$ and $\times 16$ that value, in order to boost the thermal inertia by factors 2 and 4.
\item $\bullet$ \emph{Ice abundance:} The heat capacity of dust is lower than that of water ice \citep[by a factor 2.6--3.6 at 100--$200\,\mathrm{K}$ for the 
laboratory measurements used by][]{davidsson21}. By increasing the dust/water--ice mass ratio $\mu$ the effective 
heat capacity $c$ of the medium is therefore reduced. That leads to larger temperature changes $\Delta T=\Delta E/c$ for a given energy change $\Delta E$, and 
enhances the response to solar insolation. A test was also made to reduce the nominal $\mathrm{CO_2/H_2O}=0.32$ abundance to half, again acting to reduce $c$.
\item $\bullet$ \emph{Erosion:} Nominally, $f_{\rm e}=6$ or 8 were applied. However, if $\mathrm{CO_2}$ would be 
allowed to withdraw under ground after having cleared the midnight bin, perhaps the late--night temperature would increase. 
Thus, erosion was switched off at various points on the $703.1\leq d_{\rm n}\leq 703.25$ interval. 
\item $\bullet$ \emph{Diffusivity:} Nominally  $\{L_{\rm p},\,r_{\rm p}\}=\{1,\,0.1\}\,\mathrm{cm}$ was applied. However, tests were made 
to strongly quench $\mathrm{CO_2}$ production and the associated cooling at $d_{\rm n}\geq 703.35$ by setting $\{L_{\rm p},\,r_{\rm p}\}=\{100,\,10\}\,\mathrm{\mu m}$ or 
$\{1,\,1\}\,\mathrm{\mu m}$ to boost the antenna temperature.
\end{trivlist}

Unfortunately, none of the 15 models showed signs of improvement, suggesting that the problem cannot be fixed by 
model parameter changes alone. The reasonably good fits elsewhere could be maintained at elevated $\mu$, but only if 
the thermal inertia was kept nominal. All combinations of higher $\Gamma$ and $\mu$ degraded substantially.

\begin{figure}
\scalebox{0.4}{\includegraphics{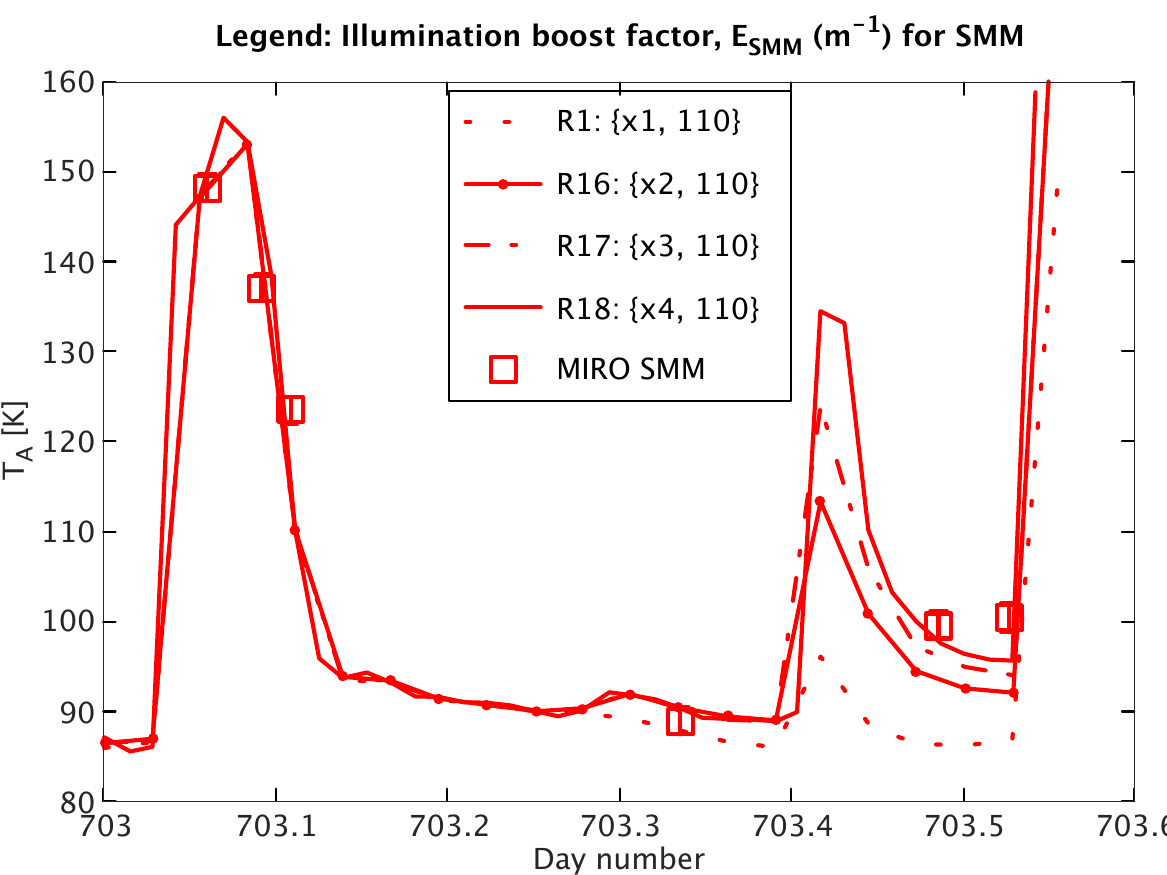}}
    \caption{2015 November and December SMM data and \textsc{nimbus} models with enhanced solar illumination 
at $d_{\rm n}\approx 703.45$--703.55. Each model has  $\mu=1$, $\nu_6=0.32$, a nominal Hertz factor, constant erosion rate with $f_{\rm e}=6$ (switching off erosion at $d_{\rm n}\geq 703.22$), and 
$\{L_{\rm p},\,r_{\rm p}\}=\{1,\,0.1\}\,\mathrm{cm}$ (changed to $\{1,\,1\}\,\mathrm{\mu m}$ at $d_{\rm n}\geq 703.35$ to quench $\mathrm{CO_2}$ net sublimation). Matching the late--night data seems to 
require a four--fold increase of illumination when the Sun is near the local horizon 
(attributed to the difference between the real--Sun having an extended disk and the model--Sun being a point source).}
     \label{fig_NIMBUS_SMM_NovDec15_02}
\end{figure}

At this point, the only option that seemed to remain was to increase the modelled solar flux near $d_{\rm n}\approx 703.45$--703.55. The justification 
for doing so is that the illumination model by \citet{davidssonandrickman14} treats the Sun like a point--source while the real Sun is an extended 
object. The model considers the Sun being above the horizon when the solar centre rises above the local nucleus topography. In most cases, when the 
Sun clears the horizon quickly and spends significant time above it, this assumption has an insignificant effect on thermophysical solutions. But during the 
short appearance of the Sun in the current problem, significant error may arise when ignoring radiation when less than half of the solar disk illuminates Aswan, 
just because the solar centre formally is below the horizon. Figure~\ref{fig_NIMBUS_SMM_NovDec15_02} shows the effect of boosting the late--night 
solar flux, assuming $\mu=1$, $\nu_6=0.32$, a nominal Hertz factor, $f_{\rm e}=6$ (switching off erosion at $d_{\rm n}\geq 703.22$), and 
$\{L_{\rm p},\,r_{\rm p}\}=\{1,\,0.1\}\,\mathrm{cm}$ (reduced to $\{1,\,1\}\,\mathrm{\mu m}$ at  $d_{\rm n}\geq 703.35$). As seen in the figure, 
the entire antenna temperature curve can in principle be reproduced if the late--night insolation has been under--estimated by a factor $\sim 4$. 

The assumption about a fixed erosion rate is not particularly realistic, and was only introduced in order to develop some 
understanding of how the solution reacts to changes in other model parameters. In order to investigate whether the 
currently available data could be used to place any constraints on the properties of Aswan, another series of 
34 \textsc{nimbus} models was considered. Here, the erosion rate of solids $\mathcal{E}$ was assumed to be proportional to the combined 
$\mathrm{H_2O}$ and $\mathrm{CO_2}$ gas production rates, $\mathcal{E}=g_{\rm e}\left(Q_{\rm H_2O}+Q_{\rm CO_2}\right)$, where 
$g_{\rm e}$ is a proportionality constant. The set of simulations systematically varied $g_{\rm e}$, the mass fraction of $\mathrm{CO_2}$ in 
the cometary material, and the thermal inertia (via the Hertz factor). The bulk density $\rho_{\rm bulk}=535\,\mathrm{kg\,m^{-3}}$ and the 
molar $\nu_6=\mathrm{CO_2/H_2O}=0.32$ ratio were held constant at all times (being the parameters that are relatively well--determined). 
Doing so, each $\mathrm{CO_2}$ mass fraction corresponds to a specific combination of $\mu$ and $\psi$. It is relevant to consider the mass fraction of 
$\mathrm{CO_2}$ as the main free parameter, because $\mathrm{CO_2}$ is driving the observed night--time temperature drop. 
A higher $\mathrm{CO_2}$ mass fraction makes the $\mathrm{CO_2}$ ice front withdraw slower, and in combination with a higher erosion rate $\mathcal{E}$ when $\mathrm{CO_2}$ is shallow, 
this tends to drive the supervolatile towards the surface. A lower $\mathrm{CO_2}$ mass fraction makes it withdraw faster, potentially leading to insignificant 
erosion and a continuous withdrawal of the $\mathrm{CO_2}$ sublimation front. 

\begin{figure}
\scalebox{0.47}{\includegraphics{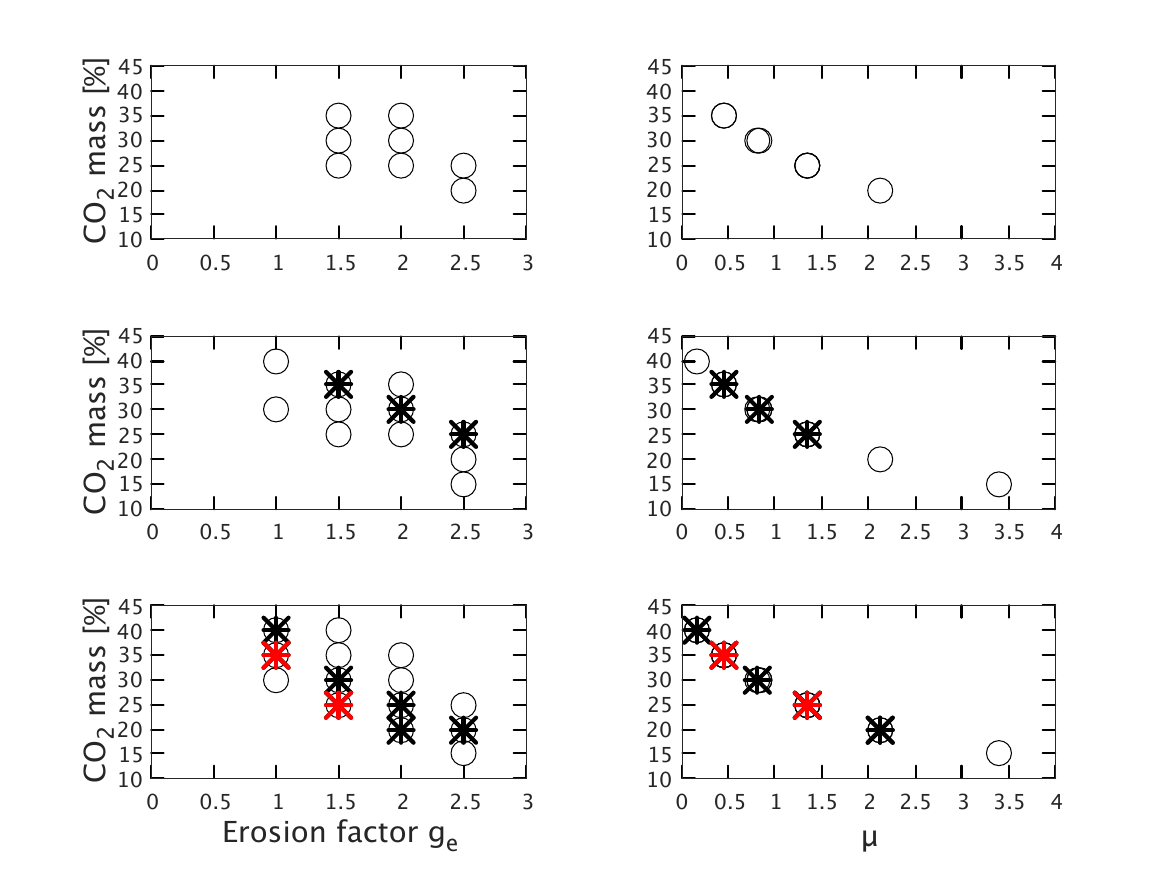}}
    \caption{This figure shows a graphic representation of the parts of parameter space covered by models 
assuming proportionality between erosion rate and vapour production rate, $\mathcal{E}=g_{\rm e}(Q_{\rm H_2O}+Q_{\rm CO_2})$, for Nov+Dec 2015. The top 
panels consider the nominal Hertz factor $h$, while in the middle and bottom panels $h$ is reduced by factors 4 and 16, respectively, thereby 
lowering the thermal inertia. The left panels show the assumed $\mathrm{CO_2}$ mass percentage (out of $\rho_{\rm bulk}=535\,\mathrm{kg\,m^{-3}}$) versus $g_{\rm e}$. 
The right panels show the $\mathrm{CO_2}$ mass percentage versus the dust/water--ice mass ratio $\mu$ (obtained by requiring a nucleus molar $\mathrm{CO_2/H_2O}=0.32$ abundance). 
Rings show considered model cases, black asterisks mark solutions that fulfil $\sqrt{\sum_i\chi_i^2/N}\leq 3$ (the first four antenna temperature bins are reasonably well reproduced), 
and red asterisk models additionally have a total July 10 -- December 31 erosion ($\sim 2\,\mathrm{m}$) being closest to the estimates from nucleus imaging (certainly $\leq 3\,\mathrm{m}$, 
probably $\leq 1.2\,\mathrm{m}$).}
     \label{fig_NIMBUS_SMM_NovDec15_03}
\end{figure}

Figure~\ref{fig_NIMBUS_SMM_NovDec15_03} is a graphic representation of the considered parameter space. Each panel row 
corresponds to an assumed level of thermal inertia: top row with nominal Hertz factor $h$ (diurnal minima ranging $\Gamma_{\rm min}=36$--$116\,\mathrm{MKS}$ and 
diurnal maxima ranging $\Gamma_{\rm max}=59$--$138\,\mathrm{MKS}$); middle row with $h/4$ ($\Gamma_{\rm min}=15$--$59\,\mathrm{MKS}$ and 
$\Gamma_{\rm max}=32$--$81\,\mathrm{MKS}$); bottom row with $h/16$ ($\Gamma_{\rm min}=8$--$38\,\mathrm{MKS}$ and $\Gamma_{\rm max}=24$--$45\,\mathrm{MKS}$). The maximum 
values occur at day when it is warmer, the minimum values occur at night when it is colder. The left column shows the considered combinations of $g_{\rm e}$ and $\mathrm{CO_2}$ mass percentage 
as rings. The right column shows the relations between applied $\mathrm{CO_2}$ mass percentages and derived $\mu$--values for the same models.

All these simulations focused on the first four bins, considering that the last two likely could be fitted if including an illumination boost 
that would not help placing any meaningful constraints on the physical properties of the surface material. Because of the small number of 
bins ($N=4$) it is not meaningful to consider the $Q$ parameter. Instead an acceptable fit was defined as having $\sqrt{\sum_i\chi_i^2/N}\leq 3$ ($\chi_i^2$ being the chi--squared residual of bin $i$ and $N$ the 
number of bins) and additionally demanding that the model passed to within $\pm 2.5\,\mathrm{K}$ of the late--night bin. Allowing for 
$\stackrel{<}{_{\sim}} 7\,\mathrm{K}$ differences between models and bins on average may be 
overly generous. However, in most cases the larger discrepancy concerns the third bin, where the modelled temperature plummets rapidly and 
easily misses the targeted temperature. The worst of the accepted cases has the model curve passing within the formal $\pm 2.5\,\mathrm{K}$ 
error bars for bins \#1, \#2, and \#4, but misses bin \#3 by $15\,\mathrm{K}$, after having had the correct temperature just $\sim 8\,\mathrm{min}$ 
earlier. 

The models that pass the quality criterion are marked with black asterisks in Fig.~\ref{fig_NIMBUS_SMM_NovDec15_03}. 
Interestingly, there are no fits for the nominal Hertz factor, and the number of successful models increase as the thermal 
inertia is reduced. The full $\min(\Gamma_{\rm min})$--$\max(\Gamma_{\rm max})$ range enveloping all successful models at all rotational phases 
is 11--$67\,\mathrm{MKS}$, or $\Gamma=40\pm 30\,\mathrm{MKS}$. There is no strong constraint on the $\mathrm{CO_2}$ abundance, in the sense that for any tested 
value in the 15--45 per cent range it was possible to find an erosion coefficient $g_{\rm e}$ that would have the model pass the quality test (except for 15 per cent $\mathrm{CO_2}$, but 
only one such model was run). 

The successful models have levels of erosion ranging 8--$26\,\mathrm{mm\,rot^{-1}}$. If such erosion takes place from the July collapse to the last 
considered observations in December (except during the two months of polar night) it would correspond to the loss of a 2--$6\,\mathrm{m}$ thick layer. 
Are such values realistic? \citet{pajolaetal17} estimate that $2.0\cdot 10^4\,\mathrm{m^3}$ worth of boulders was added to the talus under the Aswan 
cliff as a result if its collapse. The volume loss at the cliff wall was first estimated as $3.37\cdot 10^4\,\mathrm{m^3}$, based on the available pre-- and post--collapse 
Digital Terrain Models (DTMs). Subtracting the boulder volume means that $1.37\cdot 10^4\,\mathrm{m^3}$ may have eroded after the collapse due to sublimation. 
With the collapsed region constituting a surface area of $\sim 4.4\cdot 10^3\,\mathrm{m^2}$, that would suggest erosion of a $\sim 3\,\mathrm{m}$ thick layer. 
However, \citet{pajolaetal17} point out that the pre--collapse DTM has rather low resolution and does not fully capture concavities seen in images. They therefore 
propose a reduction of the volume change to $(2.20\pm 0.34)\cdot 10^4\,\mathrm{m^3}$, which would constrain the eroded layer to a thickness of $\leq 1.2\,\mathrm{m}$. 
In view of this, the models suggesting $\sim 2\,\mathrm{m}$ erosion are more attractive than the ones suggesting $\sim 6\,\mathrm{m}$ erosion. 

An additional complication is that the images used to produce the post--collapse DTM were acquired in early 2016 June \citep{pajolaetal17}, i.~e., five months 
after the last MIRO observation considered in this section. If the high level of erosion suggested by the \textsc{nimbus} and \textsc{themis} modelling 
of MIRO SMM data persisted for another five months, they could rightfully be questioned, as the removed volume would be too high. However, in section~\ref{sec_results_feb16} 
it is demonstrated that the erosion rate fell from $8$--$26\,\mathrm{mm\,rot^{-1}}$ in 2015 November and December, to $\sim 0.2\,\mathrm{mm\,rot^{-1}}$ in 2016 February. 
Therefore, the 2--$6\,\mathrm{m}$ erosion suggested by the black/red--asterisk models in Fig.~\ref{fig_NIMBUS_SMM_NovDec15_03} should be considered close to 
the terminal values. The added uncertainty regarding the timing of erosion--rate reduction further motivates placing a larger weight on the lower--erosion cases. It is noteworthy that the two models with the lowest erosion rates 
($\sim 8\,\mathrm{mm\,rot^{-1}}$) also have the smallest temperature residuals with respect to the observations, $\sqrt{\sum_i\chi_i^2/N}\approx 2.2$. Those solutions have been marked with red asterisks in 
Fig.~\ref{fig_NIMBUS_SMM_NovDec15_03}. Those models have a $\min(\Gamma_{\rm min})$--$\max(\Gamma_{\rm max})$ envelope of 12--$38\,\mathrm{MKS}$, or 
$\Gamma=25\pm 15\,\mathrm{MKS}$. The depth of the $\mathrm{CO_2}$ sublimation front in these models is $0.4\pm 0.2\,\mathrm{cm}$, which is midways between the 
failing $0.8\,\mathrm{cm}$ case and the succeeding surface--$\mathrm{CO_2}$ case in Fig.~\ref{fig_NIMBUS_SMM_NovDec15_01}. Thus, visibility of a dust and water ice 
mixture near the intrinsic nucleus mixing ratio is predicted (no dust mantle forms), but little to no exposed $\mathrm{CO_2}$ ice. The $\mathrm{CO_2}$ mass fraction is 
constrained to $30\pm 5$ per cent (out of the assumed $\rho_{\rm bulk}=535\,\mathrm{kg\,m^{-3}}$, i.~e., the absolute $\mathrm{CO_2}$ mass concentration is $160\pm 30\,\mathrm{kg\,m^{-3}}$).  
With the nucleus $\mathrm{CO_2/H_2O}=0.32$ assumption \citep[needed to explain the total nucleus $\mathrm{CO_2}$ gas production rate according to][]{davidssonetal22}, this translates 
to a dust/water--ice mass ratio of $\mu=0.9\pm 0.5$ \citep[it is reassuring that the $\mu$--value is consistent with another independent estimate for Comet 67P 
obtained by fitting the $\mathrm{H_2O}$ water production rate;][]{davidssonetal22}. The erosion coefficient is $1 \stackrel{<}{_{\sim}} g_{\rm e}\stackrel{<}{_{\sim}} 1.5$, 
i.~e., roughly equal amounts of $\mathrm{CO_2}+\mathrm{H_2O}$ vapours and dust+$\mathrm{H_2O}$ solids are being ejected. The rotationally averaged 
$\mathrm{CO_2/H_2O}$ mass ratio of the outgassed vapour is $1.6\pm 0.4$, translating to a molar ratio of $0.7\pm 0.2$. Thus, the molar ratio of the emerging gas is 
roughly twice as high as the $\nu_6=0.32$ molar ratio applied for the nucleus ices \citep[see, e.~g.,][for risks taken when equating coma and nucleus abundance ratios]{marboeufschmitt14, davidssonetal22}.

These estimates are intended to summarise the properties of the two most successful models, but it is admittedly uncertain if they apply directly to Comet 67P 
(particularly regarding the error margins). The signal originating from Aswan may be diluted with radiation originating from the surrounding dust mantle, because of 
the rather poor SMM resolution. However, the unique characteristics of the Aswan collapse site clearly did not drown in signal from the surroundings, otherwise 
the solution would just have resembled that of a simple dust mantle (as in section~\ref{sec_results_nov14}). It is also unfortunate that the SMM solution cannot be 
verified or disputed by MM observations.

\begin{figure}
\scalebox{0.40}{\includegraphics{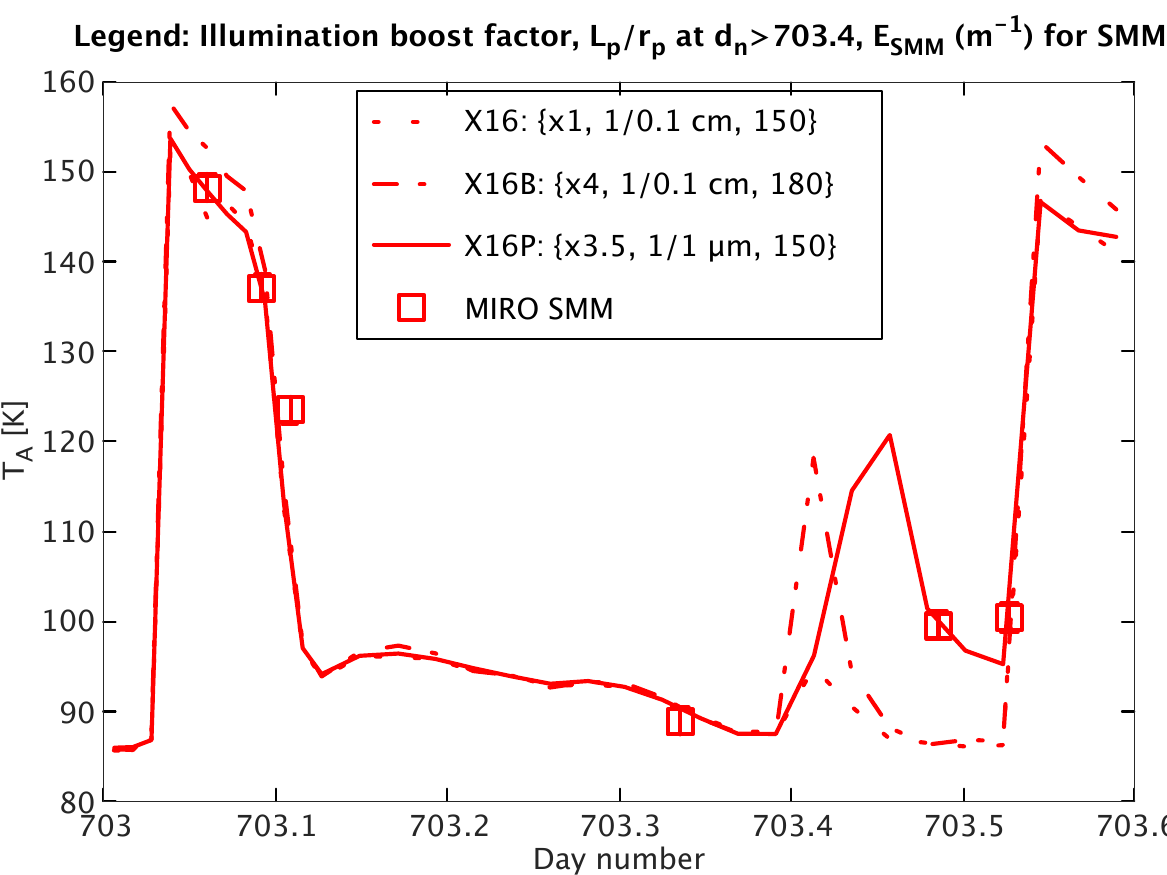}}
    \caption{November and December 2015 SMM data and \textsc{nimbus} models with erosion rate proportional to the outgassing rate, $\mathcal{E}=2\left(Q_{\rm H_2O}+Q_{\rm CO_2}\right)$. 
Each model has  $\mu=0.83$, a $\mathrm{CO_2}$ mass fraction of 30 per cent out of $\rho_{\rm bulk}=535\,\mathrm{kg\,m^{-3}}$, a reduced Hertz factor $h/4$, and 
$\{L_{\rm p},\,r_{\rm p}\}=\{1,\,0.1\}\,\mathrm{cm}$ at most rotational phases. They differ regarding level of solar illumination enhancement at $d_{\rm n}\approx 703.45$--703.55 
and diffusivity reduction at $d_{\rm n}\approx 703.4$--703.53.}
     \label{fig_NIMBUS_SMM_NovDec15_04}
\end{figure}

Finally, 5 \textsc{nimbus} models were run to test the effect of reducing the baseline diffusivity. These models applied parameters that had resulted in the smallest residuals with respect to 
the data in Fig.~\ref{fig_NIMBUS_SMM_NovDec15_03}: $h/16$, 25 per cent $\mathrm{CO_2}$ by mass in the $\rho_{\rm bulk}=535\,\mathrm{kg\,m^{-3}}$ 
material (corresponding to $\mu=1.35$ for $\nu_6=0.32$). The diffusivity was lowered an order of magnitude by considering $\{L_{\rm p},\,r_{\rm p}\}=\{1,\,0.1\}\,\mathrm{mm}$. 
Using $g_{\rm e}=1.5$ (as in Fig.~\ref{fig_NIMBUS_SMM_NovDec15_03}) resulted in the same $\mathrm{CO_2}$ sublimation front depths as in the previous 
model ($\sim 3.4\,\mathrm{mm}$ at day and $\sim 4.5\,\mathrm{mm}$ during the coldest time of night). However, because of the intrinsically lower cooling rate caused 
by the lower diffusivity, the model was too warm at the lowest $T_{\rm A}$ (for $E_{\rm SMM}=100\,\mathrm{m^{-1}}$, required to fit the day--time data). By increasing 
the erosion rate to $g_{\rm e}=2.5$, the $\mathrm{CO_2}$ front could be forced to within $\sim 1\,\mathrm{mm}$ of the surface at day, while it alternated between that 
depth and full surface exposure at night. In these conditions the model formally fitted the data ($\sqrt{\sum_i\chi_i^2/N}=3.0$ and was sufficiently cold at night), but the erosion 
was $26\,\mathrm{mm\,rot^{-1}}$ (i.~e., probably too high). Then, the diffusivity was lowered yet an order of magnitude by using $\{L_{\rm p},\,r_{\rm p}\}=\{100,\,10\}\,\mathrm{\mu m}$. 
In that case, the erosion rate was increased to $g_{\rm e}=3.0$ (resulting in $39\,\mathrm{mm\,rot^{-1}}$), which brought $\mathrm{CO_2}$ to within $1\,\mathrm{mm}$ of 
the surface at day (reproducing the first three data points for $E_{\rm SMM}=350\,\mathrm{m^{-1}}$), and exposing the $\mathrm{CO_2}$ ice at night. Yet, the fully 
exposed $\mathrm{CO_2}$ ice was not capable of lowering the antenna temperature sufficiently to reproduce the coldest bin, because the low diffusivity strongly  
reduced the net sublimation rate and cooling efficiency. I therefore conclude that $\{L_{\rm p},\,r_{\rm p}\}=\{1,\,0.1\}\,\mathrm{cm}$ yields the most realistic nominal 
diffusivity for the considered time period. 

In order to fit the last two bins, it was necessary to boost the flux during the late--night illumination episode, but also to lower the diffusivity in order to 
reduce the $\mathrm{CO_2}$ sublimation rate and the associated cooling. Most models applied $\{L_{\rm p},\,r_{\rm p}\}=\{1,\,1\}\,\mathrm{\mu m}$ at 
$d_{\rm n}=703.4$--$703.53$, which essentially switched off $\mathrm{CO_2}$ activity completely. Figure~\ref{fig_NIMBUS_SMM_NovDec15_04} exemplifies 
such models with the erosion--rate proportional to the outgassing rate,  that applied different levels of late--night illumination boosts and diffusivities. It is clear that enhanced 
solar flux is a necessary but not sufficient condition: boosting the flux by a factor 4 but maintaining $\{L_{\rm p},\,r_{\rm p}\}=\{1,\,0.1\}\,\mathrm{cm}$ creates an antenna temperature 
increased that is far too shortlived. Here, the last two bins could be reproduced by lowering the diffusivity to $\{L_{\rm p},\,r_{\rm p}\}=\{1,\,1\}\,\mathrm{\mu m}$, which extended the 
lifetime of the antenna temperature enhancement, even at a less intense 3.5--factor boost. Further tests were performed to investigate how much that 
rather extreme diffusivity assumption could be relaxed without introducing unwanted cooling. They showed that the diffusivity could be increased to 
$\{L_{\rm p},\,r_{\rm p}\}=\{100,\,10\}\,\mathrm{\mu m}$, because the model would remain to within $\pm 2.5\,\mathrm{K}$ of the data. This is still a 
2 orders of magnitude drop with respect to the nominal $\{L_{\rm p},\,r_{\rm p}\}=\{1,\,0.1\}\,\mathrm{cm}$ applied at other rotational phases. 

If this change in diffusivity is real, one may speculate that the lower value $\{L_{\rm p},\,r_{\rm p}\}=\{100,\,10\}\,\mathrm{\mu m}$ is representative of the 
undisturbed deeper material, that is being excavated at a slow pace at night. It would imply a rather evenly distributed medium of $\mathrm{\mu m}$--sized grains, lacking 
lumpiness on millimetre size--scales and above. If so, the substantially higher diffusivity $\{L_{\rm p},\,r_{\rm p}\}=\{1,\,0.1\}\,\mathrm{cm}$ is a temporary consequence of 
the violent daytime sublimation that may cause a laterally variable degree of erosion. Such a rugged and rough surface, with lumpiness on millimetre to centimetre 
size--scales in the top $\sim 1\,\mathrm{cm}$, would have a diffusivity consistent with $\{L_{\rm p},\,r_{\rm p}\}=\{1,\,0.1\}\,\mathrm{cm}$.

\subsection{February 2016: seven months after the collapse} \label{sec_results_feb16}

\begin{figure}
\scalebox{0.40}{\includegraphics{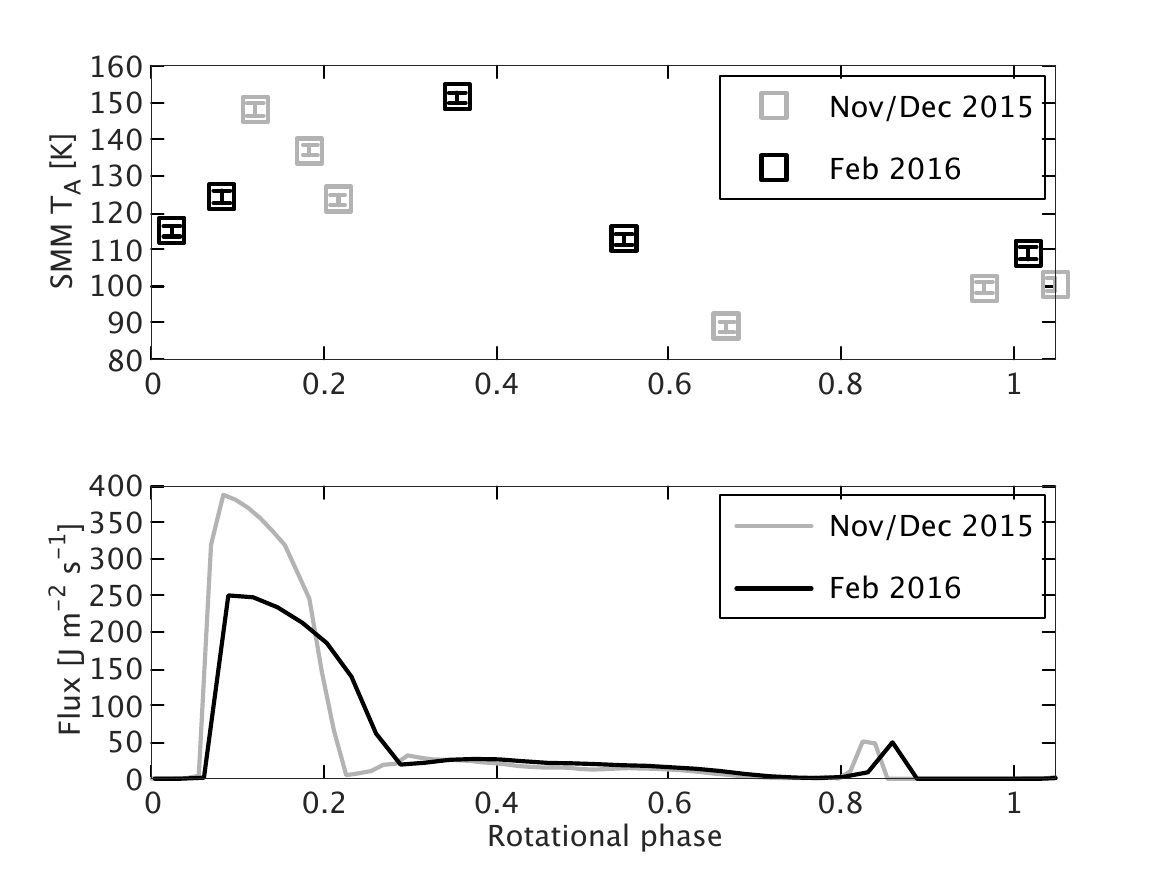}}
    \caption{\emph{Upper panel:} MIRO SMM antenna temperatures for the 2015 November--December and 2016 February periods versus rotational phase (the curves are time--shifted 
such that the daytime illumination peaks occur simultaneously). \emph{Lower panel:} The total illumination fluxes for the same periods and rotational phases. Note the non--zero fluxes 
at night due to nucleus self heating, and the late--night illumination episode caused by the complex nucleus geometry.}
     \label{fig_compare_NovDec15_Feb16}
\end{figure}

The 2016 February data set \citep[PDS website\footnote{https://pds-smallbodies.astro.umd.edu/holdings/ro-c-miro-3-ext1-\\67p-v3.0/dataset.shtml};][]{hofstadteretal18d} consists of merely nine SMM observations and the same number of simultaneous MM observations, 
acquired during a 23 day period when \emph{Rosetta} reduced its distance to the nucleus from $55.7\,\mathrm{km}$ (three observations) 
to $29.7$--$33.9\,\mathrm{km}$ (the remaining six observations). They fall into five bins, time--shifted to a master period starting 
2016 February 15, $10:35:39\,\mathrm{UTC}$, or at $d_{\rm n}=776.4414$. None of them had to be removed. Figure~\ref{fig_compare_NovDec15_Feb16} compares the 
2016 February SMM observations with those from November and December 2015, and also shows the corresponding illumination fluxes.

During the months in question the peak flux drops by 65 per cent. Yet, the early--night 2016 bin is as warm as the midday 2015 peak. 
This suggests that a significant change has taken place -- the daytime sublimation cooling ought to be weaker in 2016 than in 2015. The 2016 midnight SMM 
antenna temperature ($\sim 108\,\mathrm{K}$) is significantly warmer than the corresponding 
2015 temperature ($\sim 89\,\mathrm{K}$), despite almost identical self--heating fluxes. Yet, this bin is still below the water--cooled $\mathrm{CO_2}$--free models in 
Fig.~\ref{fig_BTM_NovDec15_dustH2O}. This suggests that $\mathrm{CO_2}$--cooling still plays a role in 2016, though smaller than in 2015. A data search for 2016 May did not 
result in enough observations to motivate a dedicated investigation, but showed that the temperature late at night was similar to that in 2016 June. It is reassuring that 
all three data sets include very low nighttime antenna temperatures, as it strengthens the notion that this is a real feature of Aswan.

\begin{figure}
\scalebox{0.40}{\includegraphics{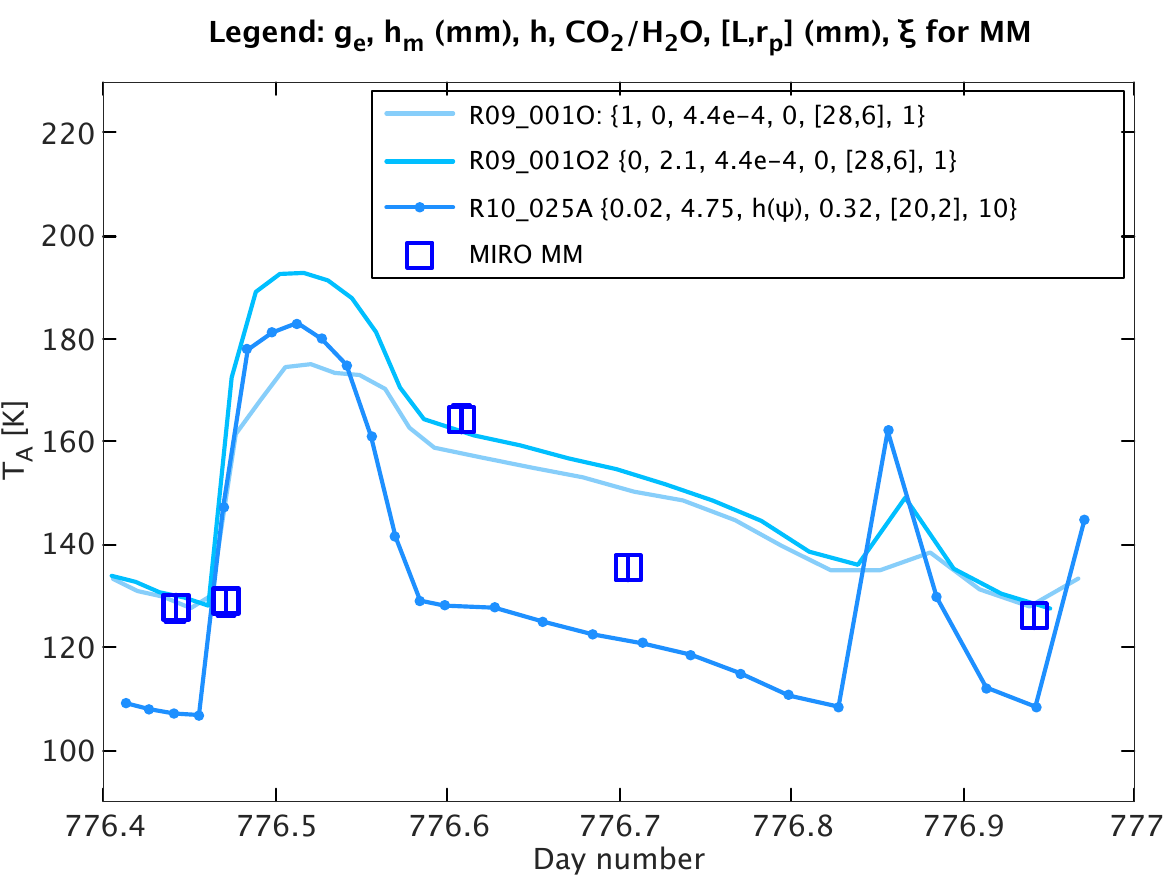}}
    \caption{In 2016 February the presence of a few--millimetres thick dust--mantle seem necessary to explain observations (R10\_001O versus R10\_001O2). 
All models have the following parameters in common: $\mu=1$, $\rho_{\rm bulk}=535\,\mathrm{kg\,m^{-3}}$, $E_{\rm MM}=70\,\mathrm{m^{-1}}$. 
R10\_001O and R10\_001O2 (both with $h=4.4\cdot 10^{-4}$ and near--surface thermal inertia ranging $15\,\mathrm{MKS}$ at night and $50\,\mathrm{MKS}$ at day) have 
differing erosion rates, $g_{\rm e}=1$ (no mantle) and $g_{\rm e}=0$ (at this point yielding a $h_{\rm m}=3.4\,\mathrm{mm}$ mantle), respectively. R10\_025A uses the 
\protect\citet{shoshanyetal02} formula with a $h\geq 4.3\cdot 10^{-4}$ ceiling (mantle thermal inertia ranging $9\,\mathrm{MKS}$ at night and $36\,\mathrm{MKS}$ at day; ice/dust mixture reaching $260\,\mathrm{MKS}$). The $\mathrm{CO_2}$ front is at 17.1--$20.1\,\mathrm{mm}$. The late--night illumination is boosted $\times 8$ at $d_{\rm n}=776.35$--776.40 
and 776.85--776.9. for R10\_025A.}
     \label{fig_NIMBUS_MM_Feb16_mantle}
\end{figure}

As before, modelling was performed in themed stages, to better understand model requirements needed to reproduce the data. Roughly 110 \textsc{nimbus} 
models and 250 \textsc{themis} models were used to investigate the role of: 1) the dust mantle; 2) the heat conductivity; 3) the surface erosion; 4) the opacity.

\emph{Role of the dust mantle.} \textsc{nimbus} and \textsc{themis} were first used to investigate whether a medium consisting of only refractories and water ice is 
capable of explaining the 2016 February MM data. Eleven models were considered, most having the same fixed \textsc{btm}--style 
constant values for dust or water--ice densities, specific heat capacities, and heat conductivities as in Table~2 of \citet{davidssonetal22b}. 
The dust always had a 0.3 filling factor and all but two models included water ice (with $\mu=4$). A fixed Hertz factor $h=4.4\cdot 10^{-4}$ 
was applied to cause a $\sim 30\,\mathrm{MKS}$ (solid--state conduction) thermal inertia for dust mantle material. Two parameters 
were varied:  the initial dust mantle thickness ($\leq 4\,\mathrm{cm}$), and the diffusivity with $\xi=1$ and $\{L_{\rm p},\,r_{\rm p}\}$ ranging 
from $\{1,\,0.1\}\,\mathrm{mm}$ (where radiative heat transport is negligible) to $\{28,\,6.6\}\,\mathrm{mm}$ (where the effective 
thermal inertia may reach $90\,\mathrm{MKS}$). In some cases, erosion (at a rate given by that of $\mathrm{H_2O}$  gas production, i.~e., $g_{\rm e}=1$) 
was employed in order to keep mantles thin. 

The models just described represent comparably dense media, with $975\,\mathrm{kg\,m^{-3}}$ mantles and $1219\,\mathrm{kg\,m^{-3}}$ 
dust/ice interiors. They were compared with two other models that instead applied $\mu=1$, mantles with density $267\,\mathrm{kg\,m^{-3}}$ and dust/ice interiors with $535\,\mathrm{kg\,m^{-3}}$, 
with \textsc{nimbus}--style temperature--dependent heat capacities and heat conductivities. The Hertz factor $h=4.4\cdot 10^{-4}$ was 
kept, yielding a thermal inertia around $15\,\mathrm{MKS}$ for this new composition, unless increased by efficient radiative heat transfer.  

Though differing slightly in details, according to their individual parameter settings, all these models were qualitatively 
similar, and suffered from the same problem. Though $E_{\rm MM}$ values could be found that reproduced the 
dawn data, there were interesting discrepancies at night. This is illustrated in Figure~\ref{fig_NIMBUS_MM_Feb16_mantle}. 
First, the early--night bin at $d_{\rm n}=776.61$ is relatively warm compared to all models that have water ice exposed on 
the surface. For example, model R09\_001O is $\sim 7\,\mathrm{K}$ too cold, while model R09\_001O2 with identical 
parameters (see the  Fig.~\ref{fig_NIMBUS_MM_Feb16_mantle} caption) but with a $h_{\rm m}=3.4\,\mathrm{mm}$ thick dust mantle 
reproduces this data point. It is therefore likely that a thin dust--mantle had developed by 2016 February. Second, all models 
are too warm at the late--night bin ($d_{\rm n}=776.71$). Even model R09\_001O with exposed water ice is $15\,\mathrm{K}$ warmer 
than the data. Such a strong cooling points to presence of $\mathrm{CO_2}$. To explore the properties of $\mathrm{CO_2}$--bearing materials 55 \textsc{nimbus} 
models with surface--absorption of solar energy ($\zeta=0$) were first considered.

\emph{The role of heat conduction.} A first group of 15 models all had significant porosity variation with depth and applied the \citet{shoshanyetal02} formula where the 
solid--state heat conductivity depends strongly on porosity. The porosity variability was obtained by considering materials with $\mu=1$ and $\mathrm{CO_2/H_2O}=0.32$, 
compress those to a fixed bulk density ($\rho_{\rm bulk}=535$--$600\,\mathrm{kg\,m^{-3}}$) and then remove all $\mathrm{H_2O}$ and/or $\mathrm{CO_2}$ in 
near--surface layers of various thicknesses. The procedure assumes that an initially uniform mixture of refractories and ices is devolatilised without any further structural changes 
after evaporation. This yielded dust mantles with porosity $\psi=0.93$--$0.94$, dust/$\mathrm{H_2O}$ layers with $\psi=0.70$--$0.73$, and  dust/$\mathrm{H_2O}$/$\mathrm{CO_2}$ 
interiors with $\psi=0.59$--$0.63$. The \citet{shoshanyetal02} formula breaks down for $\psi>0.909$, therefore either ceilings were applied where the Hertz factor was forced to remain at 
$h\geq 4.4\cdot 10^{-4}$--0.025 or the mantles were compressed to $\psi=0.81$--0.83 (equivalently, $\rho=535$--$600\,\mathrm{kg\,m^{-3}}$). Erosion rates were kept low 
($g_{\rm e}=0.02$--$0.05$) to preserve mantles. All these models had $\{L,\,r_{\rm p}\}=\{20,\,2\}\,\mathrm{mm}$ and $\xi=10$. 
This yielded a thermal inertia in the highly porous dust mantles that ranged 30--$100\,\mathrm{MKS}$ at day but 10--$70\,\mathrm{MKS}$ at night, showing the effect 
of enhanced solid--state conduction and dominating radiative heat conduction at high temperature. However, the more compact and ice--rich interior had quasi--constant 
thermal inertia in the range 270--$400\,\mathrm{MKS}$. This strong elevation of thermal inertia with depth, in combination with shallow $\mathrm{CO_2}$ sublimation 
fronts, created peculiar antenna temperature curves illustrated by model R10\_025A in Fig.~\ref{fig_NIMBUS_MM_Feb16_mantle}. Here, the nighttime antenna 
temperature is far too low, and the slope after sunset is much more shallow than indicated by the data bins.

\begin{figure*}
\centering
\begin{tabular}{cc}
\scalebox{0.43}{\includegraphics{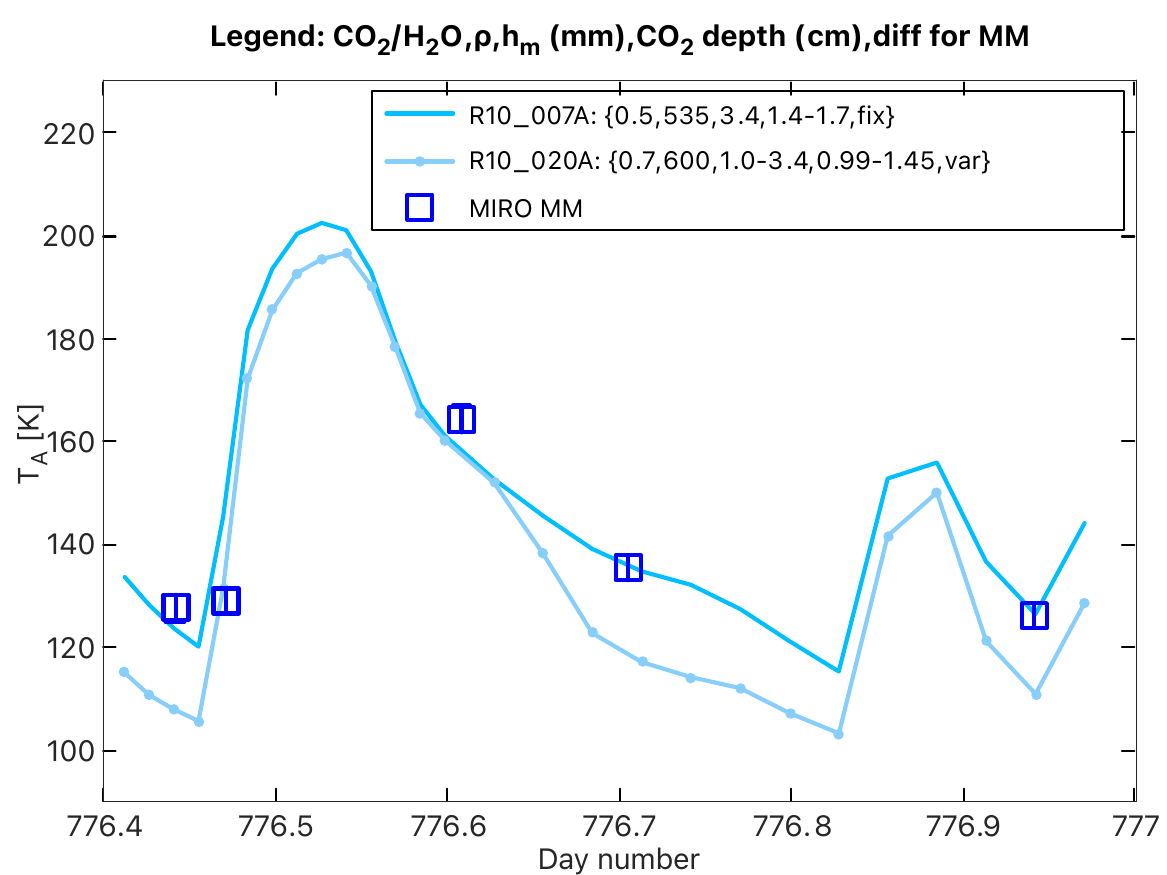}} & \scalebox{0.43}{\includegraphics{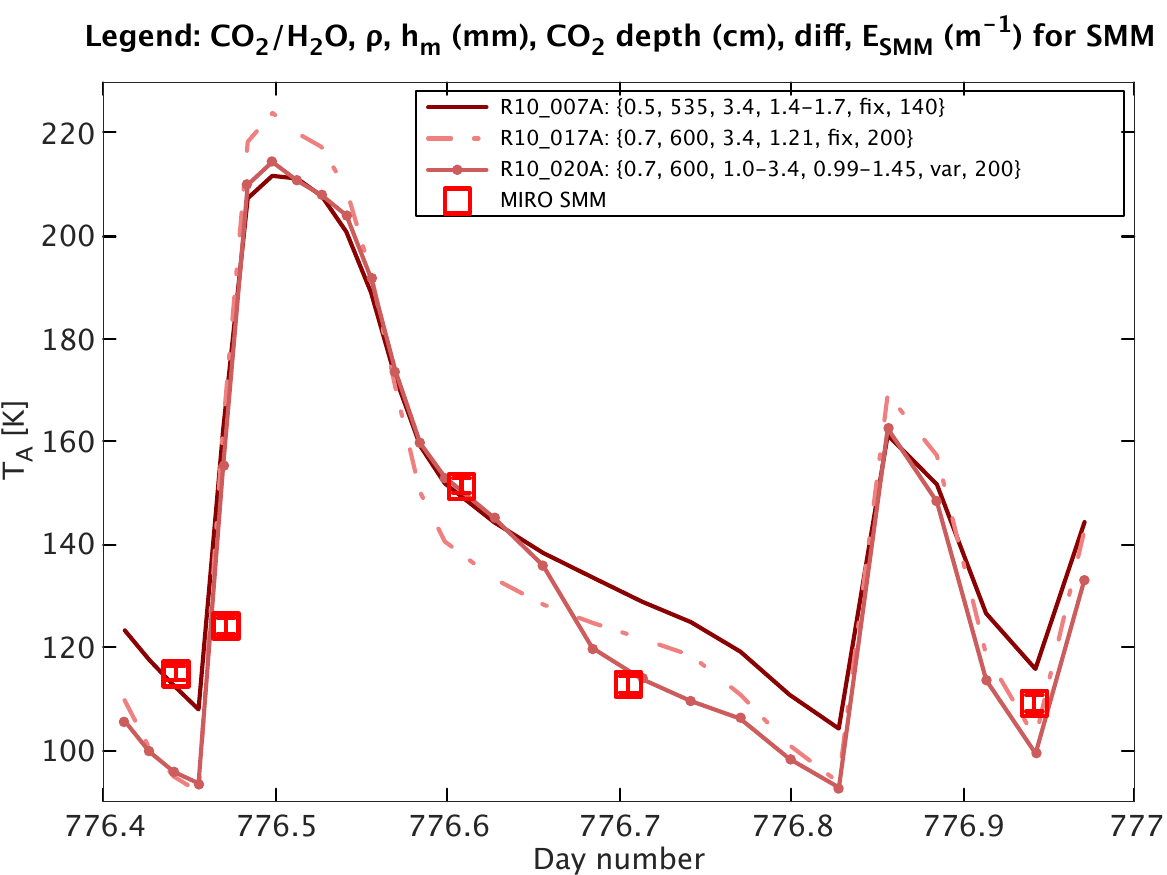}}\\
\end{tabular}
     \caption{2016 February MIRO data and examples of models that work at one wavelength channel but not the other. All models have $\mu=0.5$, $h=4.4\cdot 10^{-4}$, $g_{\rm e}=0.05$, and 
an $\times 8$ illumination flux boost at  $d_{\rm n}=776.85$--$776.9$ at the master period (and correspondingly for previous periods).  All models have a nominal diffusivity regulated by 
$\{L,\,r_{\rm p}\}=\{20,\,2\}\,\mathrm{mm}$ and $\xi=10$, except R10\_020 that switches to a 470 times higher value of $\{56,\,12,\}\,\mathrm{mm}$ and $\xi=1$ when the surface temperature is 
$120\leq T_{\rm s}\leq 150\,\mathrm{K}$, i.e., it is variable (var) instead of fixed (fix). The parameter values (abundance  $\mathrm{CO_2/H_2O}$, density $\rho$, dust mantle thickness $h_{\rm m}$ and 
$\mathrm{CO_2}$ front depth during the master period) are given in the legends. \emph{Left:}  MM data and models for $E_{\rm MM}=70\,\mathrm{m^{-1}}$. R10\_007A works at MM but not at SMM. 
\emph{Right:} SMM data and models for $E_{\rm SMM}$ given in the legend. R10\_020A works at SMM but not at MM. Model R10\_017A is identical to R10\_020A except its diffusivity is fixed (with some 
resulting differences in mantle thickness and $\mathrm{CO_2}$ front depth).}
     \label{fig_NIMBUS_MM_SMM_Feb16_falsefit}
\end{figure*}

Another 10 \textsc{nimbus} models also considered $\mu=1$ and $\mathrm{CO_2/H_2O}=0.32$, but were compressed to a uniform 
$\rho_{\rm bulk}=400$--$600\,\mathrm{kg\,m^{-3}}$ after ice removal from the surface layers. This procedure assumes that an initially uniform mixture of 
refractories and ices is devolatilised, weakened, and that the remaining material is contracting when the compressive strength is reduced. This yielded dust mantles with porosity 
$\psi=0.82$--$0.88$, dust/$\mathrm{H_2O}$ layers with $\psi=0.58$--$0.72$, and  dust/$\mathrm{H_2O}$/$\mathrm{CO_2}$ interiors with $\psi=0.59$--$0.72$. 
Despite the porosity variability, a depth--invariant Hertz--factor was applied. Nominally, $h=4.4\cdot 10^{-4}$ was used, but cases with a factor 4--8 reduction of $h$  
were considered as well. These models used either $\{L_{\rm p},\,r_{\rm p}\}=\{20,\,2\}\,\mathrm{mm}$ and $\xi=10$, or 
$\{L_{\rm p},\,r_{\rm p}\}=\{10,\,1\}\,\mathrm{mm}$ and $\xi=7.07$. In this manner, the diffusivity was held constant but different levels of radiative 
heat conduction (regulated by $r_{\rm p}$) were applied. This yielded a thermal inertia in the highly porous dust mantles that ranged 20--$60\,\mathrm{MKS}$ at 
day but 2--$16\,\mathrm{MKS}$ at night. Most models in this group reproduced the late--night MM bin for $E_{\rm MM}=50\,\mathrm{m^{-1}}$, matched the midnight bin, and had 
a steeper nighttime slope more akin to the data (though the modelled early--night MM antenna temperature still were somewhat too low). The mutual differences between 
these models were rather small, showing that the exact values for the solid--state and radiative heat conductivities are not that important, as long as the resulting thermal 
inertia remains low ($\stackrel{<}{_{\sim}} 50\,\mathrm{MKS}$). However, the drastic improvement with respect to the first group of models (represented by R10\_025A in Fig.~\ref{fig_NIMBUS_MM_Feb16_mantle}) 
demonstrates that a low quasi--constant Hertz  factor must apply throughout the top region regardless of composition and bulk density. This suggests that the physical connectivity and 
resulting solid--state heat conductivity of the near--surface material is lower (by 1--2 orders of magnitude) than given by the \citet{shoshanyetal02} formula.

\emph{The role of surface erosion.} Based on the findings above, the remaining 30 $\zeta=0$ models applied $h=4.4\cdot 10^{-4}$. The goal was to 
determine which combination of $\mathrm{CO_2}$ sublimation front depth and diffusivity that, on the one hand, reproduces the MM data, 
and on the other hand, results in an erosion rate that keeps $\mathrm{CO_2}$ ice at the desired depth while not permanently removing the dust mantle. 
A first group of 20 simulations considered erosion rates that were proportional to the total gas production rate \emph{at all times}. 

The first 13 of those models focused on reproducing the nighttime data, while ignoring discrepancies near dawn that potentially would require 
the same type of late--night illumination boost as in Sec.~\ref{sec_results_novdec15_NIMBUS}. Most model materials had $\mu=1$, $\mathrm{CO_2/H_2O=0.32}$, and 
were compressed to $535\,\mathrm{kg\,m^{-3}}$ before ices were removed from the near--surface region. Initial dust mantle thicknesses ranged 0--$3.4\,\mathrm{mm}$, 
$\mathrm{CO_2}$ front depths ranged 0.8--$6\,\mathrm{cm}$, diffusion parameters ranged $\{2,\,0.2\}\leq\{L_{\rm p},\,r_{\rm p}\}\leq \{28,\,6.6\}\,\mathrm{mm}$ and 
$1\leq \xi\leq 10$, while the erosion proportionality constant ranged $0.05\leq g_{\rm e}\leq 1$. Even with the deepest $\mathrm{CO_2}$ front and lowest diffusivity being considered, 
dust mantles would not form or remain if $g_{\rm e}=1$. Such models were too cold throughout the night, though not as extreme as R10\_025A in Fig.~\ref{fig_NIMBUS_MM_Feb16_mantle}. 
At $0.1\leq g_{\rm e}\leq 0.5$ dust mantles (2--$6\,\mathrm{mm}$ thick) would form or remain and the early--night bin would be reproduced, but the models were too warm at the late--night 
bin when the $\mathrm{CO_2}$ front was $\geq 4\,\mathrm{cm}$ from the surface. The $\mathrm{CO_2}$ cooling clearly needed to increase late at night, but in such a manner 
that the corresponding increase in outgassing rate and erosion did not remove dust from the mantle top faster than the rate by which water evaporation liberated grains at the mantle bottom. 
Increasing the diffusivity to  $\{L_{\rm p},\,r_{\rm p}\}=\{28,\,6.6\}\,\mathrm{mm}$, $\xi=1$ with $\mathrm{CO_2}$ at $4\,\mathrm{cm}$ depth while using $g_{\rm e}=0.05$ 
provided the desired late--night cooling, but then the positive effect of having a dust mantle was ruined. Only a lower diffusivity obtained by setting $\xi=10$ and 
$\{L_{\rm p},\,r_{\rm p}\}=\{20,\,2\}\,\mathrm{mm}$ (to keep the mantle thermal inertia in the 15--$60\,\mathrm{MKS}$ diurnal range), combined 
with $g_{\rm e}=0.05$ and a rather shallow $\mathrm{CO_2}$ front at 1.4--$1.7\,\mathrm{cm}$ was promising, as the two nightside bins were nearly reproduced. 
However, keeping the $\mathrm{CO_2}$ at such shallow depths during low erosion rates required a higher absolute concentration of the supervolatile, achieved by setting 
$\mu=0.5$ and $\mathrm{CO_2/H_2O}=0.5$. Furthermore, increasing the density of ice--depleted regions to $535\,\mathrm{kg\,m^{-3}}$ facilitated mantle survival 
as it further slowed its erosion rate for a given outgassing rate. Because the dawn antenna temperatures were too low, a boost of the brief late--night illumination episode 
(see the lower panel of Fig.~\ref{fig_compare_NovDec15_Feb16}) was introduced. Seven models of such type indicated that the peak flux at $d_{\rm n}=776.35$--776.40 
(prior to the master period) and $d_{\rm n}=776.85$--776.9 (master period) needed to be increased by a factor 8 
with respect to the nominal calculation of illumination conditions. Such boosts had important local effects, but did not change the model curves near the nighttime bins. As shown in the 
left panel of Fig.~\ref{fig_NIMBUS_MM_SMM_Feb16_falsefit} this model (R10\_007A) provided a reasonable fit to the MM bins. It suggested that the $\mathrm{CO_2}$ was 
$\sim 1\,\mathrm{cm}$ deeper in 2016 February than in 2015 November and December, and that the capability of a given amount of vapour to remove dust from the mantle had 
dropped by at least a factor 20. 

Once the first promising model that reproduced the MM data had been obtained, it was tested at the SMM wavelength. However, the right panel of Fig.~\ref{fig_NIMBUS_MM_SMM_Feb16_falsefit} 
shows that R10\_007A produces a SMM antenna temperature that is $\sim 17\,\mathrm{K}$ too high at the late--night bin at $d_{\rm n}=776.71$, though it fits rather well elsewhere.  Clearly, 
the physical model had to change in order to fit MM and SMM simultaneously. The key problem was to find SMM models that are sufficiently warm in the early night but substantially colder late at night.

\begin{figure*}
\centering
\begin{tabular}{cc}
\scalebox{0.43}{\includegraphics{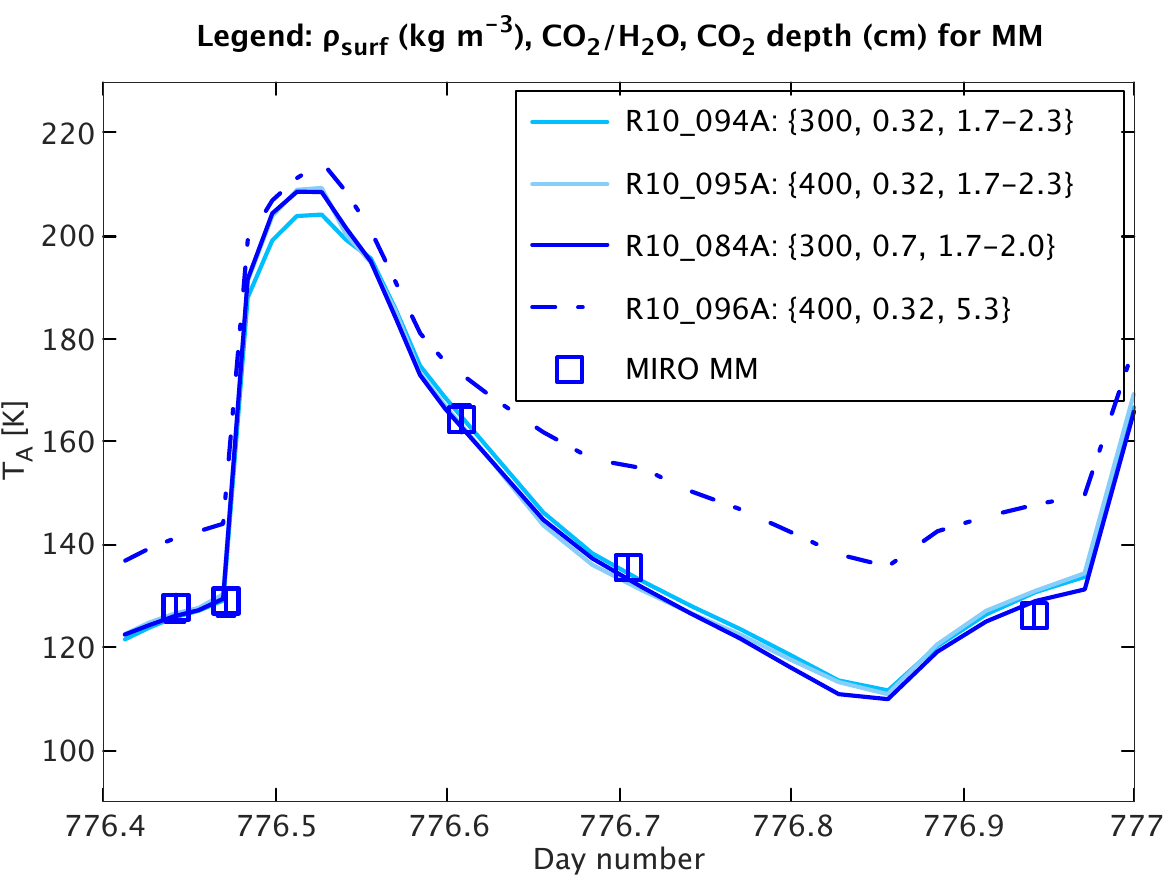}} & \scalebox{0.43}{\includegraphics{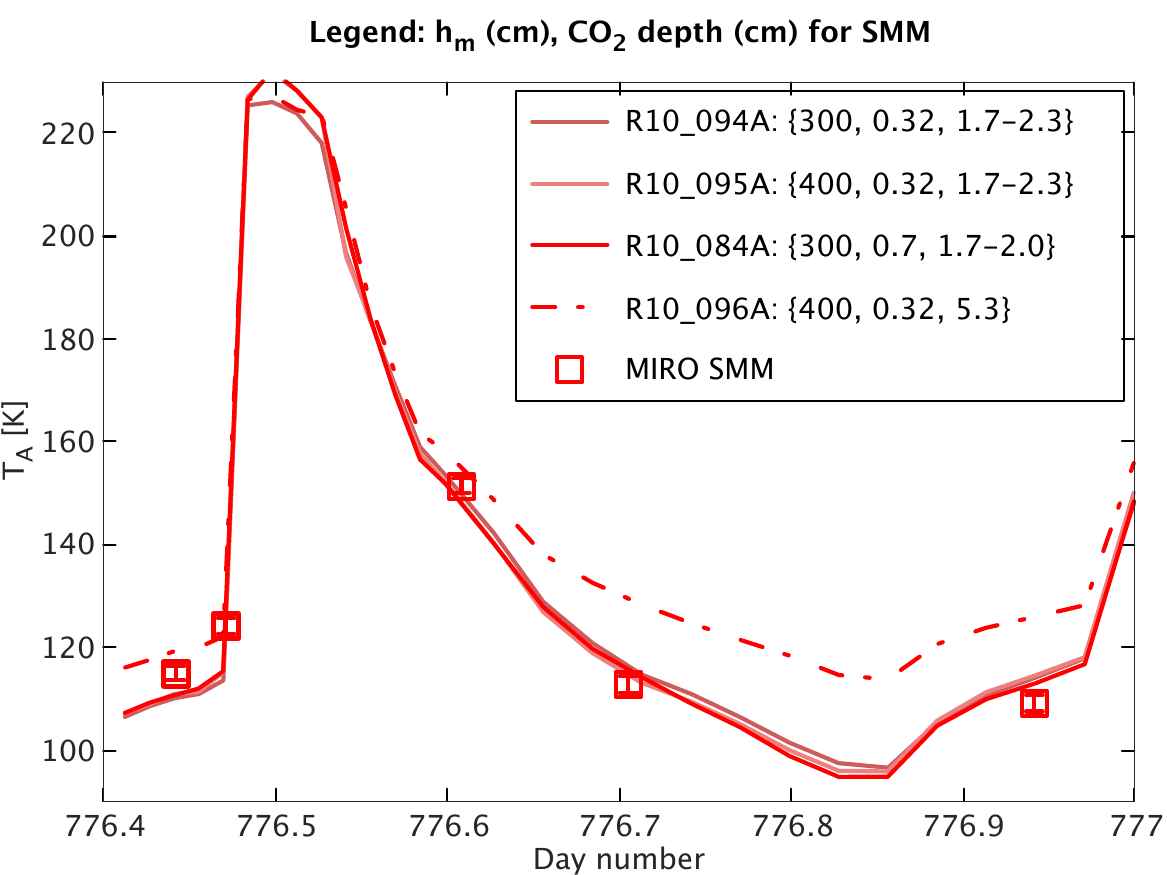}}\\
\end{tabular}
     \caption{2016 February MIRO data and examples of models that worked at both wavelength channels. The formally best MM model (R10\_095A) with $Q_{\rm MM}=0.019$ has 
$\mu=1$, $\mathrm{CO_2/H_2O}=0.32$, $\rho_{\rm bulk}=400\,\mathrm{kg\,m^{-3}}$, the $\mathrm{CO_2}$ front depth at $1.7$--$2.3\,\mathrm{cm}$, $h=4.4\cdot 10^{-4}$, 
$\{L_{\rm p},\,r_{\rm p}\}=\{20,\,2\}\,\mathrm{mm}$ and $\xi=10$,  $\zeta=5\cdot 10^{-3}\,\mathrm{m}$ at $d_{\rm n}=776.54$--776.61 and $\zeta=3.5\cdot 10^{-2}\,\mathrm{m}$ at 
$d_{\rm n}\geq 776.62$, and a late--night illumination boost regulated by $\mathcal{G}=1.1$ and  $\mathcal{F}=9.9$. The other models differ from R10\_095A as indicated in the legends. 
\emph{Left:} MM data and models for $E_{\rm MM}=70\,\mathrm{m^{-1}}$. \emph{Right:} SMM data and models for $E_{\rm SMM}=400\,\mathrm{m^{-1}}$.}
     \label{fig_NIMBUS_MM_SMM_Feb16_bestfit}
\end{figure*}

First, a larger dust/water--ice mass ratio ($\mu=3$--4) was applied in combination with stronger erosion ($g_{\rm e}\leq 4$) only when the illumination flux 
or the $\mathrm{CO_2}$ front vapour pressure exceeded given values. A larger $\mu$ (lower water abundance) facilitates a fast dust mantle thickening at times when no erosion 
is applied (thus reducing the distance between the $\mathrm{H_2O}$ and $\mathrm{CO_2}$ sublimation fronts), while vigorous erosion at specific rotational phases removes 
most of the mantle (thus bringing both $\mathrm{H_2O}$ \emph{and} $\mathrm{CO_2}$ close to the surface). The timing of erosion (or lack thereof) should ideally allow for 
some hot mantle to remain at the time of the first night bin, and for $\mathrm{CO_2}$ to be sufficiently close to the surface at the second night bin. Though the erosion mechanism remains 
elusive, the physically plausible driving force is here considered being either sunlight \citep[e.~g., electrostatic lofting; ][]{hartzelletal22} or gas drag. The illumination criterion (erosion only 
when the flux exceeds $230\,\mathrm{J\,m^{-2}\,s^{-1}}$) concentrates erosion at noon, thus allowing for mantle build--up during the morning. The pressure criterion (erosion only 
when the $\mathrm{CO_2}$ pressure exceeds $p_{\rm crit}$, with applied values in the range $0.8\leq p_{\rm crit}\leq 2\,\mathrm{Pa}$) concentrates erosion in the afternoon and 
early night because of thermal lag, thus allowing for mantle build--up throughout most of the day. None of these models were successful. Once the dust mantle is removed, it is not 
re--established in the short time available, and these models resemble R10\_025A in Fig.~\ref{fig_NIMBUS_MM_Feb16_mantle}.

Second, attempts were made to strongly increase the diffusivity when the surface temperature was within a given interval. The idea was to strongly enhance the net $\mathrm{CO_2}$ 
ice sublimation and cooling around the time of the late--night bin, though the physical reason for such an enhancement is not obvious. Model R10\_020A in the right panel of 
Fig.~\ref{fig_NIMBUS_MM_SMM_Feb16_falsefit} shows that the SMM curve can be reproduced if the diffusivity increases by a factor 470 when $130\leq T_{\rm crit}\leq 150\,\mathrm{K}$ 
(changing from a nominal $\{L_{\rm p},\,r_{\rm p}\}=\{20,\,2\}\,\mathrm{mm}$ and $\xi=10$ to $\{L_{\rm p},\,r_{\rm p}\}=\{56,\,12\}\,\mathrm{mm}$ and $\xi=1$). 
For reference, the figure also shows model R10\_017A that has the same parameters except that the diffusivity remains fixed at the smaller value. However, model R10\_020A does not 
provide an acceptable fit in the MM (Fig.~\ref{fig_NIMBUS_MM_SMM_Feb16_falsefit}, left panel). We therefore have two models (R10\_007A and R10\_020A) that work for one 
of the channels but not the other. Clearly, it is valuable if microwave instruments have capability to measure the antenna temperature at several wavelengths, as it helps identifying false solutions.

\emph{The role of opacity.} Model R10\_020A in Fig.~\ref{fig_NIMBUS_MM_SMM_Feb16_falsefit} shows that the shallow sub--surface probed by 
the SMM channel is sufficiently cool late at night, but that more energy needs to be deposited at the slightly larger depths probed by the MM channel. The 
near--surface temperature reduction cannot be achieved solely by a shallow $\mathrm{CO_2}$ sublimation front, as that would lower the temperature at depth as well. 
No combination of solid--state or radiative heat conductivity, density/porosity variation with depth, diffusivity, or compositional stratification tested thus far in $\sim 70$ models seem able to 
accomplish the necessary temperature profile. A remaining option is to relax the assumption of infinite opacity applied for illumination. All solar light and infrared self--illumination 
is thus far assumed to be absorbed at the very surface. However, \citet{davidssonetal22b} demonstrated that some MIRO data seems to require a solid--state greenhouse effect, 
which is a consequence of a finite opacity in the comet material, so that radiative energy is deposited within a near--surface layer. It is indeed contradictory to allow for transport 
of internally emitted infrared photons, while assuming complete opaqueness for visual or infrared photons arriving from the exterior. Therefore, another 47 \textsc{nimbus} models 
were run with light absorption according to Beer's law, parameterised by the $e$--folding scale $\zeta$ \citep[see, e.~g., ][]{davidssonandskorov02a,davidssonandskorov02b}. 
They all considered $f_{\rm e}=0.05$ erosion.

A first set of 12 models had fixed $\mu=1$, $\mathrm{CO_2/H_2O}=0.32$, a flat $400\,\mathrm{kg\,m^{-3}}$ bulk 
density after removal of $\mathrm{H_2O}$ (top $2.1\,\mathrm{mm}$) and $\mathrm{CO_2}$ (top 8--$20\,\mathrm{mm}$) ices, $h=4.4\cdot 10^{-4}$, 
$\{L_{\rm p},\,r_{\rm p}\}=\{20,\,2\}\,\mathrm{mm}$, $\xi=10$, and an $\times 8$ flux boost at 
$776.85\leq d_{\rm n}\leq 776.9$ for the master period (and corresponding boosts at earlier periods). Different $\zeta>0$ values, active throughout the nucleus rotation or during parts of it, were tested 
for different initial $\mathrm{CO_2}$ depths in the 8--$20\,\mathrm{mm}$ range. It was clear that $\zeta>0$ could not prevail during daytime, as temperatures were too high. 
Good results for the MM \emph{and} SMM nighttime bins were obtained if switching from $\zeta=0$ to $\zeta=5\cdot 10^{-3}\,\mathrm{m}$ at $d_{\rm n}=776.54$ 
(between the $d_{\rm n}=776.50$ illumination peak and the first nighttime bin at $d_{\rm n}=776.61$), and then increasing to $\zeta=3.5\cdot 10^{-2}\,\mathrm{m}$ at 
$d_{\rm n}=776.62$. This sequence of $\zeta$--values during the master period were then kept for the remaining models (and mirrored at previous rotation periods) while 
testing the sensitivity to other model assumptions and improving the fit for near--dawn bins. A possible mechanism that might explain night--time activation of the solid--state 
greenhouse effect is discussed later.

Specifically, 13 models with $\mu=1$, $\mathrm{CO_2/H_2O}=0.32$, initial $2.1\,\mathrm{mm}$ dust mantle thickness, $\{L_{\rm p},\,r_{\rm p}\}=\{20,\,2\}\,\mathrm{mm}$,  and $\xi=10$ tested 
various values for solid--state heat conductivity ($1.1\cdot 10^{-4}\leq h\leq 1.8\cdot 10^{-3}$), bulk density ($400\leq\rho_{\rm bulk}\leq 600\,\mathrm{kg\,m^{-3}}$),  and 
$\mathrm{CO_2}$ front depth (10--$53\,\mathrm{mm}$). The nightside MM slope steepens with $h$, matching best for $h=4.4\cdot 10^{-4}$. At SMM the slope is fixed, 
though dawn temperatures increase slightly with $h$. Increasing $\rho_{\rm bulk}$ mainly lowers the daytime MM antenna temperature (for which there is no data) and increases 
the late--night SMM antenna temperature (thus favouring $\rho_{\rm bulk}\approx 400\,\mathrm{kg\,m^{-3}}$). Both MM and SMM solutions become too warm at night when the $\mathrm{CO_2}$ 
is deeper than $20\,\mathrm{mm}$ (the effect is evident already at $23$--$27\,\mathrm{mm}$).

Another 8 models tested an alternative illumination boost mechanism, in which the peak flux was not increased but merely maintained for longer, so that the integrated 
flux increased by a factor $3\leq \mathcal{F}\leq 9$. This was done for fixed $\mu=1$, $h=4.4\cdot 10^{-4}$, $\rho_{\rm bulk}=400\,\mathrm{kg\,m^{-3}}$, 
$\{L_{\rm p},\,r_{\rm p}\}=\{20,\,2\}\,\mathrm{mm}$, $\xi=10$, and $2.1\,\mathrm{mm}$ initial dust mantles. Different supervolatile abundances 
($0.32\leq \mathrm{CO_2/H_2O}\leq 0.7$) and initial front depths (9.9--$14.5\,\mathrm{mm}$) were considered, and for some models the near--surface ($\leq 4.7\,\mathrm{mm}$) 
bulk density was lowered to $300\,\mathrm{kg\,m^{-3}}$. Such near--surface density reductions helps lowering late--night MM and SMM antenna temperatures, though 
the effect is mild ($<2\,\mathrm{K}$). A higher $\mathrm{CO_2/H_2O}$ value slows the front propagation speed when considering longer time intervals, but for a given rotation period all these 
models have $3\,\mathrm{mm\,rot^{-1}}$ (the finite spatial grid resolution prevents detection of subtle speed differences). The best results were obtained for $\mathcal{F}=9$.

The last 14 models tested yet another illumination boost mechanism, that elevated the peak flux by a factor $\mathcal{G}=1.1$--$1.5$ and maintained it until the 
integrated flux had increased by a factor $\mathcal{F}=9.0$--$9.9$. This was done for fixed $h=4.4\cdot 10^{-4}$, bulk density ($\rho_{\rm bulk}=300$--$400\,\mathrm{kg\,m^{-3}}$ in 
the top $4.7\,\mathrm{m}$ and $400\,\mathrm{kg\,m^{-3}}$ below), and initial dust mantle thickness ($2.1\,\mathrm{mm}$). These models varied the dust/water--ice abundance 
$0.25\leq \mu\leq 1$, the supervolatiles abundance ($0.32\leq \mathrm{CO_2/H_2O}\leq 0.7$) and initial front depths (9.9--$53\,\mathrm{mm}$), degree of radiative heat 
transport ($\{L_{\rm p},\,r_{\rm p}\}=\{10,\,1\}$,  $\{20,\,2\}$, or $\{40,\,4\}\,\mathrm{mm}$, with $\xi=7.071$, $10$, or $14.14$ to keep the diffusivity fixed). Variations in $\mu$ for a 
given $\rho_{\rm bulk}$ mainly changes the effective specific heat capacity, with little effect on the antenna temperatures. The change in $r_{\rm p}$ considered here adjusts the 
antenna temperatures primarily at the late--night bin, but by less than the $2.5\,\mathrm{K}$ error bars.

The formally best MM model (R10\_095A with $Q_{\rm MM}=0.019$) is shown in the left panel of Fig.~\ref{fig_NIMBUS_MM_SMM_Feb16_bestfit}. It had $\mu=1$, $\mathrm{CO_2/H_2O}=0.32$, a homogeneous bulk 
density $\rho_{\rm bulk}=400\,\mathrm{kg\,m^{-3}}$, the $\mathrm{CO_2}$--front depth varying between $1.7$--$2.3\,\mathrm{cm}$ during the master period, $h=4.4\cdot 10^{-4}$, 
diffusivity regulated by $\{L_{\rm p},\,r_{\rm p}\}=\{20,\,2\}\,\mathrm{mm}$ and $\xi=10$, a solid--state greenhouse effect regulated by $\zeta=5\cdot 10^{-3}\,\mathrm{m}$ at $d_{\rm n}=776.54$--776.61 
and $\zeta=3.5\cdot 10^{-2}\,\mathrm{m}$ at $d_{\rm n}\geq 776.62$, and a late--night illumination boost regulated by $\mathcal{G}=1.1$ and  $\mathcal{F}=9.9$. The solid--state and radiative conductivities 
combine to an effective thermal inertia of $\sim 45\,\mathrm{MKS}$ at day and $\sim 15\,\mathrm{MKS}$ at night. I note, that model R10\_084A (with just a 
slightly lower $Q_{\rm MM}=0.014$) performs somewhat better at the last bin. That is because the $\mathrm{CO_2}$ front is somewhat shallower at that point ($2.0\,\mathrm{cm}$ instead of $2.3\,\mathrm{cm}$; it 
withdraws slower because of a higher supervolatile abundance, $\mathrm{CO_2/H_2O}=0.7$). In addition, that model has a reduced bulk density ($300\,\mathrm{kg\,m^{-3}}$) in the top $5\,\mathrm{mm}$. 
However, that property alone is of little importance (illustrated by R10\_094, that also has reduced near--surface density but is virtually inseparable from R10\_095A). 

The corresponding SMM curves are shown in the right panel of Fig.~\ref{fig_NIMBUS_MM_SMM_Feb16_bestfit}. Although none of the three curves formally fit the data ($Q<4.1\cdot 10^{-6}$) it 
is evident that these curves are much better fits than those in Fig.~\ref{fig_NIMBUS_MM_SMM_Feb16_falsefit}. The key property that enables simultaneous reproduction of the MM and SMM 
2016 February data therefore seems to be the existence of a solid--state greenhouse effect. It is surprising that it only seems to be required at night. If real, this could mean that the 
medium `puffs up' at night, allowing for the self--heating photons from the surroundings to enter near--surface pore spaces that are larger than during day. Potentially, that is caused by 
the $\mathrm{CO_2}$ outgassing that peaks around sunset because of the thermal lag. 

Fig.~\ref{fig_NIMBUS_MM_SMM_Feb16_bestfit} also shows model R10\_096A, with $\mathrm{CO_2}$ moved to a depth of $5.3\,\mathrm{cm}$. It is evident from the high nighttime MM and SMM 
antenna temperatures that the $\mathrm{CO_2}$ front must be substantially shallower than that. However, a shallow front also means that it moves relatively fast, here at $4\pm 1\,\mathrm{mm\,rot^{-1}}$. 
In order to remain near $\sim 2\,\mathrm{cm}$ depth, the nucleus needs to erode at a similar rate. Yet, these models have $\sim 0.2\,\mathrm{mm\,rot^{-1}}$ erosion, which seems necessary to 
allow the slowly sublimating water to withdraw quickly enough to enable dust mantle formation. If the diffusivity is sufficiently reduced at day by some sort of compaction (which would be consistent 
with the apparent switch--off of the solid--state greenhouse effect), that could perhaps slow the $\mathrm{CO_2}$ front rate to match that of erosion, thus keeping the stratification quasi--stationary.

\subsection{June 2016: eleven months after the collapse} \label{sec_results_june16}

The first time period with a relatively large number of close--range MIRO observations \citep[PDS website\footnote{https://pds-smallbodies.astro.umd.edu/holdings/ro-c-miro-3-ext2-\\67p-v3.0/dataset.shtml};][]{hofstadteretal18e} took place in 2016 June, almost a year after the 
collapse. From June 1--26 a total of 79 SMM/MM observation pairs were acquired, during which the \emph{Rosetta}--comet distance 
ranged 15.5--$31.1\,\mathrm{km}$. The data were time--shifted to a master period starting at $d_{\rm n}=900.5316$ (2016 June 18, $12:45:28\,\mathrm{UTC}$) 
and grouped into 15 bins. Only one bin had to be rejected, because the illuminated small lobe intercepted parts of the beams when observing 
the dark Aswan in the background. In this case, all rotational phases were reasonably represented. 

The peak illumination is down from $250\,\mathrm{J\,m^{-2}\,s^{-1}}$ in 2016 February to $140\,\mathrm{J\,m^{-2}\,s^{-1}}$ in 
2016 June. Also, the midnight self--heating flux is down from $\sim 20$ to $10\,\mathrm{J\,m^{-2}\,s^{-1}}$. Yet, the midnight 
temperatures have increased slightly, from $135$ to $138\,\mathrm{K}$ (MM) and from 
$110$ to $120\,\mathrm{K}$ (SMM). The late 2015 to early 2016 trend of raising midnight antenna 
temperature, despite of reduced illumination fluxes, therefore continues in mid--2016. As demonstrated in the following, 
this behaviour is apparently caused by a steadily reduced (but not eliminated) level of $\mathrm{CO_2}$ cooling as the supervolatile  
withdraws further underground.

\begin{figure}
\centering
\begin{tabular}{c}
\scalebox{0.43}{\includegraphics{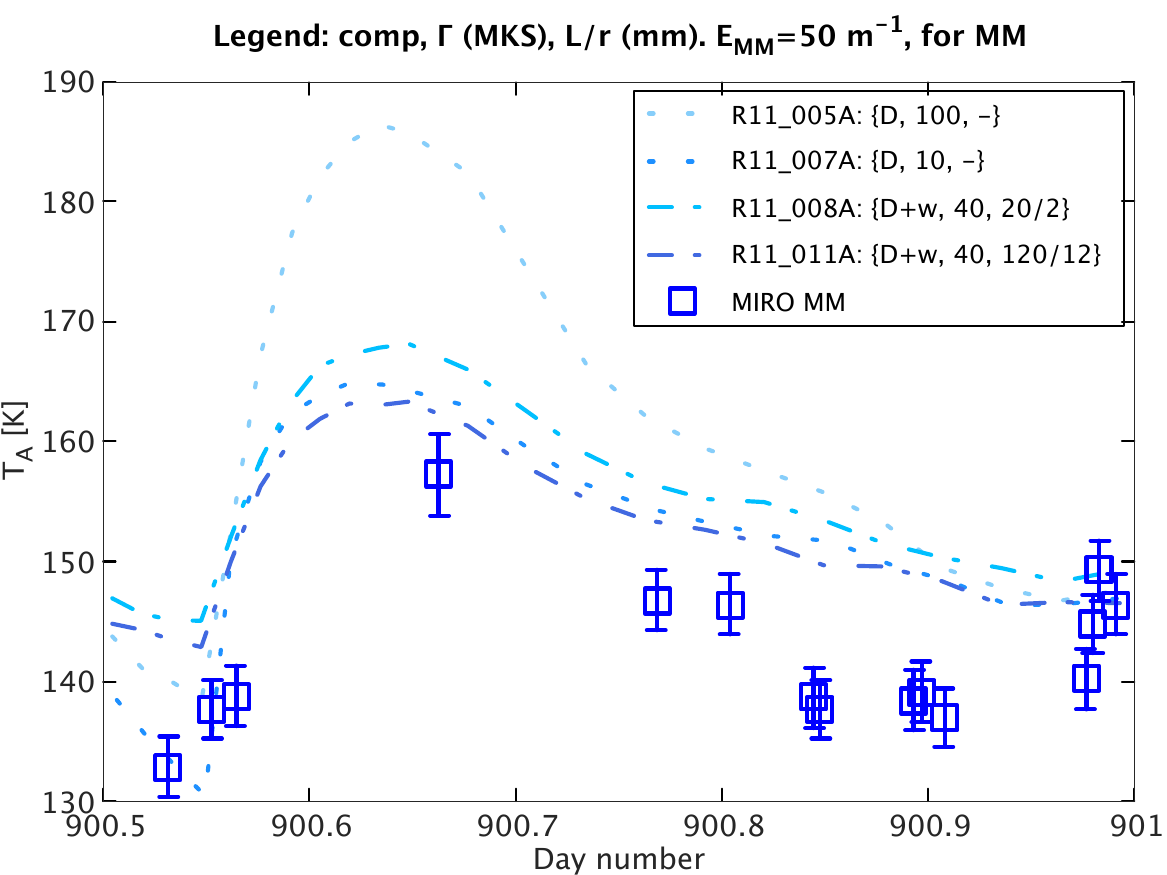}}\\
\end{tabular}
     \caption{2016 June MIRO MM data and \textsc{nimbus/themis} models. Media consisting only of dust or dust$+\mathrm{H_2O}$ are not capable of matching the data.}
     \label{fig_NIMBUS_MM_Jun16_failed}
\end{figure}

\begin{figure*}
\centering
\begin{tabular}{cc}
\scalebox{0.43}{\includegraphics{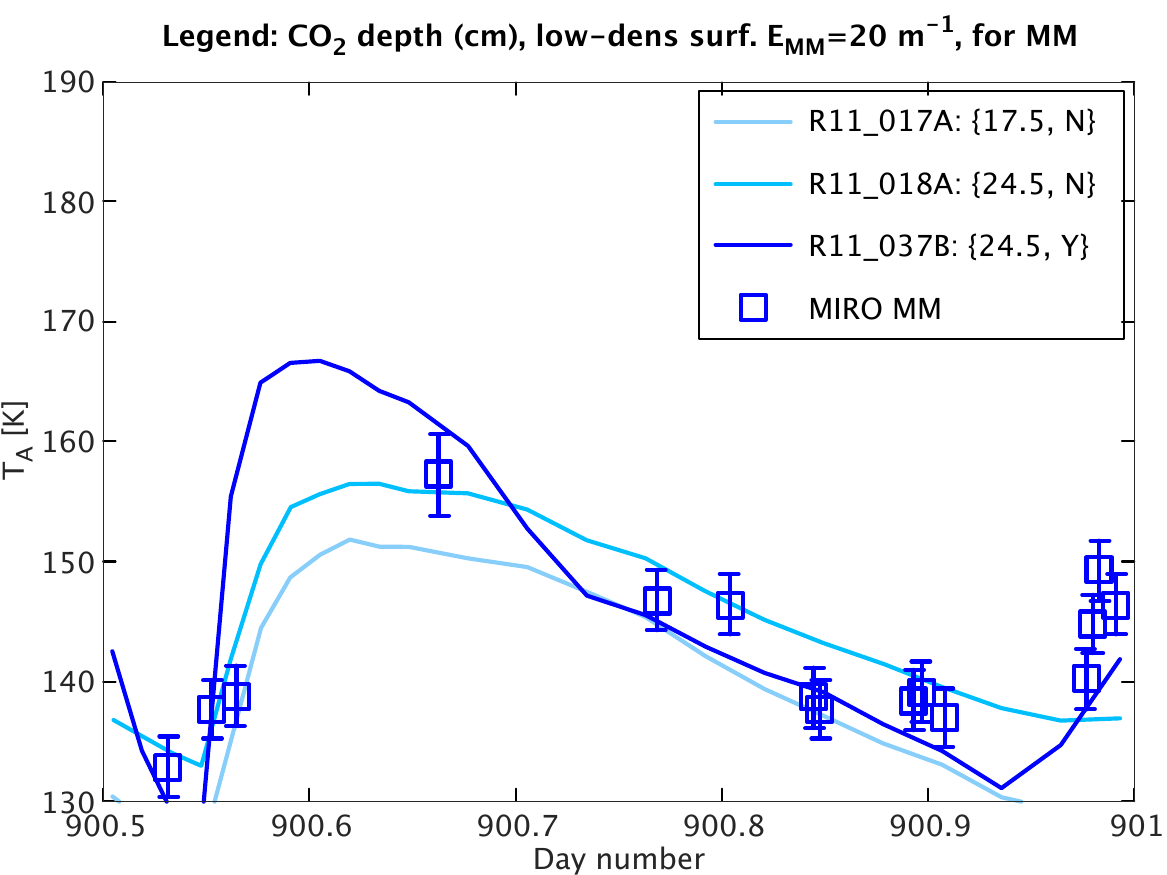}} & \scalebox{0.43}{\includegraphics{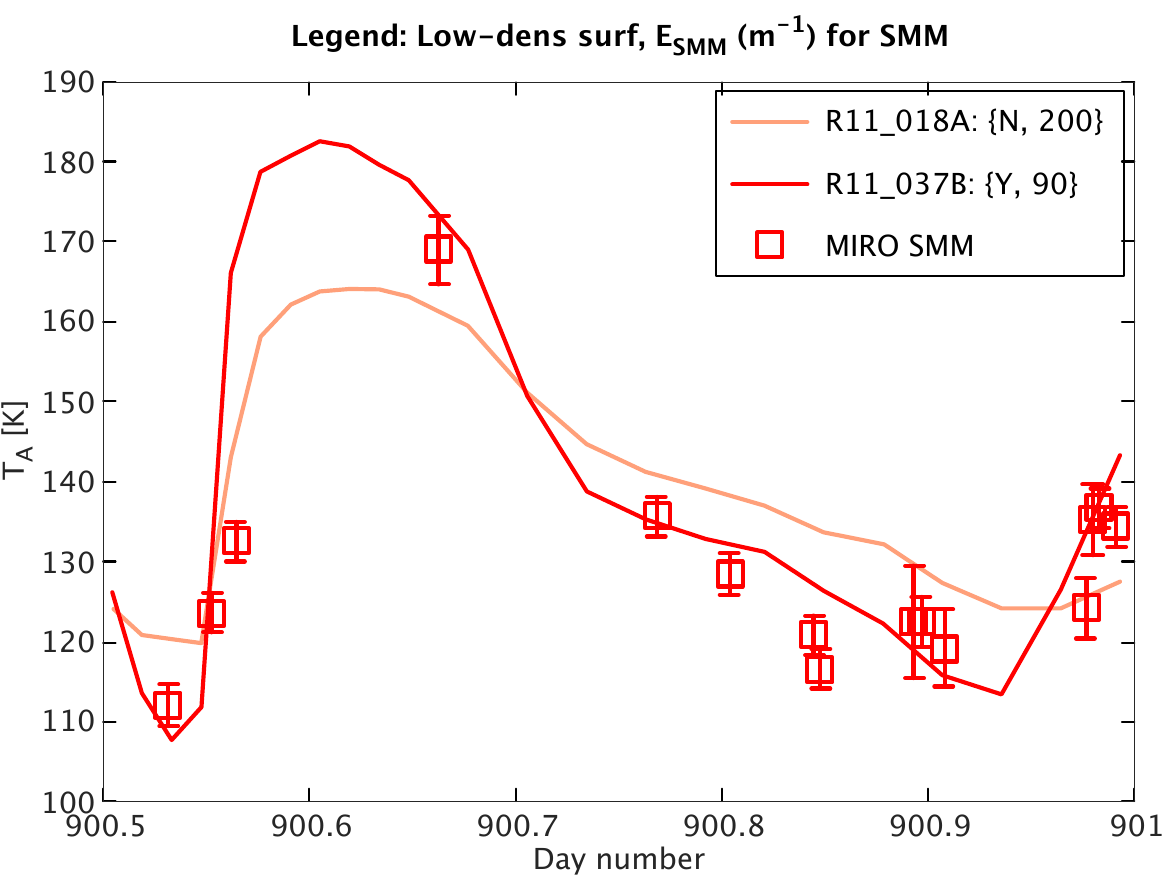}}\\
\end{tabular}
     \caption{2016 June MIRO data and \textsc{nimbus} models. The best achievable model at MM and SMM is R10\_037B with $\mu=1$, $\mathrm{CO_2/H_2O}=0.30$, $0.5\,\mathrm{mm}$ dust 
mantle, the $\mathrm{CO_2}$ front at $24.5\,\mathrm{cm}$, density $73\,\mathrm{kg\,m^{-3}}$ at $\leq 3.1\,\mathrm{cm}$, $390\,\mathrm{kg\,m^{-3}}$ at 3.1--$24.5\,\mathrm{cm}$, 
$535\,\mathrm{kg\,m^{-3}}$ at $\geq 24.5\,\mathrm{cm}$, thermal inertia 10--$15\,\mathrm{MKS}$ on the surface increasing to $190\,\mathrm{MKS}$ at depth, diffusivity from 
$\{L_{\rm p},\,r_{\rm p}\}=\{100,\,10\}\,\mathrm{\mu m}$ and $\xi=1$, and $\times 3$ flux boost at $d_{\rm n}=900.93$--$901$. The other models have $\{L_{\rm p},\,r_{\rm p}\}=\{20,\,2\}\,\mathrm{mm}$ and 
$\xi=1$, different $\mathrm{CO_2}$ depths as indicated in the legend, no flux boost, and yes/no (Y/N) on near--surface density reduction. \emph{Left:} MM data and models for $E_{\rm MM}=20\,\mathrm{m^{-1}}$. 
\emph{Right:} SMM data and models with $E_{\rm SMM}$ 
according to the legend.}
     \label{fig_NIMBUS_MM_SMM_Jun16_fit}
\end{figure*}

Initially, attempts were made to fit the MM data with refractory material only, considering seven \textsc{nimbus} models with 
$\rho_{\rm bulk}=535\,\mathrm{kg\,m^{-3}}$ and thermal inertia in the range $10\leq\Gamma\leq 100\,\mathrm{MKS}$. 
The two extreme examples are shown in Fig.~\ref{fig_NIMBUS_MM_Jun16_failed} as dotted curves, assuming $E_{\rm MM}=50\,\mathrm{m^{-1}}$ 
(extinction $e$--folding scale of $2\,\mathrm{cm}$). Apart for dawn bins, these models are far too warm. Reducing $E_{\rm MM}$ lowers the 
peak temperature slightly, leading to a match for the $\Gamma=10\,\mathrm{MKS}$ model when $E_{\rm MM}=20\,\mathrm{m^{-1}}$ 
(extinction $e$--folding scale of $5\,\mathrm{cm}$, probably on the verge of what is realistic). However, the amplitude is reduced substantially, 
making the antenna temperatures elsewhere far too high. In short, a refractory non--sublimating medium is not capable of reproducing the 2016 June data.

Later, water ice was introduced (applying $\mu=1$) and four \textsc{nimbus} models tested the effect of varying diffusivity when 
water ice was exposed at the surface. Two examples for $E_{\rm MM}=50\,\mathrm{m^{-1}}$ are shown as dashed--dotted curves in 
Fig.~\ref{fig_NIMBUS_MM_Jun16_failed}. Because of partial ice removal in the top few millimetres, the near--surface thermal inertia is 
$\Gamma\approx 40\,\mathrm{MKS}$, growing to $\Gamma\approx 300\,\mathrm{MKS}$ at depth. In such conditions, the antenna 
temperature can be brought near the ice--free $\Gamma=10\,\mathrm{MKS}$ case when $\{L_{\rm p},\,r_{\rm p}\}=\{20,\,2\}\,\mathrm{mm}$. 
Yet, it is not possible to lower the model antenna temperature sufficiently to match the data, even when $\{L_{\rm p},\,r_{\rm p}\}=\{120,\,12\}\,\mathrm{mm}$, 
which would imply centimetre--wide and decimetre--long channels through which water vapour efficiently could escape to space. Tests were made to lower 
the bulk density (hence, the thermal inertia), or replacing the standard \citet{shoshanyetal02} porosity--dependent Hertz factor with a fixed low value (resulting in e.~g. $\Gamma\approx 30\,\mathrm{MKS}$ at 
all depths). This increased the curve amplitudes, but yet did not yield sufficiently cool temperatures at night while being even higher compared to the data at day. Despite optimal 
conditions for net sublimation and cooling, water ice is not capable of lowering the model to match the data.

No option seemed to exist but to include sublimating $\mathrm{CO_2}$ ice once more (using $\mu=1$, $\mathrm{CO_2/H_2O}=0.32$, $\rho_{\rm bulk}=535\,\mathrm{kg\,m^{-3}}$, and $g_{\rm e}=1$). 
A first series of eight \textsc{nimbus} simulations tested the effect of different initial $\mathrm{CO_2}$ front depths, assuming a relatively high diffusivity of 
$\{L_{\rm p},\,r_{\rm p}\}=\{20,\,2\}\,\mathrm{mm}$. Fronts as shallow as $\sim 2\,\mathrm{cm}$, obtained for 2016 February, yield far too cold model curves. 
Figure~\ref{fig_NIMBUS_MM_SMM_Jun16_fit} shows that a model with $\mathrm{CO_2}$ at $17.5\,\mathrm{cm}$ barely reaches the lower end of the data, 
but that a model with the front at $24.6\,\mathrm{cm}$ matches the data fairly well (using $E_{\rm MM}=20\,\mathrm{m^{-1}}$ in both cases to reproduce the curve shape). 
Note that the model does not reproduce the pre--dawn antenna temperature rise, that coincides with a brief glimpse of direct solar illumination enabled by the irregular nucleus topography.

In Section~\ref{sec_results_novdec15_NIMBUS}, I speculated that a relatively low diffusivity of $\{L_{\rm p},\,r_{\rm p}\}=\{100,\,10\}\,\mathrm{\mu m}$ possibly is 
more representative of undisturbed comet material, and that the millimetre--centimetre--sized pores consistent with best--fit diffusivities perhaps are consequences of 
particularly violent erosion caused by near--surface $\mathrm{CO_2}$ ice sublimation. Such effects might be less important in 2016 June, as it seems like the $\mathrm{CO_2}$ 
is deeper and less active. Therefore, a second series of four $\mathrm{CO_2}$--including \textsc{nimbus} simulations were made, assuming $\{L_{\rm p},\,r_{\rm p}\}=\{100,\,10\}\,\mathrm{\mu m}$. 
With the net sublimation rate reduced accordingly, the $\mathrm{CO_2}$ front needs to be located closer to the surface to have the same cooling effect. These simulations showed that a lower--diffusivity 
model with a $\mathrm{CO_2}$ front depth of $14\,\mathrm{cm}$ is similar to the higher--diffusivity model R11\_18A with $24.6\,\mathrm{cm}$ front depth shown in the left panel of Fig.~\ref{fig_NIMBUS_MM_SMM_Jun16_fit}. 
Thus, judging from the MM data alone, they do not allow for a determination of the diffusivity, but the $\mathrm{CO_2}$ front depth is constrained to $20\pm 6\,\mathrm{cm}$.

Model R11\_18A (with a relatively large diffusivity and $\mathrm{CO_2}$ at $\sim 25\,\mathrm{cm}$), that performed well at MM, did not reproduce the SMM data 
convincingly, as shown in the right panel of Fig.~\ref{fig_NIMBUS_MM_SMM_Jun16_fit}. At $E_{\rm SMM}$ the overall level of the curve is about right, but the amplitude is too small. 
That improved when focusing on the lower--diffusivity option (here $\{L_{\rm p},\,r_{\rm p}\}=\{100,\,10\}\,\mathrm{\mu m}$ and $\xi=1$) and using shallower $\mathrm{CO_2}$ fronts. 
Yet, the low nighttime SMM temperatures could not be fully matched. A possible remedy that was envisioned involved a thin surface layer with lower density than elsewhere. The local 
reduction in heat capacity and heat conductivity would tend to increase the temperature amplitude, and if the layer was sufficiently thin it might mostly improve the SMM fit without ruining the available MM fit. 
To test this hypothesis, ten additional \textsc{nimbus} models were considered, having different layer thicknesses and densities. 

The bulk density at depth was $\rho_{\rm bulk}=535\,\mathrm{kg\,m^{-3}}$, reduced to $390\,\mathrm{kg\,m^{-3}}$ in regions that lost all $\mathrm{CO_2}$ 
ice, as in all other models considered for 2016 June. The most promising candidate amongst models with a low--density surface region had the bulk 
density reduced by a factor $\sim 5$ (from 390 to $70\,\mathrm{kg\,m^{-3}}$) in a $2\,\mathrm{cm}$ layer. At that point, three additional models were run to 
understand how much the pre--dawn illumination spike had to be boosted to best fit the data. It turned out that a threefold flux increase is sufficient for the 
model to resemble the data. The SMM antenna temperature of the final model, R18\_037B, is seen in the right panel of Fig.~\ref{fig_NIMBUS_MM_SMM_Jun16_fit}. 
As can be seen, the increased amplitude significantly improves the fit both during day and night (also note the pre--dawn enhancement due to the flux boost). 
The corresponding MM antenna temperature is also plotted in the left panel of Fig.~\ref{fig_NIMBUS_MM_SMM_Jun16_fit}. Compared to R11\_018A, this 
model has a larger amplitude, making it somewhat too warm at day and too cold at night, barely managing to reach the pre--dawn bins. Despite $Q_{\rm MM}$ and 
$Q_{\rm SMM}$ not reaching $\geq 0.01$, R10\_037B is the best achievable match, and is considered acceptable at both MM and SMM. Note that these models do have a 
dust mantle, though it is very thin ($0.5\,\mathrm{mm}$) because of a 1--$2\,\mathrm{mm\,rot^{-1}}$ erosion rate. The preferred solution has 
$\{L_{\rm p},\,r_{\rm p}\}=\{100,\,10\}\,\mathrm{\mu m}$ and $\xi=1$, the $\mathrm{CO_2}$ front at $\sim 25\,\mathrm{cm}$ (a combination enabled by the low--density near--surface region).

\section{Discussion} \label{sec_discussion}

The MIRO measurements combined with the model analysis and interpretation above, provide a first unique insight on how comet 
nucleus stratification forms and develops. Admittedly, data are sparse and the results are potentially model--dependent to some degree. However, it is 
reassuring that solutions for different data sets are mutually consistent and that evolutionary trends appear systematic. For example, all 
three post--collapse data sets require the presence of shallow $\mathrm{CO_2}$ ice and some level of illumination boost when the 
Sun briefly emerges late at night prior to the proper dawn.  The inferred withdrawal of $\mathrm{CO_2}$ ice beneath the surface progresses monotonically 
over time. Furthermore, the independent thermophysical solutions (requirements of exposed water ice in 2015 November and December, but 
of a dust mantle in 2016 February and June) are consistent with changes in albedo over time documented by OSIRIS. The best--fit values of physical 
parameters are plausible, given previous estimates and theoretical expectations. This lends credibility to the results.

The most likely interpretations of the MIRO observations of the Aswan cliff wall can be summarised as follows:
\begin{enumerate}
\item \emph{November 2014 (8 months pre--collapse):} a $\stackrel{>}{_{\sim}}3\,\mathrm{cm}$ dust mantle with $\sim 30\,\mathrm{MKS}$ 
thermal inertia, extinction coefficients $E_{\rm MM}\approx 65$--$70\,\mathrm{m^{-1}}$ and $E_{\rm SMM}\approx 600\,\mathrm{m^{-1}}$ (for 
an assumed mantle density of $330\,\mathrm{kg\,m^{-3}}$), with a submillimetre single--scattering albedo of $w_{\rm SMM}\approx 0.17$--$0.20$.
\item \emph{November -- December 2016 (five months post--collapse):} an exposed dust/water--ice mixture with $\Gamma=25\pm 15\,\mathrm{MKS}$ 
and $\mu=0.9\pm 0.5$, eroding with $1 \stackrel{<}{_{\sim}} g_{\rm e}\stackrel{<}{_{\sim}} 1.5$, overlaying material that includes $\mathrm{CO_2}$ ice with a  
mass fraction $30\pm 5$ per cent (absolute concentration $160\pm 30\,\mathrm{kg\,m^{-3}}$) at a depth of $0.4\pm 0.2\,\mathrm{cm}$, predicted to 
give rise to a coma mixing ratio of $\mathrm{CO_2/H_2O}=1.6\pm 0.4$ (mass) $=0.7\pm 0.2$ (molar), which is twice the applied molar 
$\mathrm{CO_2/H_2O}=0.32$ mixing ratio of the nucleus ice. Models with daytime diffusivity $\{L_{\rm p},\,r_{\rm p}\}=\{1,\,0.1\}\,\mathrm{cm}$, dropping 
temporarily to $\{L_{\rm p},\,r_{\rm p}\}=\{100,\,10\}\,\mathrm{\mu m}$ late at night (always using $\xi=1$), are consistent with the data.  
\item \emph{February 2016 (seven months post--collapse):} a $2\pm 1\,\mathrm{mm}$ dust mantle with $\Gamma\approx 45\,\mathrm{MKS}$ at day and $\Gamma\approx 15\,\mathrm{MKS}$ 
at night, eroding with $g_{\rm e}\approx 0.05$, with the $\mathrm{CO_2}$ front at $2.0\pm 0.3\,\mathrm{cm}$. Models with $\mu=1$, $\mathrm{CO_2/H_2O}=0.32$, a homogeneous 
bulk density $\rho_{\rm bulk}=400\,\mathrm{kg\,m^{-3}}$, diffusivity regulated by $\{L_{\rm p},\,r_{\rm p}\}=\{20,\,2\}\,\mathrm{mm}$ and $\xi=10$ and a nighttime solid--state greenhouse 
effect with $\zeta=3.5\cdot 10^{-2}\,\mathrm{m}$ are consistent with the data. 
\item \emph{June 2016 (eleven months post--collapse): }  a $\sim 0.5\,\mathrm{mm}$ dust mantle with $\Gamma\approx 10$--$15\,\mathrm{MKS}$, a thin, low--density zone ($70\,\mathrm{kg\,m^{-3}}$ in 
the top $\sim 2\,\mathrm{cm}$) overlaying a denser  interior with $390$ or $535\,\mathrm{kg\,m^{-3}}$ above/below the $\mathrm{CO_2}$ front, located at $20\pm 6\,\mathrm{cm}$ (with a preference for 
the largest values). The diffusivity is regulated by $\{L_{\rm p},\,r_{\rm p}\}$ on the range $\{100,\,10\}\,\mathrm{\mu m}$ and $\{20,\,2\}\,\mathrm{mm}$ ($\xi=1$) with a preference for the lower values. 
Models with $\mu=1$, $\mathrm{CO_2/H_2O}=0.32$,  are consistent with the data. 
\end{enumerate}

In the following, six aspects of these findings are discussed: 1) dust mantle formation; 2) the temporal evolution of the $\mathrm{CO_2}$ front depth; 
3) a re--interpretation of the measured albedo; 4) abundances; 5) thermophysical properties of the near--surface material; 6) implications for cryogenic sample--return missions.

\emph{Dust mantle}. Prior to \emph{Rosetta}, no empirical information was available on the time--scale of formation or the thickness of comet nucleus dust mantles. In the absence of erosion, 
thermophysical models suggest rapid dust mantle formation at Aswan (e.~g., formation and growth to $\sim 8\,\mathrm{mm}$ in just one month, see section~\ref{sec_results_novdec15_NIMBUS}). However, 
the Aswan collapse, and the albedo evolution documented by OSIRIS \citep{pajolaetal17}, suggested that the cliff may have needed up to 10--13 months to fully grow back its mantle (see below). 
Unfortunately, only a handful of OSIRIS images show Aswan in the preceding 2016 January -- May timeframe, merely providing highly oblique views of small fractions of the collapse site. 
The current independent MIRO--based results therefore further constrain the dust mantle formation time -- the mantle appears to have been absent in 2015 November and December, but present as of 
2016 February, thus formed 5--7 months post--collapse.  This relatively long (half--year) formation timescale emphasises the importance of erosion, that has delayed the process. In this case, erosion from the 
mantle top is likely driven by drag from outflowing $\mathrm{H_2O}$ and $\mathrm{CO_2}$ vapour, perhaps coupled with some fatigue process that reduces the tensile force between dust grains 
over time, see the \citet{davidssonetal21} discussion of the \citet{ratkekochan89} experiments. Dust mantle growth at its base is necessarily driven by water sublimation. It is not immediately clear 
why the mantle growth rate started to exceed the erosion rate at this particular point in time. However, the peak solar flux fell from $\sim 400\,\mathrm{J\,m^{-2}\,s^{-1}}$ to $\sim 250\,\mathrm{J\,m^{-2}\,s^{-1}}$ around this time 
(see Fig.~\ref{fig_compare_NovDec15_Feb16}), and according to the best fit models the following changes took place in key parameters: 1) peak surface temperature 
$187\rightarrow 236\,\mathrm{K}$; 2) $\mathrm{CO_2}$ diurnal peak pressure $0.26\rightarrow 45\,\mathrm{Pa}$; 3) $\mathrm{CO_2}$ production rate 
$1.8\cdot 10^{21}\rightarrow 1.6\cdot 10^{21}\,\mathrm{molec\,s^{-1}}$; 4) $\mathrm{H_2O}$ diurnal peak pressure $0.02\rightarrow 0.65\,\mathrm{Pa}$; 5) $\mathrm{H_2O}$ production 
rate $2.9\cdot 10^{21}\rightarrow 3.5\cdot 10^{20}\,\mathrm{molec\,s^{-1}}$. Despite the solar flux reduction the peak surface temperature increases (due to the mantle formation), and the pressures 
are strongly elevated (because the model diffusivity dropped a factor 500). Interestingly, the $\mathrm{H_2O}$ and $\mathrm{CO_2}$ production rates are mutually similar (falling due to the smaller 
availability of energy) while $\mathrm{CO_2}$ pressures strongly exceeds those of $\mathrm{H_2O}$ because of their difference in volatility. In case these numbers are close to the actual ones, it 
may suggest that the onset of dust mantle formation is more sensitive to reductions in solar flux and/or production rates, than changes in gas pressures.

Section~\ref{sec_intro} summarises previously established estimates of the dust mantle thickness on 67P. The current paper adds a new mantle thickness estimate of at least $3\,\mathrm{cm}$ 
for the pre--collapse Aswan site. It is also found that the newly formed post--collapse dust mantle remained thin (a few millimetres or fractions thereof) for at least 2--3 months. It shows, for the 
first time, that comet nucleus mantle formation and thickening both are slow processes, and that the thicker mature mantles discussed in section~\ref{sec_intro} likely would have taken months or years to form. This has 
implications for our understanding of activity following events that causes major exposure of deep nucleus material, such as nucleus fragmentation, splitting, and perhaps outbursts.

\emph{The $CO_2$ front depth.} The proposed exposure of $\mathrm{CO_2}$ ice at Aswan does not come as a surprise. A previous $3.8\,\mathrm{m}$ estimate of the average northern 
hemisphere $\mathrm{CO_2}$ front depth by \citet{davidssonetal22} was based on \textsc{nimbus} model reproduction of the global $\mathrm{CO_2}$ production rate curve inferred 
from ROSINA measurements. Additionally, \citet{davidssonetal22b} demonstrated that MIRO measurements are consistent with $\mathrm{CO_2}$ ice at $\sim 0.5\,\mathrm{m}$ depth at a 
specific location in Hapi. The sudden exposure of material at a depth of $12\,\mathrm{m}$ \citep{pajolaetal17} at Aswan, or perhaps merely 6--$10\,\mathrm{m}$ if the post--collapse 
erosion suggested in section~\ref{sec_results_novdec15_NIMBUS} indeed took place, therefore ought to have brought $\mathrm{CO_2}$ ice to the surface. However, it does come as 
a surprise that the $\mathrm{CO_2}$ managed to remain superficial for so long -- at depths of $0.4\pm 0.2\,\mathrm{cm}$ after 5 months, $2.0\pm 0.3\,\mathrm{cm}$ after seven months, 
and $20\pm 6\,\mathrm{cm}$ after 11 months, according to the current investigation. The problem here is not so much the potential burial under airfall material. It should be 
remembered that surfaces with slopes above the angle of repose \citep[$45\pm 5^{\circ}$ for Comet 67P;][]{groussinetal15} do not accumulate airfall, and the Aswan cliff is far steeper. Instead, 
the problem is how surface erosion manages to keep up with the $\mathrm{CO_2}$ front withdrawal during so many months, and to identify the exact mechanism that determines when and 
why the front withdrawal speed suddenly becomes significantly higher than the erosion speed. Is there a simple mechanism related to threshold illumination or outgassing levels that always works 
similarly on all comets, or is the process stochastic and unpredictable (e.~g., related to local variations in tensile strength of the near--surface material)? Hopefully, a better understanding of the 
comet erosion process \citep[currently not well understood due to the `cohesion bottleneck';][]{jewittetal19} can be achieved when aided by the unique empirical input on dust mantle 
net growth and $\mathrm{CO_2}$ front movements on an actual comet nucleus, presented in this paper.

\emph{Re--interpreting the Aswan albedo}. According to \citet{pajolaetal17} the normal albedo at Aswan was $A_{\rm max}>0.4$ on 2015 July 15, it was down to $A_{\rm obs}=0.17\pm 0.02$ on 2015 December 25,  and reduced further 
to $A<0.12$ by 2016 August 6 \citep[note that the normal albedo of the mature dust mantle is $A_{\rm d}=0.07\pm 0.02$;][]{fornasieretal15}. \citet{pajolaetal17} state that the initial high $A_{\rm max}>0.4$ value is due to 
`exposure of pristine material enriched in water ice' and they explain the gradual albedo reduction solely as due to water--ice sublimation. However, according to the \textsc{nimbus} model that reproduces the 2015 November and 
December MIRO data, a pristine dust and water--ice mixture was exposed at night during this time, due to an efficient erosion caused by near--surface $\mathrm{CO_2}$. During day, the surface water ice abundance 
temporarily decreased by a factor $\delta=0.4$, which is what OSIRIS likely would have imaged. One may estimate the fractional surface area coverage of water ice $f_{\rm w}''$ and dust $f_{\rm d}''=1-f_{\rm w}''$ at the 
time of observation, assuming that the two substances are physically separate, from $A_{\rm w}f_{\rm w}''+A_{\rm d}f_{\rm d}''=A_{\rm obs}$. Assuming an albedo $A_{\rm w}=0.9$ of pure water ice grains, this 
yields $f_{\rm w}''=0.12$ and $f_{\rm d}''=0.88$. The pristine mixture of water ice and dust would then have $f_{\rm w}'=f_{\rm w}''/\delta=0.30$ and $f_{\rm d}'=1-f_{\rm w}'=0.70$. The corresponding albedo of 
the pristine mixture would then be $A_{\rm mix}=A_{\rm w}f_{\rm w}'+A_{\rm d}f_{\rm d}'\approx 0.32$. It means that the even higher $A_{\rm max}$,  observed less than a week after the collapse, possibly 
requires an additional exposure of $\mathrm{CO_2}$ ice. Requiring that the final mixture has $A_{\rm sv}f_{\rm sv}+(1-f_{\rm sv})A_{\rm mix}>A_{\rm max}$ yields a surface coverage of the supervolatile (sv) 
of $f_{\rm sv}\geq 0.12$ (if applying $A_{\rm sv}=1$), whereby the surface coverage of the other components modify to $f_{\rm w}\leq 0.27$ and $f_{\rm d}\leq 0.62$. Based on the surface conditions suggested by the 
thermophysical modelling required to match the MIRO data, I therefore postulate that: 1) the highest albedo $A_{\rm max}>0.4$ represents the nucleus internal mixture of dust, $\mathrm{H_2O}$, and $\mathrm{CO_2}$; 
2) that removal of $\mathrm{CO_2}$ would have lowered the albedo to $A_{\rm mix}\approx 0.32$ for water ice and dust; 3) that lower values observed subsequently was due to partial ($A_{\rm obs}\approx 0.17$) or 
full ($A\rightarrow 0.07$) removal of surface water ice.

 \emph{Abundances.}  Reproduction of the MIRO 2015 Nov/Dec data, for models with the most acceptable (lowest) erosion rates suggests $30\pm 5\%$ $\mathrm{CO_2}$ by mass. For 
$\rho_{\rm bulk}=535\,\mathrm{kg\,m^{-3}}$ that means $160\pm 30\,\mathrm{kg\,m^{-3}}$ $\mathrm{CO_2}$. A previous estimate of $\mathrm{CO_2/H_2O}=0.32$ by \citet{davidssonetal22}, 
needed to fit the $\mathrm{CO_2}$ production rate curve inferred from ROSINA measurements, thereby implies $\mu=0.9\pm 0.5$. That value is also consistent with the fit of the $\mathrm{H_2O}$ 
production rate curve \citep{davidssonetal22}. Under what conditions are these values consistent with the previously estimated fractional area coverages (that equal the fractional volumes of the material)? 
By definition $\mu=f_{\rm d}'\rho_{\rm d}/f_{\rm w}'\rho_{\rm w}$, where $\rho_{\rm d}$ and $\rho_{\rm w}$ are the bulk densities of dust and water ice, respectively. Consistency between the proposed 
$\mu\approx 0.9$, $f_{\rm w}'\approx 0.30$ and $f_{\rm d}'\approx 0.70$ therefore requires that $\rho_{\rm d}/\rho_{\rm w}=0.39$. Such a bulk density ratio can be achieved if considering dust as 
a $\psi=0.91$ porosity assemblage of $3250\,\mathrm{kg\,m^{-3}}$ grains (yielding $\rho_{\rm d}=295\,\mathrm{kg\,m^{-3}}$) and water as a $\psi=0.17$ porosity assemblage of $917\,\mathrm{kg\,m^{-3}}$ 
grains (yielding $\rho_{\rm w}=760\,\mathrm{kg\,m^{-3}}$). These specific values (combined with a $\psi=0.17$ porosity for assemblages of $1500\,\mathrm{kg\,m^{-3}}$ grains of $\mathrm{CO_2}$, yielding 
$\rho_{\rm sv}=1245\,\mathrm{kg\,m^{-3}}$) where chosen just to give the total mixture a bulk density of $\rho_{\rm bulk}=540\,\mathrm{kg\,m^{-3}}$, similar to that of Comet 67P. Finally, consistency with 
the area (volume) fractions derived above based on $A_{\rm max}>0.4$ requires that the molar abundance is $\mathrm{CO_2/H_2O}> f_{\rm sv}\rho_{\rm sv}m_{\rm H_2O}/  f_{\rm w}\rho_{\rm w}m_{\rm CO_2}=0.30$. 
This is the case for the applied $\mathrm{CO_2/H_2O}=0.32$. Note, that these examples are not meant to prove that the solutions are correct, but merely to illustrate under what conditions the preferred abundance 
derived from the thermophysical analysis of MIRO measurements would be compatible with the albedo measurements. Such compatibility would be achieved if rather compact icy particles of $\mathrm{H_2O}$ 
and $\mathrm{CO_2}$ are dispersed within a medium of highly porous dust grains.

\emph{Thermophysical parameters.} The nucleus physical parameters that regulate its thermophysical behaviour, and thus cometary activity, are still poorly known. For recent reviews on this 
topic, see e.~g., \citet{groussinetal19} and \citet{guilbertlepoutreetal23}. All attempts to determine such parameters, whether it be through remote sensing \citep[e.~g.,][]{groussinetal07,groussinetal13,davidssonetal13,
davidssonetal22,davidssonetal22b,schloerbetal15,marshalletal18} or \emph{in situ} measurements \citep[e.~g.,][]{spohnetal15} are therefore immensely important. The current paper contributes to these efforts 
by constraining the gas diffusivity, opacity, and the relative importance of solid--state and radiative conduction to the thermal inertia.

\textsc{nimbus} applies a diffusivity $\mathcal{D}$ (units $\mathrm{m^2\,s^{-1}}$) on the format
\begin{equation} \label{eq:01}
\mathcal{D}=\frac{20L_{\rm p}+8L_{\rm p}^2/r_{\rm p}}{20+19L_{\rm p}/r_{\rm p}+3(L_{\rm p}/r_{\rm p})^2}\frac{\psi}{\xi^2}\sqrt{\frac{k_{\rm B}T}{2\upi m_k}}.
\end{equation}
When multiplying $\mathcal{D}$ with the spatial gas density gradient $d\rho/dx$ according to Fick's law, one obtains the mass flux expressed by the Clausing formula \citep[equation 46 in ][]{davidsson21}. 
For simplicity and better physical intuition, this paper reports the lengths $L_{\rm p}$, radii $r_{\rm p}$, and tortuosity $\xi$ of hypothesised tubes applied in various models to calculate the gas mass flux. 
However, here those parameters are used to calculate $\mathcal{D}$ (assuming $\psi=0.7$, $T=150\,\mathrm{K}$, and the $\mathrm{CO_2}$ molecular mass $m_k=7.3\cdot 10^{-26}\,\mathrm{kg}$ for reference), to ease comparison 
with other measurements. For example, KOSI experiments with pores reaching $r_{\rm p}\approx 1\,\mathrm{mm}$ \citep{lammerzahl95} reported diffusivities of $\mathcal{D}\approx 0.1\,\mathrm{m^2\,s^{-1}}$ 
\citep{benkhoffandspohn91}, which would be achieved if applying $L_{\rm p}\approx 10r_{\rm p}$ and $\xi=1$ in equation~\ref{eq:01}.

In their study of 67P, \citet{davidssonetal22} found that the northern hemisphere had $\{L_{\rm p},\,r_{\rm p}\}=\{100,\,10\,\mathrm{\mu m}\}$ and $\xi=1$ inbound, corresponding to $\mathcal{D}=0.001\,\mathrm{m^2\,s^{-1}}$. This drastically 
changed to $\{10,\,1\,\mathrm{cm}\}$ and $\xi=1$, or $\mathcal{D}=0.9\,\mathrm{m^2\,s^{-1}}$ for the same hemisphere outbound, presumably because the addition of coarse (mm--dm) airfall material around 
perihelion, as evidenced by resolved \emph{Philae}/ROLIS and \emph{Rosetta}/OSIRIS imaging \citep{mottolaetal15,pajolaetal17b}. Though the water production from the (airfall--free) south was characterised by 
$\mathcal{D}=0.001\,\mathrm{m^2\,s^{-1}}$ post--perihelion, a much smaller diffusivity of $\mathcal{D}=3\cdot 10^{-6}\,\mathrm{m^2\,s^{-1}}$ was required for the deeper sourced $\mathrm{CO_2}$ 
(resulting from $\{10,\,1\,\mathrm{\mu m}\}$ and $\xi=5$). \citet{davidssonetal22} hypothesised that this was due to sub--surface compression and pore--size reduction caused by the strong 
$\mathrm{CO_2}$ gas pressure gradient near perihelion, a mechanism that was discussed in more depth by \cite{davidssonetal22b}.

The analysis of MIRO observations of a pit--forming region in Hapi by \citet{davidssonetal22b} found a diffusivity corresponding to $\mathcal{D}=0.5\,\mathrm{m^2\,s^{-1}}$ for aged and partially 
devolatilised airfall material, that was abruptly reduced to $\mathcal{D}=10^{-4}\,\mathrm{m^2\,s^{-1}}$ in what appears to be an episode of crumbling and collapse that additionally 
increased the thermal inertia, optical opacity, and microwave extinction coefficients, introduced SMM scattering, and lowered the visual albedo observed by OSIRIS \citep{davidssonetal22c}. Such crumbling was also 
indicated by \citet{davidssonetal22} to explain the drop from $\mathcal{D}=0.9\,\mathrm{m^2\,s^{-1}}$ to $\mathcal{D}=0.001\,\mathrm{m^2\,s^{-1}}$ around aphelion to bridge the post-- 
and pre--perihelion branches. 

The current investigation finds that the daytime diffusivity was $\mathcal{D}=0.1\, \mathrm{m^2\,s^{-1}}$ five months post--collapse ($\{1,\,0.1\,\mathrm{cm}\}$  and $\xi=1$), but that 
it may have dropped to $\mathcal{D}=0.001\,\mathrm{m^2\,s^{-1}}$ at night ($\{100,\,10\,\mathrm{\mu m}\}$  and $\xi=1$). The nighttime value is close to the north--hemisphere inbound average, 
while the daytime value is more reminiscent of the airfall diffusivity, though not as extreme. As mentioned in section~\ref{sec_results_novdec15_NIMBUS}, this could be the result of some 
`roughening' or `puffing up' of the material during the strong $\mathrm{CO_2}$ sublimation at day, while the diffusivity returns to values representative of the deeper and less disturbed material 
at night (when the top $\sim 1\,\mathrm{cm}$ has eroded off). That notion is enforced when considering the other results. Seven months post--collapse (at calmer conditions), the diffusivity is permanently 
$\mathcal{D}=0.002\,\mathrm{m^2\,s^{-1}}$ (based on $\{20,\,2\,\mathrm{mm}\}$ and $\xi=10$), i.e., close to the 2015 November and December nighttime value. The diffusivity is less 
easy to constrain in 2016 February, with $\mathcal{D}=0.001$--$0.2\,\mathrm{m^2\,s^{-1}}$ providing acceptable MM solutions depending on the $\mathrm{CO_2}$ front depth 
($\{100,\,10\,\mathrm{\mu m}\}$ to $\{20,\,2\,\mathrm{mm}\}$, for $\xi=1$), though the SMM additionally gives preference to the smaller value.

Though these values represent the first attempts to estimate comet nucleus diffusivities, the notion currently emerging based on the current work and that of 
\citet{davidssonetal22,davidssonetal22b} is that $\mathcal{D}\approx 0.001\,\mathrm{m^2\,s^{-1}}$ seems to represent the nominal and fairly pristine nucleus 
material. However, the full range of 67P behaviour includes factor $\sim 100$--1000 increases when material is either expanding due to activity or re--assembles through 
deposition, and factor $\sim 100$-1000 decreases due to compaction and compression events. That corresponds to a baseline void space typically being on the 
$\sim 0.1\,\mathrm{mm}$ level, making `excursions' on the full range stretching from $\sim \mathrm{\mu m}$ (typical of constituent monomer grain sizes) to $\sim \mathrm{dm}$ \citep[among the 
largest structural sizes seen among debris in the coma;][]{rotundietal15,davidssonetal15b}, depending on level and type of comet activity. Independent verification of these numbers would be highly desirable.

The \emph{solid--state greenhouse effect} arises when the opacity for incoming radiation is finite, i.~e., is absorbed over some characteristic depth scale $\zeta$. 
With few exceptions \citep[e.~g.][]{komleetal90,davidssonandskorov02b}, thermophysical models of comets have normally assumed $\zeta=0$ (i.~e., all incoming radiation 
is absorbed at the very surface). Previous analysis of MIRO data by \citet{davidssonetal22b} revealed a significant solid--state greenhouse effect in aged airfall debris, suggesting 
$\zeta=2.15\cdot 10^{-2}\,\mathrm{m}$ for a specific location in Hapi in 2014 September. No measurable signatures of such an effect were detected for the same location in 
2014 October, which was interpreted as a consequence of the compaction suggested by the previously mentioned changes in thermal inertia, extinction coefficients, and albedo.

The current investigation found that MM and SMM observations in 2016 February could only be fitted simultaneously if a solid--state 
greenhouse effect was active at night, having $\zeta=3.5\cdot 10^{-2}\,\mathrm{cm}$. This rejuvenates the interest in this mechanism and 
its potential importance in comet nucleus thermophysics. A well--known consequence of $\zeta>0$ is that the temperature maximum moves 
sub--surface, making the surface itself several $10\,\mathrm{K}$ cooler than when $\zeta=0$. In this context, \emph{Rosetta}/VIRTIS observations are 
of special interest and importance, because unlike MIRO, VIRTIS is only sensitive to the surface temperature by operating at a much shorter wavelength. 
\citet{ciarnielloetal23} present histograms of nucleus surface temperatures on the range $205\leq T_{\rm s}\leq 235\,\mathrm{K}$ derived from VIRTIS--M--IR observations 
acquired in 2014 August 15. They are compared to histograms obtained by a `grey model' where local illumination conditions are converted to temperature by assuming 
zero thermal inertia, no sublimation cooling, and no surface roughness. Interestingly, the real nucleus has a substantial excess of cool ($205\leq T_{\rm s}\leq 215\,\mathrm{K}$) terrain 
(presumably corresponding primarily to regions with morning or afternoon illumination), compared to the grey model. A non--zero thermal inertia does lower the mean and 
near--noon temperatures, but significantly increases the temperatures at afternoon, night, and morning. Surface roughness significantly increases the emitted radiance, hence retrieved temperatures. 
Had the model included thermal inertia and roughness, then both effects would have elevated the lowest model temperatures, and the number of cool model pixels would have been reduced even further. 
The comet surface is essentially devoid of water ice \citep{capaccionietal15,filacchioneetal16b}, except in small isolated bright spots \citep{baruccietal16}. Though surface ice would have reduced 
model temperatures, that effect is significantly suppressed when sublimation in taking place under ground. In short, the grey model is already significantly biased towards low temperatures, 
yet large areas of the real nucleus are even colder. Even in the warmer ($215\leq T_{\rm s}\leq 235\,\mathrm{K}$) regions the signatures of radiance--enhancing roughness are described as 
having a `limited contribution'. A possible explanation for this is a reduction of surface temperatures due to the solid--state greenhouse effect. I encourage the VIRTIS 
team to consider this possibility.

The situation in 2014 September 12--15 is quite different, as evidenced by the analysis of VIRTIS--M--IR data by \citet{marshalletal18}. They find that the nucleus radiance 
is `almost always' brighter (thus, temperatures are higher) than models, even when including some thermal inertia. They find that residuals between models and measurements 
are reduced as the thermal inertia is increased up to $\sim 80\,\mathrm{MKS}$. Furthermore, the effect of roughness is described as `significant'. Between 2014 August and 
September, the nucleus therefore appears to shift from having an excess of low surface temperatures to more frequently displaying high surface temperatures, compared to a grey model. If this 
interpretation is correct, it could signify at transition $\zeta\rightarrow 0$, whereby surface temperatures are gradually increasing. This transition is now placed in temporal context 
along with other evidence. The fit of the water production rate requires a substantial drop in diffusivity in northern smooth terrains from $\mathcal{D}\approx 0.9\,\mathrm{m^2\,s^{-1}}$ outbound 
to $\mathcal{D}\approx 0.001\,\mathrm{m^2\,s^{-1}}$ inbound. This process (presumably due to pulverisation of airfall chunks, that simultaneously would suppress the solid--state 
greenhouse effect) must have affected a substantial fraction of northern smooth terrains already in 2014 August, based on the water production curve \citep{davidssonetal22}. However, some high--$\mathcal{D}$ and 
high--$\zeta$ terrain may have remained and been sufficiently common to cause a visible effect on the 2014 August VIRITS--M--IR observations. Such material would mostly be gone 
by 2014 September \citep[according to my interpretation of global VIRTIS--M--IR surveillance; ][]{marshalletal18,ciarnielloetal23}, though at least one location in Hapi observed by 
MIRO went through the transition between 2014 September and October \citep{davidssonetal22b}.

The last thermophysical parameter to be discussed is the thermal inertia. According to the current work, Aswan is characterised by $\Gamma\approx 30\,\mathrm{MKS}$ in 2014 November, 
$\Gamma=25\pm 15\,\mathrm{MKS}$, $\Gamma\approx 45\,\mathrm{MKS}$ at day but $\Gamma\approx 15\,\mathrm{MKS}$ at night in 2016 February, and  $\Gamma\approx 10$--$15\,\mathrm{MKS}$ 
in 2016 June. These are mutually similar, and consistent with previous values inferred from MIRO observations \citep[see, e.~g., the compilation in Fig.~13 by][]{marshalletal18}. Note, however 
that \citet{davidssonetal22b} suggested an increase from $\Gamma=30\pm  10\,\mathrm{MKS}$ to $\Gamma=65$--$110\,\mathrm{MKS}$ at a Hapi location between 2014 September and October. 
The latter value is closer to those derived from VIRTIS and MUPUS observations \citep[again, see Fig.~13 in][]{marshalletal18}. These differences may reflect a real variety in the level of material 
compaction (thus heat conductivity and capacity) at different locations and/or at different times, or partially be model/instrument--dependent. The current results do not indicate drastically different 
conditions between the surface and freshly exposed deeper--seated nucleus material. That suggests that global thermophysical modelling with thermal inertia values derived from surface observations 
can be performed more confidently than previously.

 \emph{ Implications for cryogenic sample return missions.} The purpose of such missions is to retrieve, preserve, and transport ice--rich comet material to Earth, preferably containing 
both super-- and hypervolatiles. Knowing at what depth $\mathrm{CO_2}$ and CO are present is necessary in order to design the retrieval system. Collecting ice from $0.1\,\mathrm{m}$, 
$1\,\mathrm{m}$, or $10\,\mathrm{m}$ depth requires very different types of technical solutions.  The engineering complexity and cost grow quickly with the depth that needs to be reached. 
Not being able to convincingly demonstrate that a proposed system is guaranteed to encounter ice is a potential major weakness that easily leads to mission proposal rejection in extremely 
competitive environments. For such missions to be selected, i.~e., if ice--rich comet samples ever are going to reach Earth laboratories, it is extremely important that as much quantitative 
information as possible are assembled on comet nucleus stratification and composition \emph{prior} to mission proposal composition.

In this context, the current paper provides two important pieces of information: 1) the Aswan collapse exposed both $\mathrm{H_2O}$ and $\mathrm{CO_2}$ ice, probably located $\leq 6$--$10\,\mathrm{m}$ below 
the previous surface; 2) the $\mathrm{CO_2}$ remained to within $\sim 0.2\,\mathrm{m}$ of the surface for at least 11 months after the collapse. The analysis of MIRO observations of exposed 
ice is therefore consistent with the suggestion by \citet{davidssonetal22} that $\mathrm{CO_2}$ ice is typically located $\sim 4\,\mathrm{m}$ below the surface on the northern hemisphere of 67P 
\citep[though pit--forming regions in smooth terrain may have $\mathrm{CO_2}$ ice as shallowly as $\sim 0.5\,\mathrm{m}$;][]{davidssonetal22b}. A spacecraft designed to land on horizontal 
surfaces (with respect to the local gravity field) must therefore be capable of drilling to depths of several meters to encounter $\mathrm{CO_2}$ ice (we still do not know if $\mathrm{CO_2}$ to within 
$\sim 0.5\,\mathrm{m}$  of the surface at pit--forming terrain is typical or an exception). However, if sampling can be done while 
hoovering near a cliff side, such $\mathrm{CO_2}$ ice can be available just centimetres--decimetres within the wall. Beyond the engineering challenges of sampling while hoovering, an 
additional caveat is that the cliff must have experienced a recent collapse. Though large--scaled ($\sim 100\,\mathrm{m}$) collapse may be rare and unpredictable, the significant amount of 
talus observed under cliffs \citep{pajolaetal15} suggests that smaller events may happen rather frequently. Because $\mathrm{CO_2}$ ice may remain near the surface for a year or more 
once it has surfaced, this relative longevity adds positively to the notion of cryogenic sampling at cliffs.

\section{Conclusions} \label{sec_conclusions}

This paper presents \emph{Rosetta}/MIRO observations of the Aswan cliff on Comet 67P/Churyumov--Gerasimenko, that collapsed on 2015 July 10. It describes the analysis 
of the data, using two thermophysical models (\textsc{btm} and \textsc{nimbus}) and a radiative transfer code (\textsc{themis}), that 
allows for the estimate of composition, stratification, and physical properties of the near--surface material, and how they evolved over time. This is done 
both before (2014 November) and after (2015 November/December, 2016 February, and 2016 June) the event. The Aswan cliff collapse 
provides a unique glimpse of the deeper interior of a comet nucleus. It is found that the collapse exposed not only $\mathrm{H_2O}$ 
ice, but $\mathrm{CO_2}$ ice as well. This is the first example of an empirical study that reveals the time--scales and dynamics of dust 
mantle formation and $\mathrm{CO_2}$ sublimation front withdrawal, after exposure of relatively pristine comet material to solar heating. 
The main results are summarised as follows:

\begin{enumerate}
\item The dust/water--ice mass ratio is estimated as $\mu=0.9\pm 0.5$, and the model solutions 
are consistent with a molar $\mathrm{CO_2}$ abundance of $\sim 30$ per cent relative to water.
\item Dust mantle formation was initiated $\sim 7$ months after the collapse, and the mantle remained thin (millimetres or fractions thereof) for 
at least another  $\sim 4$ months. 
\item The $\mathrm{CO_2}$ front remained very shallow ($\leq 0.4$--$2.0\,\mathrm{cm}$) during the first $\sim 7$ months 
after the collapse, and almost one year after the event it was still to within $20\pm 6\,\mathrm{cm}$ of the surface.
\item The gas diffusivity seems to have decreased over time, peaking at $\mathcal{D}=0.1\,\mathrm{m^2\,s^{-1}}$ and falling 
by two orders of magnitude in the following $\sim 6$ months. There are indications that a solid--state greenhouse effect may 
have been active parts of the time. 
\item Collapsed cliffs are promising locations for cryogenic comet nucleus sampling and return of icy material to Earth, as 
they offer availability of water ice and supervolatiles with relatively small requirements on drilling. 
\end{enumerate}

\section*{Acknowledgements}

The author is indebted to Dr. Pedro J. Guti\'{e}rrez from Instituto de Astrof\'{i}sica de Andaluc\'{i}a, Granada, Spain, for invaluable discussions and comments. 
This research was carried out at the Jet Propulsion Laboratory, California Institute of Technology, under a 
contract with the National Aeronautics and Space Administration. The author acknowledges funding from 
NASA grant 80NM0018F0612 awarded by the \emph{Rosetta} Data Analysis Program.\\

\noindent
\emph{COPYRIGHT}.  \textcopyright\,2023. California Institute of Technology. Government sponsorship acknowledged.

\section*{Data Availability}

The data underlying this article will be shared on reasonable request to the corresponding author.

\bibliographystyle{mnras}
\bibliography{/Users/bjorndavidsson/Documents/Latex/mylit}

\bsp	
\label{lastpage}
\end{document}